# Multifield phonon spectrometrics of structured crystals and liquids

39 k Words, 41 figs, 371 refs

Chang Q. Sun[1,2*], Xuexian Yang[3], Yi Sun[4], Yongli Huang[5,*]

Highlight

- Development of regulations for the multifield bond oscillation dynamics and energetics.
- Bond vibration frequencies vary with programmed perturbation of the crystal potentials.
- Frequency shift correlates the perturbation-relaxation-property of a substance.
- Phonon spectrometrics offers ever-unexpected information on the performance of bonds.

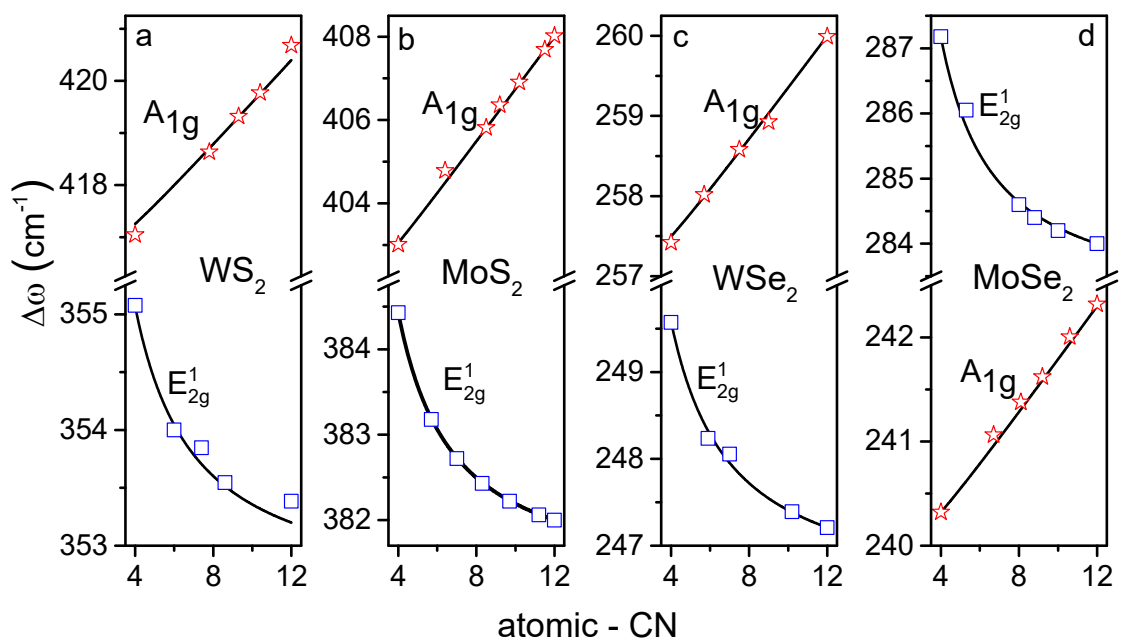

## Key Points

Atomic undercoordination, charge injection, mechanical and thermal activation mediate material's properties intrinsically by bond relaxation from one equilibrium to another while the phonon spectrometrics probes the ever-unexpected information of the binding energy density, atomic cohesive energy, single bond force constant, crystal elastic modulus, Debye temperature and the abundance-length-stiffness transition of bonds under perturbation.

[1] EBEAM, Yangtze Normal University, Chongqing 408100, China (corresponding author: ecqsun@gmail.com)
[2] NOVITAS, School EEE, Nanyang Technological University, Singapore 639798 (ecqsun@ntu.edu.sg)
[3] Key Laboratory of Mineral Cleaner Production and Green Functional Materials, Jishou University, Hunan 416000, China (yangxuexiand@163.com)
[4] Harris School of Public Policy, University of Chicago, Chicago, Illinois 60637, United States (yisun2@uchicago.edu)
[5] School of Materials Science and Engineering, Xiangtan University, Hunan 411105, China (corresponding author: huangyongli@xtu.edu.cn）

## Abstract

Bond relaxation from one equilibrium to another under perturbation matters uniquely the performance of a substance and thus it has enormous impact to materials science and engineering. However, the basic rules for the perturbation-bond-property correlation and efficient probing strategies for high-resolution detection stay yet great challenge. This treatise features recent progress in this regard with focus on the multifield bond oscillation notion and the theory-enabled phonon spectrometrics. From the perspective of Fourier transformation and the Taylor series of the potentials, we correlate the phonon spectral features directly to transition of the characteristic bonds in terms of stiffness (frequency shift), number fraction (integral of the differential spectral peak), structure fluctuation (linewidth), and the macroscopic properties of the substance. A systematic examination of the spectral feature evolution for group IV, III-V, II-VI crystals, layered graphene nanoribbons, black phosphor, (W, Mo)($S_2$, $Se_2$) flakes, typical nanocrystals, and liquid water and aqueous solutions under perturbation has enabled the ever-unexpected information on the perturbation-bond-property regulations. Consistency between predictions and measurements of the crystal size-resolved phonon frequency shift clarifies that atomic dimer oscillation dictates the vibration modes showing blueshift while the collective vibration of oscillators formed between a certain atom and its nearest neighbors governs the modes of redshift when the sample size is reduced. Theoretical matching to the phonon frequency shift due to atomic undercoordination, mechanical and thermal activation, and aqueous charge injection by solvation has been realized. The reproduction of experimental measurements has turned out quantitative information of bond length, bond energy, single bond force constant, binding energy density, vibration mode activation energy, Debye temperature, elastic modulus, and the number and stiffness transition of bonds from the mode of references to the conditioned upon perturbation. Findings prove not only the essentiality of the multifield lattice oscillating dynamics but also the immense power of the phonon spectrometrics in revealing the bond-phonon-property correlation of solid and liquid substance.

# 1. Wonders of Multifield Lattice Oscillation

Highlights

- *Substance evolves its structures and properties when subjecting to perturbation.*
- *Bond relaxation and the associated electronic energetics dictate material's performance.*
- *Correlating and resolving the bond-phonon-property of a substance is a high challenge.*
- *Phonon spectrometrics resolves ever-unexpected information and deepens the physical insight.*

## 1.1. Significance of Multifield Lattice Oscillation

Much attention has been paid to bond formation and dissociation by chemical reaction that revolves the materials properties in an abrupt way [1]. For instance, nitrogenation turns the metallic Gallium into the semiconductive GaN for intense blue light emission [2]. Oxidation transits Zn and Al into the wide bandgap ZnO semiconductors for electronic optical devices and into $Al_2O_3$ insulator for fast thermal energy dissipation [3]. However, bond gradual relaxation has been less attended from one equilibrium to another or from formation to dissociation by external perturbation such as compressing, heating, stretching, atomic undercoordination by defect, nanostructure and surface formation, contamination by charge injection, doping and impurities. Bond relaxation and the associated electronic localization, energetics, entrapment, and polarization of electrons in various energy bands dictate the performance of substance under perturbation [4].

Variation of the size and shape of a crystal has created tremendous fascinations, which has formed the foundations for nanoscience and nanotechnology being recognized as a thrust to the science and technology of the concurrent century and future generations [5, 6]. Nanostructured materials perform differently from their bulk counterparts as the quantities like elastic modulus, dielectric constant, work function, band gap, critical temperatures for phase transition, keep no longer constant but vary with the size and shape of nanostructures. Bond order deficiency shortens and stiffens the bonds between the fewer-coordinated atoms (called confinement in occasions) [7-9]. Atomic undercoordination strengthens the nanocrystals leading to the inverse Hall-Patch effect - hardest at the 10s nanometer scale [7, 10] but depresses or raises the critical temperatures for phase transition [11]. Hetero-coordination may harden the twin grain boundaries [12] by energy densification or soften some other materials at the interfaces by polarization [13].

Multifield lattice oscillation of the sized crystals have received extensive attention [14-16] because the phonon behavior influences directly on the electrical and optical transport dynamics in semiconductors [17, 18], such as electron-phonon coupling, photoabsorption, photoemission and waveguide devices for light transportation. The Raman-active modes of $Bi_2Se_3$ nano-pallets shift a few wavenumbers lower as the thickness is decreased in the vicinity of ~ 15 nm [19], similar to that of the D and 2D modes in the number-of-layer resolved graphene [20, 21]. The LO mode softening has also been observed in a CdS film thinner than 80 nm [22]. The frequency of the LO mode for a 9.6 nm-sized CdSe dot is slightly lower than that of the corresponding CdSe bulk at room temperature. As the CdSe crystal size is reduced to 3.8 nm, the peak frequency shifts to a lower frequency by about 3 cm$^{-1}$ (ref [23]).

Considerable attention has also been paid to the study of bulk compounds due to their intriguing thermal and mechanical properties. Efforts have also made to potential applications in optoelectronic devices such as waveguides, laser frequency folding, high capacity memory, sensors, actuators, etc. [24, 25]. Materials under mechanical and thermal perturbation vary their structures and properties such as phase transition or mechanical hardness [26]. Compression hardens a substance and raises the vibration frequency and the critical pressure for phase transition or regular substance. Heating and stretching have the opposite effect of compression to narrow the bans gap and lower the work function. The volume concentration of nanopores below a certain value can harden the substance but above the critical value causes detrimental to the yield strength of the porous materials [7].

In conjunction to the Raman shift, the bulk modulus *B* or the inverse of expansibility are correlated to the material's performance such as acoustic transmission, Debye temperature, specific heat capacity, and thermal conductivity, which keep constant at the ambient atmospheres. However, the modulus turns to be tunable with the variation of the *T* and *P* [27-30]. Atomistic simulations have revealed that the *B* of a substance is softened under elevated temperature and stiffened under increased pressure [28].

The macroscopic properties of a substance depend functionally on the bond length, bond energy, and valence electron configuration. For instance, the band gap and dielectrics varies with the interatomic bond energy and the electronic occupancy in the conduction and in the valence bands [31]. Likewise, the local bond energy density governs the elastic modulus and the yield strength at failure [32]; the atomic cohesive energy [26] dictates the phase transition temperatures, catalytic activation, interatomic diffusion, etc. The competition of binding energy density and atomic cohesive energy determines the inverse Hall-Petch relationship and the maximal hardness of a crystal at the nanometer scale [33].

The curvature of the bond potentials at equilibrium r = d determines the frequency of vibration ω in the form of $\mu\omega^2 x^2 = [U''(d) + U'''(d)x/3]x^2$ where the nonlinear term can be omitted in the harmonic approximation that is proper at the equilibrium. According to the dimensional analysis, the ω is proportional to the bond length *d* and energy *E* as formulated as $(\Delta\omega)^2 \propto E/(\mu d^2)$ [34] with μ being the reduced mass of the vibrating dimer. Any perturbation will relax the bond and shift the phonon frequency. Therefore, external stimulus changes the material's properties by relaxing its chemical bond and the crystal potentials, which provides one with opportunities to calibrate and control the property change of a substance.

1.2. Outline of Experimental Observations

1.2.1. Size Matter – Atomic Undercoordination

The Raman-active modes and the IR-active modes for the sized and the layered two-dimensional (2-D) structures show various vibration frequency-shift trends [35-39]:

1) The transverse (TO) and the longitudinal optical (LO) vibrational modes shift toward either higher or lower frequencies.
2) The $E_{2g}$ mode for $WX_2$ and $TiO_2$ and the G mode for graphene undergo blueshift as the feature size is reduced.
3) The $A_{1g}$ mode and the D mode shift to lower frequency when the features size decreases.
4) Low-frequency Raman (LFR) acoustic vibration presents at a few or a few tens $cm^{-1}$ wave numbers, or in the THz range (~ 33 $cm^{-1}$), and this mode undergoes a blueshift when the feature size is reduced. The LFR disappears at infinitely large crystal size.

The Raman spectra for (a) $CeO_2$ [40] and (b) Si [41] nanoparticles in Figure 1 show consistently that crystal size reduction softens the phonons, broadens the linewidth, and asymmetrizes the line-shape of the characteristic phonons. The presence of oxygen vacancies or other impurities also affects the spectral line shapes and peak frequencies. The shape of a peak features the intrinsic population and its maximum corresponds to the highest probability. The integration of the population function within a certain frequency range represents for the number of phonons contributing to the spectral features [6]. In place of the peak maximum normalization crossing all the spectra for the same specimen of different sizes, as depicted in Figure 1, the peak area normalization is physically meaningful as this process minimizes the experimental artifacts such as scattering due to surface roughness.

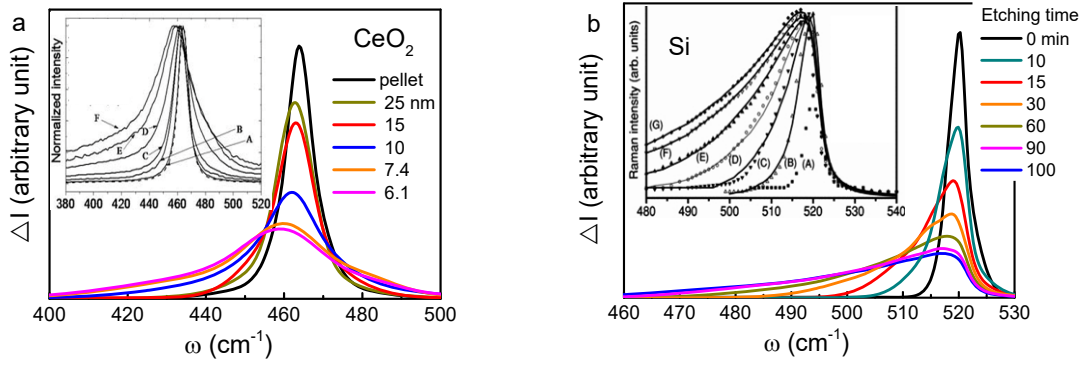

Figure 1. Raman shifts for the size-resolved (a) CeO$_2$ (featured at 465 cm$^{-1}$) [40] and (b) Si (521 cm$^{-1}$) [41] nanoparticles being peak area normalized. Insects compare the widely-used peak maximal intensity normalized spectra. Size reduction softens the phonon stiffness, broadens the linewidth, and asymmetrizes the line shape, being attributed to quantum confinement, nonlinear effect, or inhomogeneous strains (reprinted with permission from [40, 41]).

The Raman shifts are very sensitive to the feature size of the examined substance at the nanometer scale because of the raised ratio of the undercoordinated atoms [36]. For the layered graphene instance [42], the D and 2D bands undergo a redshift but the G band blueshifted from 1582 to 1587 cm$^{-1}$ when its number-of-layer (n) is decreased from 20 to single [21, 43]. The Raman peaks transit from the dominance of the monolayer graphene component to the dominance of the bulk graphite component when the number-of-layer n increases from a few to the bulk value [21]. The opposite trends of number-of-layer suggest that different yet unclear mechanisms govern the G mode and the D/2D modes. One can also estimate the exact number of layers of a graphene [44, 45] and the diameter of a single-walled CNT [46] as the frequency of the radial breathing mode $\omega_{RBM}$ is inversely and empirically proportional to the thickness of the graphene and the CNT diameter.

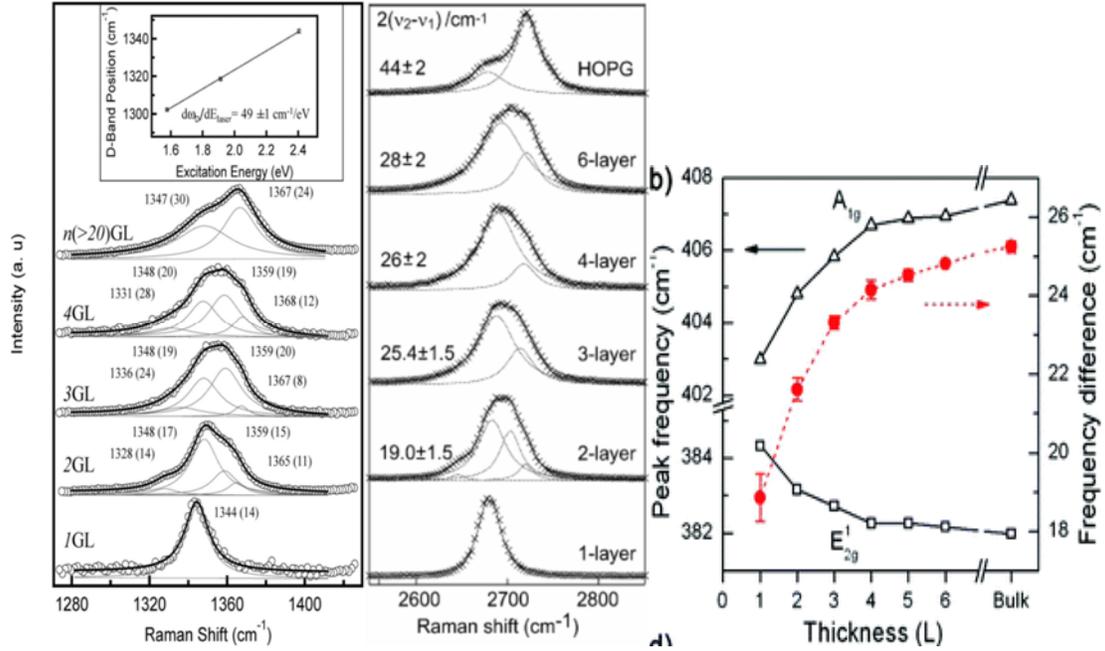

Figure 2. Number-of-layer resolved Raman (a) D mode and (b) 2D band for the few-layered GNR compared with bulk HOPG reference [21], compared with (c) the $E^1_{2g}$ and $A_{1g}$ peak frequency shift of the layered $MoS_2$ films [39]. The 2D peak is centered at 2678 cm$^{-1}$ for the monolayer GNR. Inset (a) shows the D-band $d\omega_D/dE_{ex}$ dispersion as a function of excitation energy of the incident light [20] and the red line in (c) shows the net frequency shift between the two modes (Reprinted with copyright permission from [20, 21, 39]).

The $G$ mode and the $D$ mode in carbon allotropes is conventionally attributed to the resonant excitation of the π states that have the long-range polarizability feature [47, 48]. Raman data on disordered and diamond-like carbon are often categorized as arising from factors that control the peak identities. The Raman spectral features vary with the facilitation the $sp^2$ sites in the $sp^2$-bonded atomic clusters. In cases where a fraction of $sp^3$ hybridization is involved in the $sp^2$ clustering, such as in the hydrogenated amorphous carbon (*a*-C:H) or the tetrahedral-amorphous carbon (*ta*-C) films, we may derive the $sp^3$ fraction from the the Raman spectra. However, the Raman modes and their shifts are directly related to the oscillation of lattices in different geometries under perturbation.

From the dispersion of Raman phonon frequencies and peak intensities with excitation wavelength, Ferrari et al [47, 48] derived the local bonding and structural disorder of graphene. They found three basic features. Under visible light excitation, graphene shows the $D$ mode centered at 1350 cm$^{-1}$ and the $G$ mode at 1600 cm$^{-1}$; however, under UV excitation, an extra $T$ peak appears at 1060 cm$^{-1}$ for the H-free carbons and at 980 cm$^{-1}$ for the hydrogenated ones. The *G* peak shows structural disorder being attributed to the stretching motion of sp$^2$ pairs. This *G* peak disperses only in the amorphous networks. The *D* peak disperses swiftly in the ordered structure but it is weak for the amorphous carbon.

Figure 2 shows the D/2D and the G mode frequency evolution with the GNRs thickness compared with bulk highly oriented pyrolytic graphite (HOPG) reference [20, 21]. When the bulk graphite evolves into the monolayer GNR, the D/2D peaks shift down from 1368/2714 to 134/2678 cm$^{-1}$. In contrast, the G band shifts up when the number-of-layer turns to be fewer [43]. The G-mode blueshift follows the empirical relations [49]: $\omega_G(n) = 1581.6 + 5.5/n$, or $\omega_G(n) = 1581.6 + 11/(1+n^{1.6})$.

Likewise, the layered MX$_2$ (X = S, Se; M = W, Mo) semiconductors show the same trends of phonon frequency relaxation to graphene [50-59]. The E$^1_{2g}$ phonon bandf undergoes a blueshift but the A$_{1g}$ band a redshift as the MoS$_2$ number-of-layer is decreased [39]. Along with the phonon frequency shift, the number-of-layer reduction deepens the surface-potential-well of the MoS$_2$ [60], which evidences the BOLS prediction of the surface bond contraction and local bond potential depression [61].

1.2.2. Compression and Directional Uniaxial-stain

Mechanical compression stiffens Raman phonons in general, as shown in Figure 3 [62], while the uniaxial stretching softens and splits the Raman phonons of graphene and WX$_2$, see Figure 4 [63]. The velocity of phonon stiffening varies with not only the bond nature of the substance but also the specific mode of the same material.

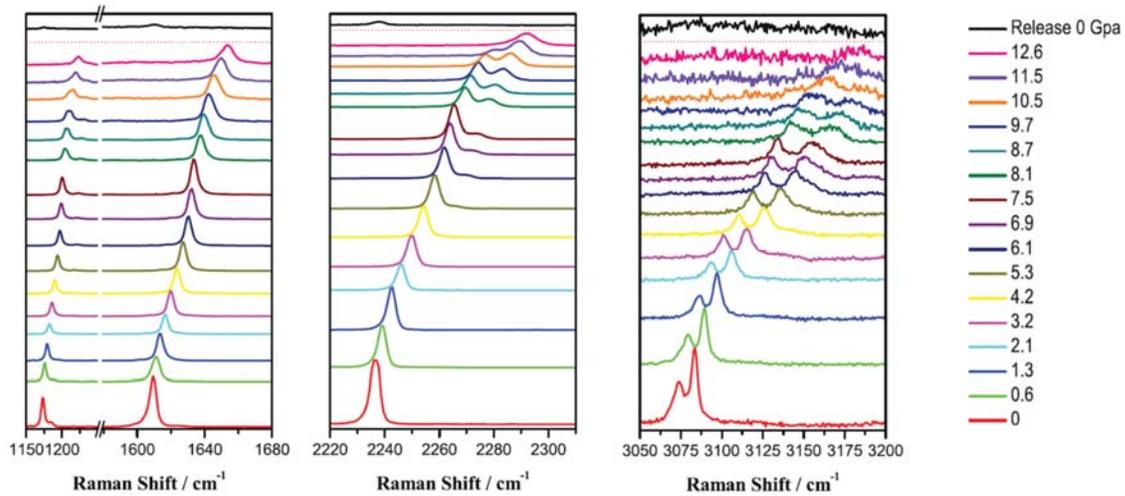

Figure 3. Pressure dependence of the Raman phonon shift for a TPN (Terephthalonitrile) crystal. (Reprinted with permission from [62])

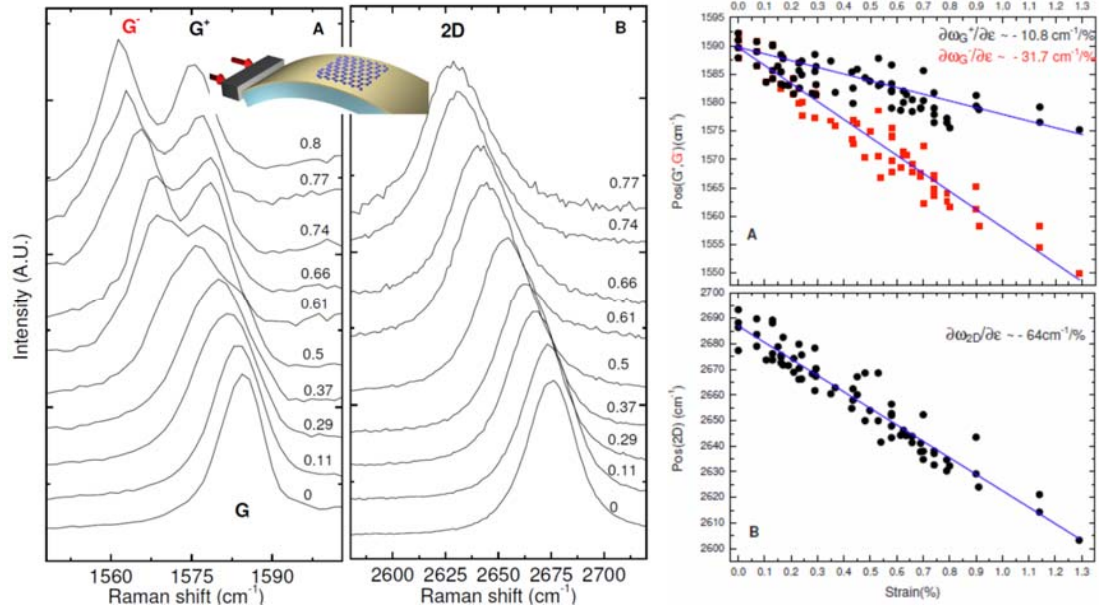

Figure 4. Uniaxial strain softened Raman (a) G mode and (b) 2D mode of graphene collected using polarized light. The G peak splits into the $G^+$ and the $G^-$ subbands. The strain-induced phonon softening is approximated in terms of different Grüneisen constants, $\partial\omega/\partial\varepsilon$ (reprinted with permission from [63, 64]).

Mohiuddin and co-workers [63] suggested that a uniaxial stretching strain can degenerate the $E_{2g}$ or the G optical mode into two components: one is perpendicular to the strain and the other is polarized along the strain. Increasing the strain further softens and separates the $G^+$ and the $G^-$ peaks, in agreement with first-principles calculations. Small strains also soften the 2D bands but do not split them.

Figure 5 shows compression induced phonon shift described with a linear $\gamma = \partial\omega/\partial P$ Grüneisen parameter but at higher pressure, greater than 1.5 GPa, the measurement curve deviates much from the γ constqant. Meanwhile, atomic undercoordination and mechanical compression enhance each other on the Raman blueshift of graphene [65]. The monolayer graphene shifts most, and the graphite shifts least their G mode compared with other layered graphene samples under the same pressure. The γ =

∂ω/∂P deviates from its linear form to the measured nonlinear form as the graphite turns into the monolayer graphene.

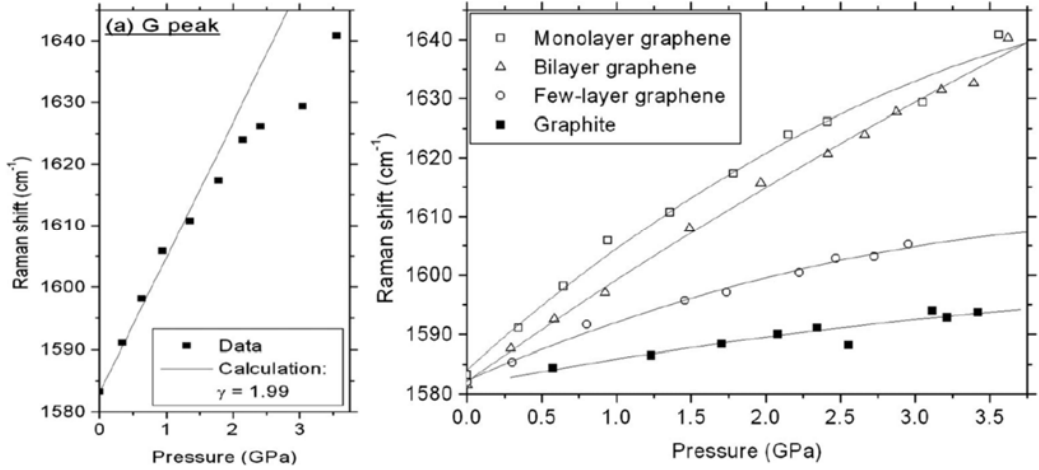

Figure 5. Pressure-resolved (a) G mode frequency shift approximated in γ = ∂ω/∂P Grüneisen constant at low pressures (b) for the monolayer, bilayer, few-layer and bulk graphite (reprinted with permission from [65]).

1.2.3. Debye Thermal Decay

Figure 6 shows the thermal evolution of the frequency shift (Δω) and the fluctuation order (full-width-at-half-maximum, FWHM or $\Gamma$) of the Raman characteristic phonons for GeSe [66]. The $\Gamma$ describes the structure thermal fluctuation and the Δω is the stiffness standing for the bond stretching vibration. The thermal fluctuation does not contribute to the systems energy on average, but the fluctuation is associated with the phonon thermal softening – Debye thermal decay.

The Raman phonon frequency thermal decay follows the general trend displayed by GeSe, diamond [67-70], and GaN [71] with or without involvement of interfaces or impurities, and even the AlN and InN vibration modes in the InAlN alloy [72] albeit different high-temperature Grüneisen slopes. Within the temperature range of −190 and 100 °C, the G peak position and its temperature coefficient of graphene shows the number-of-layer resolved manner. The Δω transits from 1582 to 1580 cm$^{-1}$ and the linear slope χ shifts from −0.016 to −0.015 cm$^{-1}$/°C when the graphene turns from single- to bilayer under 488 nm laser excitation [45].

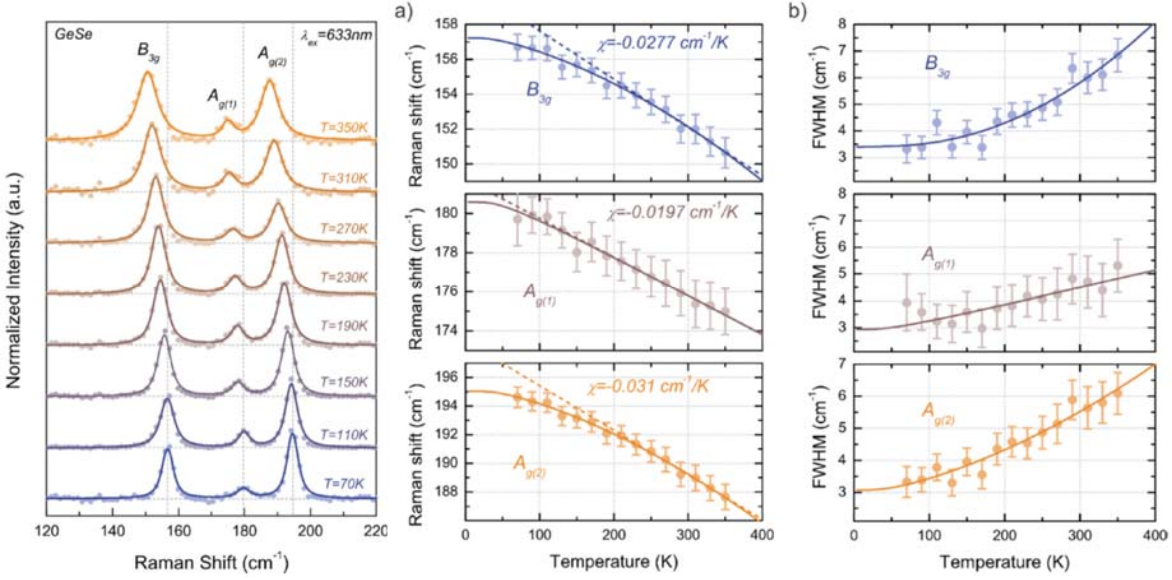

Figure 6. Debye thermal decay of the mode-resolved (a) Raman spectra, (b) peak frequency, and (c) $\Gamma$ for GeSe thin flakes measured using $\lambda$ = 633 nm light excitation. The peak shift is approximated with the Grüneisen constant, $\chi = \partial\omega/\partial T$, at high temperatures (Reprinted with permission from [66]).

These experimental observations show the following common features of the Debye thermal decay:

1) Heating softens the $\Delta\omega$ and broadens the $\Gamma$ without exception.
2) The $\Delta\omega$ and $\Gamma$ change slowly at low temperatures and then transit into a linear slope at higher temperatures toward melting.
3) The slope $\chi$ of the linearly thermal decay and the transition temperature vary from mode to mode for a specific specimen.
4) The overall temperature trends of the $\omega(T)$ follows the $\omega(T)/\omega(T_0) = 1 - U(T)$ with $U(T)$ being thermal integration of the Debye specific heat curve.

1.3. Overview on Theoretical Progress
1.3.1. Quantum Size Trends
1.3.1.1. Empirical Formulation

Compared with the huge volume of Raman database, the physical insight into the size dependent Raman shift is limited. Interpretations are often based on the continuum dielectric mechanism [73, 74], microscopic lattice dynamics [38, 75], multi-phonon resonant scattering [36], strain or phonon confinement [40-42]. Dynamic information could not be available on the bond formation, bond dissociation, bond relaxation and bond vibration under given perturbations. Particularly, for the size-reduction induced phonon frequency blueshift, redshift, and LFR creation are beyond capacity of available theories.

One often describes the size-dependent Raman shifts in the following hypothetic relation with multiple freely adjustable parameters [35, 38],

$$\omega(K) = \omega(\infty) + A_f (d_0/K)^\kappa ,$$

where $A_f$ and $\kappa$ are adjustables. The $d_0$ is the bond length that contracts with the solid dimension or size [76]. $K$ is the dimensionless size of the crystal. For the LFR in the THz frequencies, $A_f > 0$, $\kappa = 1$. When the particle size approaches infinity, the LFR mode disappears, $\omega(\infty) = 0$. For the redshift, $A_f < 0$. For Si example, $\omega(\infty) = 520$ cm$^{-1}$, $\kappa$ varies from 1.08 to = 1.44, depending on experimental conditions.

Turning the number-of-layer from a monolayer to the bulk can soften the phonon frequency ω and lower the photon energy or the band gap $E_G$ of a semicopndctor [77, 78]. The photon energy $E_G$ and phonon frequency shift ωunder perturbation of size N, pressure P, and temperature T follow the hypothetic empirical relationships with numerous adjustable parameters [25, 35, 79, 80]:

$$\left. \begin{array}{l} \omega(N) - \omega_0 \\ E_G(N) - E_0 \\ \omega(T) - \omega_0 \\ \omega(P) - \omega_0 \end{array} \right\} = \begin{cases} -d(a/N)^q \\ A N^{-2} - B N^{-1} - C \\ \Delta\omega_e(T) + \Delta\omega_d(T) \\ kP + lP^2 \end{cases}$$

(1)

where $\omega_0$ is the reference frequency and $E_0$ the photon energy for the bulk. $N$ is the number of atomic layers. The $a$ is the lattice constant. A, B, C, and d, q, k, and l are hypothetic fitting parameters. The terms of $N^{-1}$ stands for the potential energy and the $N^{-2}$ for the exciton (electron−hole pair) kinetic energy in the quantum confinement scheme. These hypothetic models can reproduce observations albeit the unclear physical meanings.

### 1.3.1.2. Hwang's Scheme

Hwang and co-workers [81] presented a sophisticated theory describing the size effect on the Raman shift of the optical and the LFR mode. The LFR is attributed to the quadruple vibration, lattice contraction, and the size-induced optical softening due to quantum confinement. The LFR mode at the THz regime is associated with the dimer oscillation between nanoparticles. The phonon frequencies are crystal size and host-matrix dependent. The LFR for $Al_2O_3$ [82] and $SiO_2$ [83] embedded Ag nanoclusters was attributed to being arisen from the quadrupolar vibration modes. The surface plasmas of the encapsulated Ag particles enhanced quadrupolar modes. The stronger plasmon-phonon coupling was deemed to stem the LFR scattering.

One can also ascribe the polarized LFR peaks as the confined LA-like and the depolarized LFR as the TA acoustic phonons [84]. The interaction between the support matrix and the nanoparticle softens the depolarized and polarized LFR modes.

Hwang et al also suggested that the size-reduction induced lattice contraction stems the LFR blueshift. For example, $B_2O_3$-$SiO_2$ embedded $CdS_xSe1_{-x}$ nanocrystals are subject to the size-induced compressive strain [85]. The surface tension increases with bond strain. It is suggested that the compressive stress overcomes the redshift of the confined phonon due to negative dispersion and thus drives the LFR blue shift. The LFR blueshift was also related to the bond relaxation to the latent heat of fusion and entropy in the classical thermodynamic manner [86].

The high-frequency optical modes shift in opposite directions. The Raman blueshift is usually suggested to be activated by surface disorder [87], surface stress [88, 89], surface chemical and phonon quantum confinement [90, 91]. $TiO_2$ particles' Raman shifts were attributed to the effects of particle size reduction. Size reduction varies the force constants and amplitudes of the vicinal bonds [92]. In contrast, the stress effect is insignificant for the hydrogenated [93, 94].

Hwang et al [95] deals with phonon redshifts of nanostructured CdSe embedded by considering the effect of bond contraction,

$$\omega(K) = \omega_L + \Delta\omega_D(K) + \Delta\omega_C(K)$$

(1)

where $\omega_L$ is the bulk value of the LO phonon frequency. $\Delta\omega_D(K)$ and $\Delta\omega_C(K)$ are the frequency shifts due to dispersion and lattice contraction, respectively. The $\Delta\omega_D(K)$ follows,

$$\Delta\omega_D(K) = \left[\omega_L^2 - \beta_L^2\left(\frac{\mu_{np}}{Kd_0}\right)^2\right]^{1/2} - \omega_L \cong -\left(\frac{\beta_L^2}{2\omega_L}\right)\left(\frac{\mu_{np}}{Kd_0}\right)^2$$

(2)

where the $\beta_L$ defines the dispersivity. The $\mu_{np}$ is the $n_p \neq 0$ root of the $\tan(\mu_{np}) = \mu_{np}$ equation. The $\Delta\omega_C(K)$ is formulated as [85]:

$$\Delta\omega_C(K) = \omega_L\left[\left(1 + \frac{3\Delta d(K)}{d}\right)^{-\gamma} - 1\right] \cong -3\gamma\omega_L\frac{\Delta d(K)}{d}$$

where,

$$\begin{aligned}\frac{\Delta d(K)}{d} &= (\alpha' - \alpha)(T - T_g) - \frac{2\beta_c}{3}\left(\frac{\sigma_\infty}{Kd_0} + \frac{b}{2(Kd_0)^2}\right) \\ &\cong (\alpha' - \alpha)(T - T_g) - \frac{\beta_c b}{3(Kd_0)^2}\end{aligned}$$

(3)

The $\alpha$ and $\alpha'$ are the nanocrystal thermal-expansion coefficient and the embedding matrix, respectively. $\gamma$ represents the Grüneisen parameter. $T_g$ *is* the annealing temperature and $T$ the testing temperature. The $\beta_c$ *is* the compressibility and the $\sigma_\infty$ is the bulk value of surface tension. The *b* describes the size-effect on surface tension that contributes to the frequency shift insignificantly. The first term in eq (3) describes the crystal-matric interface thermal-mismatching. The second term features the crystal size reduction enhancement of the surface tension. Combining eq (3) and (2), yields,

$$\begin{aligned}\frac{\Delta\omega(K)}{\omega_L} &= -3\gamma(\alpha' - \alpha)(T - T_g) - \left[\frac{1}{2}\left(\frac{\beta_L\mu_{n_p}}{\omega_L}\right)^2 - \gamma\beta_C b\right](Kd_0)^{-2} \\ &= A - BK^{-2}\end{aligned}$$

(4)

For a free surface, $\alpha' = \alpha$, and $b = 0$. As remarked on by Hwang et al [95], limitations exist to use this formulation, because thermal-expansion coefficient in the $T - T_g$ range is hard to detect. The *B* value in eq (4) is determined by the phonon dispersion and the surface tension. Therefore, a *B* >0 value suggests that the negative phonon dispersion overtones the size-effect on the surface tension to causes a phonon frequency redshift. On the contrary, phonon frequency blueshift will occur. If the two effect are comparable, $B = 0$, the size dependence disappears.

1.3.2. Grüneisen Notion for Compression and Thermal Excitation

Besides the hypothetically polynomial expressions [96, 97], one often uses the Grüneisen parameter, $\gamma = -\partial\omega/\partial\varepsilon$ or $\gamma_E = -\partial Ln\omega/\partial Ln\varepsilon$ to describe the effect of strain or compression on the phonon frequency shift. The Grüneisen parameter is the slope of the experimental ω-ε curve. The following addresses the $E_{2g}$ mode shift of graphene in terms of the Grüneisen parameter and the shear deformation potential $\beta_{E2g}$ [63, 65]:

$$\gamma_{E2g} = -\frac{1}{\omega_{E2g}^0}\frac{\partial \omega_{E2g}^h}{\partial \varepsilon_h}; \quad \beta_{E2g} = \frac{1}{\omega_{E2g}^0}\frac{\partial \omega_{E2g}^s}{\partial \varepsilon_s}$$

(5)

where $\varepsilon_h = \varepsilon_{ll} + \varepsilon_{tt}$ is the hydrostatic component of the applied uniaxial strain and $\varepsilon_s = \varepsilon_{ll} - \varepsilon_{tt}$ is the shear component of the strain, $l$ is along the strain direction, and $t$ is the transverse direction; $\omega_{E2g}^0$ is the referential G peak position under zero strain.

On the other hand, the thermal expansion coefficient α(t) varies non-linearly in the low temperatures and then increases with temperature toward a constant. The α(t) also varies with the feature size of nanostructures [98, 99]. X-ray absorption fine structure spectroscopy (XAS) investigations [100] suggested that within the 20-300 K range the first Au-Au neighbor distance in the 2.4-50.0 nm sized gold nanoparticles is different from that of the bulk counterpart. Atomic undercoordination has an opposite effect to compensate for thermal expansion on the lattice constant [9].

According to Cardona [101], the bond thermal expansion could be formiulated in terms of the lattice vibration frequency $\omega_q$ and the Grüneisen parameter, $\gamma_q$,

$$\frac{\Delta d}{d_0} = \alpha T = \frac{\hbar}{3BV}\sum_q \gamma \omega_q \left[ n_B(\omega_q) + \frac{1}{2} \right]$$

$$\propto \begin{cases} \dfrac{2kT}{BV_C}<\gamma_q> & (T > \theta_D) \\ \int_0^{\omega_D} <\gamma_q> \omega^3 \left\{ \left[\exp(\hbar\omega/kT)-1\right]^{-1} + 1/2 \right\} d\omega & (else) \end{cases}$$

$$\gamma_q = -\frac{\partial Ln(\omega_q)}{\partial Ln(V)}$$

(6)

where $V$ is the volume and $B$ is the bulk modulus. The $n_B(\omega_q)$ is the Bose-Einstein population function. $V_C$ is the unit cell volume and the $<\gamma_q>$ is an average of $\gamma_q$. Grüneisen [102] derived the volumetric thermal expansion coefficient as a function of the specific heat $C_v$, $\gamma$, $V$, and $B$,

$$\alpha = \frac{\gamma C_v}{V B_T}$$

(7)

One may note that the bandgap, elastic modulus, and the phonon frequency follow the same trend of Debye thermal decay. The following equation empirically describes the thermal evolution of the photonic bandgap $E_g$ [80, 103], with $\beta$ being a fitting parameter $\theta_D$ the Debye temperature.

$$E_g(T) = E_{g0} - \frac{\beta T^2}{T + \theta_D}$$

(8)

Typically, the thermal evolution of the elastic modulus $Y$ or the bulk modulus $B$ follows the empirical relationships [104, 105],

$$Y = Y_0 - b_1 T \exp(-T_0/T),$$

$$Y = Y_0 - \frac{3R\gamma\delta T}{V_0} H\left(\frac{T}{\theta_D}\right)$$

$$B_T^0 = B_{T=0}^0 \times \exp\left[\int_{T=0}^T \alpha_V^0(T)\delta^0(T)dT\right]$$

where, $H\left(\dfrac{T}{\theta_D}\right) = 3\left(\dfrac{T}{\theta_D}\right)^3 \int_0^{\theta_D/T} \dfrac{x^3 dx}{e^x - 1}$

(9)

where $Y_0$ is the Young's modulus and $B^0_{T=0}$ the bulk modulus at $T = 0$ K. The $B$ is correlated to $Y$ by $Y/B = 3 \times (1 - 2\nu) \approx 3B$, as the Poisson ratio $\nu$ is negligible. The parameters $b_1$ and $T_0$ are arbitrary constants for data fitting. $\gamma$ is Grüneisen parameter and $\delta$ is Anderson constant. The $\alpha_V^0$ and $\delta^0(T)$ are respectively the volume thermal expansion coefficient, the Anderson-Grüneisen parameter.

Numerically, the first expression could fit the linear part and the last two could reproduce measurements over the entire temperature range despite the freely adjustable parameters such as the $\gamma$ and $\delta$ that are hardly experimentally available. The physical origin for the thermally driven elastic softening is still open for exploration.

1.3.3. Phonon Optical-Acoustic Thermal Degeneration

The thermal evolution of the $\Gamma$ and $\Delta\omega$ are often attributed to the anharmonic phonon–phonon interactions, lattice-mismatch, volume thermal expansion, and the optical phonon degeneration into multiple acoustic phonons [66, 106]. On the base of the extended Klemens–Hart–Aggarwal–Lax premise [107, 108], Balkanski et al described the $\Delta\Gamma(T)$ and $\Delta\omega(T)$ in the following forms [109]:

$$\begin{pmatrix} \Delta\omega(T) \\ \Delta\Gamma(T) \end{pmatrix} = \begin{pmatrix} A & B \\ C & D \end{pmatrix} \begin{pmatrix} 1 + \dfrac{2}{e^{x/2} - 1} \\ 1 + \dfrac{3}{e^{x/3} - 1} + \dfrac{3}{\left(e^{x/3} - 1\right)^2} \end{pmatrix}$$

(10)

where $x = \hbar\omega_0/k_B T$ with $\hbar$ being the Plank constant and $k_B$ the Boltzmann constant. The $\hbar\omega_0$ is the phonon energy at $T = 0$ K from which the $\Delta\Gamma(T)$ and $\Delta\omega(T)$ shift. $A$, $B$, $C$ and $D$ are adjustable parameters. This notion attributes the $\Delta\Gamma(T)$ and $\Delta\omega(T)$ to the cubic and quartic anharmonicity of lattice potential. This approach turns the optical branch into two (A) or three (B) components of acoustic branches.

Comparatively, Kolesov considered an alternative on the process of phonon excitation in a $\Delta\omega(T)$ function of anharmonic vibration of the chemical bonds [110]:

$$\begin{cases} \omega(T) = \omega_0 + c\langle n \rangle + c_i \langle n_i \rangle \\ \langle n \rangle = \dfrac{1}{e^x - 1} \end{cases}$$

(11)

where $\omega_i < \omega$. The $\langle n_i \rangle$ is the population function of $\omega_i$ selected from the thermal bath. The $c$ and $c_i$ are adjustable weighting factors.

From the perspective of continuum thermodynamics, Calizo et al [111] suggested that both the thermal anharmonicity and the thermal expansion determine the Raman frequency shift in the following experimental-based expression,

$$\Delta\omega \equiv (\chi_T + \chi_V)\Delta T = \left(\frac{d\omega}{dT}\right)_V \Delta T + \left(\frac{d\omega}{dV}\right)_T \Delta T = \left(\frac{d\omega}{dT}\right)_V \Delta T + \left(\frac{d\omega}{dV}\right)_T \left(\frac{d\omega}{dT}\right)_V \Delta T$$

(12)

where $\chi_T$ is self-energy shift due to the phonon modes coupling and $\chi_V$ is the volume change arising from thermal expansion.

Although eqs (10)-(12) reproduce equally well the $\Delta\omega(T)$ for diamond and silicon, the physical indication of the adjustable weighting factors is unclear. Eq (12) correlates the $\Delta\omega(T)$ directly to the $P$, $V$, and $T$ with the respective slopes of the measurements.

1.4 Motivation and Objectives

Phonon spectroscopy is a widely-used tool of detection but the conventional approaches of spectral peak Gaussian decomposition or empirical simulation of the spectral feature evolution under perturbation prevented its advantage in revealing the bonding dynamics. One often simply decomposes a spectral peak into multiple Gaussian components with limited constraints albeit physical indications or simulates the spectral peak evolution in an empirical manner with multiple freely adjustable variables. Conventional phonon spectroscopy analysis has delivered information far less than it is supposed to be because of lacking the phonon-bond-stimulus correlation functions. The Grüneisen constant is an experimental derivative showing the trend of frequency shift with the stimulus. The optical phonon degeneration and the multiple phonon resonant scatting are extrinsic artifacts exist throughout the processes of experiments.

The key challenge is a theory to reproduce observations because of bond relaxation in length and energy and the bond-phonon-property correlation of the examined substance to reconcile as many as perturbations to the phonon frequency shift in terms of excited bond relaxation. One needs to correlate the $\Delta\omega$, $\Delta\Gamma$, $\Delta A$(abundance) intrinsically to the bond length an energy relaxed by external stimuli. Therefore, bond relaxation is profoundly and ubiquitously important to engineering of materials and thus should receive deserved attention.

We devoted to modeling the multifield lattice oscillation in the past decade to meet the following targets:

1) Theoretical reproduction of the phonon spectroscopy measurements with physically meaningful parameters.
2) Provision of atomistic, local, quantitative, elemental information on bonding dynamics from the measurements.
3) Comprehension of the physical mechanism and basic rules governing observations.
4) Correlation of the detectable quantities, $\Delta Q(x_i)/Q(x_{i0}) = f(x_i, d(x_i), E(x_i))$, to the bond length and energy that relax upon perturbation by the degree of freedom $x_i$.

This article deals with the multifield lattice oscillation dynamics to amplify the *Coordination-resolved Electron Spectrometrics* [6, 112] and its analytical strategies to the current *Multifield Phonon Spectrometrics* [113-115] for consistent insight into phonon relaxation with derivative of conventionally-unexpected quantitative information. From the viewpoint of local bond averaging (LBA) approach and Fourier transformation [7], we will be focused on formulating the Raman shifts of structured crystals a functions of the length, order, and strength of the representative bond responding to the externally applied perturbations.

Theoretical reproduction of the experimental observations clarifies the mechanism behind the observations with ever-unexpected information. The information includes the frequency $\omega(1)$ of reference and its bulk shift, atomic cohesive energy $E_{coh}$, bond energy $E_b$, binding energy density $E_{den}$, compressibility $\beta$, Debye temperature $\theta_D$, elastic modulus $B$, force constant $k$, and the effective atomic CN for the structured crystals.

1.5 Scope

This work begins with an overview on the significance of bond relaxation, lattice oscillation, available experimental database, and theoretical approaches to naming challenges and potential opportunities. Focus is given on generalizing the trends of phonon frequency shift due to activation of multifield in terms of atomic undercoordination, mechanical activation, thermal excitation and available theoretical descriptions. Thus, we can discriminate the intrinsic effects from those extrinsic artifacts and specify rooms for theoretical unification to gain consistent insight and the ever-unexpected information on bond relaxation by perturbations, in terms of degrees of freedom. It is exciting to find out that crystal size reduction creates three types of phonon frequency shifts unseen in the bulk – size-reduction induced $E_{2g}$ blueshift, $A_{1g}$ redshift, and the emerging of the low-frequency Raman (LFR) mode in the THz frequencies that undergoes blueshift with the inverse of size. The Raman shift, bandgap, and elastic modulus follows a Debye thermal decay – drops nonlinearly and then linearly as temperature rises in a way of $1-U(T)/E_{con}$ with $U(T)$ and $E_{con}$ being the integral of the Debye specific heat and the atomic cohesive energy. Compression stiffens the phonon frequency nonlinearly and the mechanical compression enhances the effect of atomic undercoordination on the phonon relaxation. However, it is inspiring to note the large gap to be filled for the bond-phonon-property correlation. Overwhelming debating approaches exist from various, intrinsic and extrinsic, perspectives for a specific phenomenon. The conventional spectral peak decomposition and evolution simulation with freely-adjustable parameters with limited information hindered the progress in understanding the nature of observations. It is urgent to develop a theory to reconcile the perturbation-bond-property correlation and to derive information on bonding dynamics, which drove the presented dedication made in past decade.

Section 2 is dedicated to developing the average-bond oscillating dynamics notion to incorporate the quantum approaches and Fourier transformation. Conventionally, the phonon relaxation is described from the perspective of Gibbs free energy, Grüneisen parameters, $\partial\omega/\partial x_i$, or their similarities, with $x_i$ being the specific degree of freedom. In contrast, we consider the Hamiltonian, Schrödinger equation, Lagrangian oscillation mechanics, Fourier transformation, and Taylor series of potentials with focus on the bond relaxation by external perturbation. The function dependence of a detectable quantity $Q(x_i)$ on the bond length and energy is necessary. To seeks for the relative change of a detectable quantity, $\Delta Q(x_i)/Q(x_{i0}) = f(x_i, d(x_i), E(x_i))$ with $x_i$ being any stimulus taken as a hidden factor driving bond relaxation. The $Q(x_i)$ can be phonon frequency, bandgap, elastic modulus, etc. This way of approach reconciles the perturbations of atomic and molecular undercoordination, temperature, strain, pressure into one equation.

Section 3 examined the two-dimensional layered graphene, black phosphorus, and (W, Mo)(S, Se)$_2$ structures by theoretically reproducing their Raman shifts due the number-of-layers, orientational strain, compression, and thermal activation with derivative of ever-unexpected information. The information includes the bond length and energy, band nature index, the dimer vibration frequency of reference, Debye temperature, atomic cohesive energy, thermal expansion coefficient, compressibility, elastic modulus, binding energy density, which should be the capability of the spectrometrics. It is uncovered that the single dimer-vibration and the collective dimer-vibration govern the number-of-layer reduction induced blue and red phonon frequency shift. The slope of the Debye thermal decay of the ω(T), $\partial\omega/\partial T$, at higher temperatures is the inverse of atomic cohesive energy and the pressure slope of the ω(P) profile, or the $\partial\omega/\partial P$, approaches the inverse of binding energy density.

Section 4 proved first the core-shell structures of water droplets and nanocrystals with derive of the skin thickness of 0.09 nm for the water droplets and two atomic diameters, 0.5 nm for CeO$_2$ nanocrystals. The skins follow the universal bond order-length-strength (BOLS) notion that specify the shorter and stiffer bonds between undercoordinated atoms. Examination of the sized crystals from nanoscale to the bulk of group IV, III-nitride, II-oxide under the perturbation of size reduction, pressure and temperature derived the same kind information described in section 3.

Section 5 introduces briefly the recent work [116-120] focusing on the effect of pressure, temperature, molecular undercoordination, and charge injection by acid, base, and salt solvation on the performance of water ice. Aqueous charge injection in forms of anions, cations, electrons, lone pairs, molecular dipoles, and protons modulates the hydrogen bonding (O:H–O) and properties of a solution through O:H formation, H↔H and O:⇔:O repulsions, electrostatically screened polarization, solute-solute interaction, and H–O bond contraction due to bond order deficiency. Polarization by charge injection and molecular undercoordination modify the critical pressures and temperatures for the confined ice-quasisolid and the salted water-ice phase transition under heating and compression. Consistency between theoretical predictions and measurements confirms the ever-unaware issues such as quasisolid phase of negative thermal expansion (NTE) due to O:H–O bond segmental specific heat disparity. Molecular coordination deficiency and electrostatic polarization result in the supersolid phase. Excessive protons and lone pairs form the H$_3$O hydronium and HO$^-$ hydroxide, which turns an O:H–O bond into the H↔H anti–hydrogen-bond upon acid solvation and O:⇔:O super–hydrogen-bond on base solvation. The aqueous molecular nonbonding can extend to other molecular crystals and to the negative thermal expansion of other solid substance.

The last section 6 summarizes the attainment, limitation, and forward-looking directions. The phonon spectrometrics detects directly the bond responding to perturbation but the electron spectrometrics probes the behavior of electrons in various bands/levels due to bond relaxation, which are totally different but complement each other. This set of spectrometrics provides a powerful technique applicable to situations when electrons and phonons are involved and under any external perturbation – for atomistic, dynamic, local, and quantitative information on bond and electron performance and consistent insight into the nature of observations, an essential theoretical and experimental strategies for functional materials devising.

## 2. Theory: Multifield Oscillation Dynamics

Highlights

- *Bond order, length, and strength stem the phonon frequency, bandgap, and elasticity.*
- *Perturbation relaxes the bond and evolves the phonon and related properties.*
- *Phonon relaxation fingerprints the intrinsic manner of bond response to perturbation.*
- *Spectrometrics enables the ever-unexpected information on bond-phonon-property cooperativity.*

2.1. Lattice Oscillation Dynamics

2.1.1. Single-Body Hamiltonian

There are three major approaches for the oscillation dynamics of a bond that is relaxed in length and energy by perturbing its crystal potential or by applying an external force: resolution to the Schrödinger equation [121], monatomic and diatomic chain dispersion [122], and the Lagrangian oscillation mechanics for the coupled oscillators [123, 124].

Bond relaxation and associated electronic redistribution in the real and energy spaces mediate the structure and properties of a substance [4]. An electron in a solid or in a liquid is subject to its intra-atomic potential of $v_{atom}(r)$ and the superposition of all interatomic potentials, $U(r)$, involved in the single-body Hamiltonian of the Schrödinger equation [121]:

$$i\hbar \frac{\partial}{\partial t}|v,r,t> = \left[-\frac{\hbar^2 \nabla^2}{2m} + v_{atom}(r) + U(r,t)\right]|v,r,t>$$

$$\text{where, } U(r) = \sum_i u_i(r) = \sum_{n=0} \left(\frac{d^n U(r)}{n! dr^n}\right)_{r=d_z} (r-d_z)^n$$

(13)

the $|v,r,t>$ is the Bloch wave function that describes the electronic behavior at site $r$ in the $v_{th}$ energy level [125]. The single-body approach approximates the long-range interactions and the many-body effects as a background of mean field.

For electrons in a core level follows the tight-binding approach [122]. The first term in eq (13) is the electronic kinetic energy. The coupling of the intra-atomic potential $v_{atom}(r, t)$ function and the related wave functions, $E_v(0) = <v, r, t|v_{atom}(r, t)|v, r, t>$, determines the $v_{th}$ energy level of an isolated atom. The $E_v(0)$ is the reference from which the specific core level shifts upon perturbation such as bond formation with different numbers or types of neighboring atoms [6]. The core level shifts when the electrons are subjecting to perturbation $U(R)(1+\Delta)$ in terms of the exchange integral dominance and the overlap integral as a secondary [6]. The exchange integral and the overlap integral are both depend on the perturbed bond energy $E_b(1+\Delta)$ with $\Delta$ being the perturbation.

The first Fourier coefficient of the crystal potential $U(r)$ determines the bandgap $E_g$ between the conduction and the valence band, according to the nearly-free electron approximation [122],

$$\begin{cases} E_g = 2|U_1| \propto \langle E_b \rangle \\ U_1 = \int U_{cry}(r) e^{ik \cdot r} dr \end{cases}$$

(14)

The bandgap is proportional to the interatomic bond energy $\langle E_b \rangle$ as well. The Bloch wave function approaches the performance of the nearly-free electrons. The intrinsic $E_g$ is different from the optical band gap with involvement of electron-phonon coupling interaction. Therefore, a perturbation changes the band gap and the core level shift in the same way $\Delta E_g(x_i) \propto \Delta E_b(x_i)$ intrinsically.

For lattice oscillation, the wave function describes the vibrating oscillator. The $V_{atom}$ is replaced with $V_{dimer}$ for the intra-dimer interaction [121]. The crystal potential $U_{cry}(r)$ adds a perturbation to the $V_{dimer}$, which expands into a Taylor series at equilibrium (r = d). The solution to the Schrödinger equation for the oscillation is generally a Fourier transformation function. Both Taylor and Fourier series are intrinsically correlated through their coefficients. A Taylor series can be considered a special case of the generalized Fourier series, with an orthonormal base of power functions and a properly defined inner product [126].

The $E_v(0)$ is replaced by $\hbar\omega_0$ that is the reference dimer vibration energy and $\omega = \omega_0(1+\Delta)$ is the shift under perturbation. The $\omega$ varies with the curvature of the potential in the form of $\mu\omega^2 = [U(d)(1+\Delta)]''$ at the relaxed equilibrium at which the nonlinear contribution is within the instrumental detection limit. The anharmonic correction shifts the H−O phonon frequency only by some 100 cm$^{-1}$ without producing any new features spectrum [127].

One can replace the integration of the coupled wave functions and interatomic potentials with the interatomic potential energy directly to simplify the discussion. At equilibrium, the coordinate $(d, E)$ for the potential curve gives directly the bond length ($d$) and bond energy ($E$). Despite the precision of the solution, this approximation allows one to focus on the nature origin behind and the varying trends of phononic measurements. Consideration of the relative frequency shift due to perturbation further improves the precision due to simplification.

## 2.1.2. Atomic Chains

Raman scattering of the radiating dipoles moment, which follows the laws of momentum and energy conservation between the incident photon and the phonons being activated. When one considers a crystal, each Bravais unit cell contains $n$ atoms, there are three acoustic modes and $3(n-1)$ optical modes. The optical modes from the relative dislocation of atoms in the unit cell and acoustic mode arises from the mass center in-phase motion of the unit cell.

For an decoupled monoatomic chain and a diatomic chain of the same spring force constant $\beta$, one can solve the lattice vibration equations to derive the phonon dispersion relations in the reciprocal $k = 2\pi/\lambda$ space with $\mu = m_1 m_2/(m_1 + m_2)$ being the reduced mass of the oscillator [122]:

$$\begin{cases} \omega^2(k) = \dfrac{2\beta}{\mu}\sin^2\left(\dfrac{ka}{2}\right) \propto \dfrac{\beta}{\mu} & (monatomic\ chain) \\ \omega_\pm^2(k) = \dfrac{\beta}{\mu}\left[1 \pm \sqrt{1 - \dfrac{4\mu^2 \sin^2\left(\dfrac{ka}{2}\right)}{m_1 m_2}}\right] \propto \dfrac{\beta}{\mu}[1 \pm \delta(k)] & (complex\ atomic\ chain) \end{cases}$$

(15)

These solutions agree with the single-bond approach shown in Table 1, irrespective of the acoustic or the optical phonon. Since the wavelength of the IR and the visible light ($500 < \lambda < 1500$ nm) and is much greater than the lattice constant in a $10^{-1}$ nm order. The single-bond approximation is valid as one is focused on the frequency shift of a specific vibration mode within a certain range near the Brillouin zone center, see Figure 7.

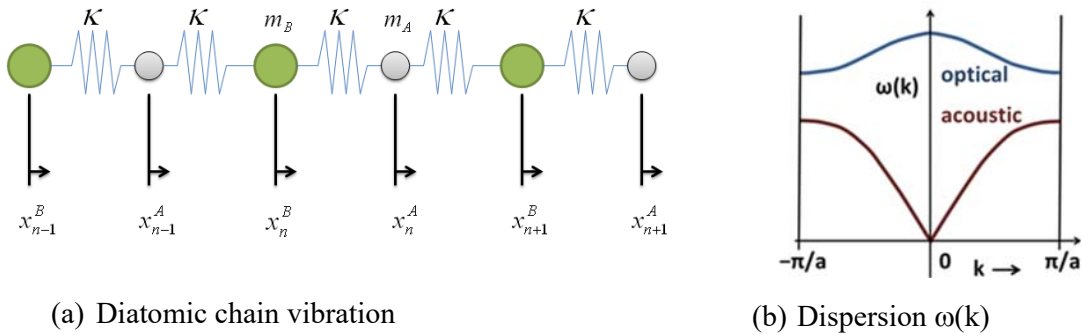

(a) Diatomic chain vibration      (b) Dispersion ω(k)

Figure 7. Illustration of (a) the diatom chain vibration (K = β is the force constant) and (b) shows the dispersion of the acoustic and optical phonons in the momentum domain. The acoustic phonon dispersion holds for monatomic chain as well. The IR and visible light is within the tiny $k = 2\pi/\lambda$ ($\lambda >>$ a) value at the Brillouin center [122].

2.1.3. Lagrangian Mechanics of Coupled Oscillators

An alternative yet efficient way to deals with oscillating system is solving the Lagrangian oscillation equation. For aqueous solutions, the segmented O–H:O bond approaches an asymmetrical oscillator pair coupled by the O-O Coulomb repulsion [124]. The oscillator pair follows Lagrangian motion equation [128]:

$$\frac{d}{dt}\left(\frac{\partial L}{\partial (dq_i/dt)}\right) - \frac{\partial L}{\partial q_i} = Q_i$$

(16)

The Lagrangian $L = T - U$ consists of the potential energy $U$ and the kinetic energy $T$ of the oscillating system. The $Q_i$ is the non-conservative forces due to perturbation. The non-conservative forces include electrification, molecular undercoordination, mechanical compression, radiation absorption, thermal excitation, etc. [129]. The action of a $Q_i$ force relaxes the bond from one equilibrium to another. The time dependent $q_i(t)$, represents the generalized variables, denoting the coordinates of an atom composed the oscillator. The kinetic energy $T$ sums vibration energies of the individual atoms, in the form of $2T_i = m_i(dq_i(t)/dt)^2$. The potential energy $U$ is composed of all interatomic interactions. The $u_i$ is the coordinate of the $i^{th}$ atom.

Resolution to the O–H:O coupled oscillator pair [128] turned out the segmental $\omega_x(k_x)$ dispersion (x = H and L for the H–O and the O:H segment, respectively). The $k_x$ and $k_C$ correspond to the segmental force constant and the O–O coupler [123, 124]. The $m_x$ is the reduced mass of the oscillator.

$$\omega_x = (2\pi c)^{-1}\sqrt{\frac{k_x + k_C}{m_x}}$$

(17)

Because of the segmental $k_x$ disparity, this dispersion specifies that under any perturbation, the O ions dislocate in the same direction by different amount along the O–H:O. The O:H relaxes always more than the H–O does. Consequently, if one segment becomes longer, its phonon turns to be softer, and vice versa. Decoupling the $k_C$, the dispersion degenerates into the isolated oscillators, which is the non-segmented A-B type bond approximation, which is equivalent to eq (15) in the Brillouin zone center and $k_x = \beta$ being the force constant.

### 2.1.4. Collective Oscillation

The vibration amplitude is the dislocation of an atom with its equilibrium position $x = r - d_0$. The high-order terms of the potentials contribute to the nonlinear behavior. For a dimer oscillator, the atomic coordination number is $z = 1$; for collective oscillation, $z > 1$, summing over all neighboring coordinating atoms. Since the dimer vibration amplitude $x \ll d_0$, the mean contribution to the force constant and to the magnitude of the atomic vibration approximates [34],

$$k_1 = k_2 = \cdots = k_z = \mu_i (c\omega)^2,$$

and, $x_1 = x_2 = \cdots = x_z = (r - d_0)/z$.

Hence, the resultant binding energy of a given atom to its $z$ neighbors,

$$u(r) = -zE_b + \left.\frac{zd^2 u(r)}{2! dr^2}\right|_{d_0} (r - d_0)^2 + \ldots$$

(18)

This relation features the phonon frequency as a function of bond order-length-energy, in terms of the curvature of the resultant potential.

### 2.2. Taylor Coefficients versus Observables

The crystal potential $U_{cry}(r)$ and the associated electron distribution uniquely determine the performance of a substance that is represented by an average of the whole pack of bonds in the substance [6, 7, 9, 36, 130]. As exemplified in Table 1, the Taylor coefficients correspond directly to the detectable quantities in terms of the dimensionality.

Table 1. Correlation between the macroscopic detectable quantities and the Taylor series coefficients.

| microscopic q | bond identities | macroscopically detectable Q | No |
|---|---|---|---|
| $E_z = U(d)$ | bond energy $E_z$ | core level shift $\Delta E_v$, band gap $E_G$, | (1) |
| $\left.\dfrac{dU(r)}{dr}\right|_{r=d} = 0 \propto \left\lfloor\dfrac{E}{d}\right\rfloor$ | bond length $d$ | mass density $d^3$, strain $\Delta d/d$ | (2) |
| $f = -\dfrac{dU(r)}{dr}$ | at non-equilibrium | force | (3) |
| $P \propto -\dfrac{dU(r)}{r^2 dr} \propto \dfrac{U(r)}{r^3}$ | | pressure $-\dfrac{\partial U(r)}{\partial r}\Big/\dfrac{\partial V}{\partial r}$ | (4) |
| $k = \left.\dfrac{d^2 U(r)}{dr^2}\right|_{r=d} \propto \left\lfloor\dfrac{E}{d^2}\right\rfloor$ | force constant $k$ | bond stiffness $Yd$, dimer vibration frequency: $\omega = k/\mu = (E/\mu d^2)^{1/2} \propto (Yd)^{1/2}$ | (5) |
| $B \propto \left.\dfrac{d^3 U(r)}{dr^3}\right|_{r=d} \propto \left\lfloor\dfrac{E}{d^3}\right\rfloor$ | energy density $E_{den}$ | elastic modulus $B$, $Y \propto -V\dfrac{\partial P}{\partial V}$ | (6) |
| $T_C \propto zE_z$ | atomic cohesive energy $E_{coh}$ | critical temperature for phase transition; energy band width $E_{v,w} \propto 2zE_z$ | (7) |

For instance, the interatomic potential determines the energy band structure including the core level shift and forbidden band gap between the conduction and the valence band [6]. One can take phonons as individual particles, so the integrals are related directly to the interatomic potential energies. This approximation avoids the combination of wave functions and simplifies the approximation, which may be subject to some precision with focus on the nature origin and the trend of change. Considering the proportional relations of the dimensionality would be adequate as one is focusing on the relative change of a known bulk property as the standard reference upon perturbation.

From the dimensionality analysis of Table 1 (line 5 and 6), one can correlate the atomic site resolved elastic modulus $B_i(z_i)$ and the vibration frequency shift $\Delta\omega(z_i)$ with $[\Delta\omega(z_i)]^2/[B_i d_i] \equiv 1$ if the $z_i$ and $\mu$ conserve. This relationship applicable to any interatomic potential $U(r)$ as the $B_i$ and $\Delta\omega(z_i)$ are related only to the bond order-length-energy at equilibrium.

### 2.3. Single Bond Multifield Oscillations
### 2.3.1. Bond Length and Energy Relaxation

The interatomic bond is the basic unit of structure and energy-storage. The bond relaxes upon perturbation, which in turn mediates the electronic energetics and the detectable quantities of a substance when subjecting to perturbation. Any perturbation will transmit the initial $U(r, t)$ into another equilibrium $U(r, t)(1 + \Delta)$ by relaxing the length and energy of the interatomic bond. If all the bond order deficiency, lattice strain, interface stress, and thermal relaxation are involved, the representative bond will change its energy $E_i(z, \varepsilon, T, P, \ldots)$ and length $d(z, \varepsilon, T, P, \ldots)$ as follows [36],

$$\begin{cases} d(z,\varepsilon,P,T,\ldots) = d_b \Pi_J(1+\varepsilon_J) = d_b \left[ (1+(C_z-1))\left(1+\int_0^\varepsilon d\varepsilon\right)\left(1+\int_{T_0}^T \alpha(t)dt\right)\left(1-\int_{P_0}^P \beta(p)dp\right)\ldots \right] \\ E(z,\varepsilon,P,T,\ldots) = E_b\left(1+\sum_J \Delta_J\right) = E_b\left[1+\dfrac{\left(C_z^{-m}-1\right)-d_z^2\int_0^\varepsilon \kappa(\varepsilon)\varepsilon d\varepsilon - \int_{T_0}^T \eta(t)dt - \int_{V_0}^V p(v)dv\ldots}{E_b}\right] \end{cases}$$

where

$$\begin{cases} d_b = d(z_b, 0, P_0, T_0) \\ E_b = E(z_b, 0, P_0, T_0) \end{cases}$$

(19)

The $\varepsilon_J$ is the strain and the $\Delta_J$ is the energy perturbation. The summation and production are over all the $J^{th}$ stimulus of all the degrees of freedom ($z, \varepsilon, T, P, \ldots$). The $C_z$ is the bond contraction coefficient depending on the atomic coordination numbers ($z$ or $CN$) that is an ever-oversighted variable grounding the defect and surface science and nanoscience. $C_z$ -1 is the undercoordination induced strain. The $m$ is the bond nature index that correlates the bond length to energy. The $\alpha(t)$ is the temperature-dependent thermal expansion coefficient. The $\eta(t) = C_V(t/\theta_D)/z$ is the Debye specific heat of the representative bond for a z-coordinated atom. The compressibility ($p < 0$), $\beta = -\partial v/(v\partial p)$, is an inversion of its elastic modulus in dimension. The $k(\varepsilon)$ is the strain-dependence of the single bond force constant. All these variables and their functional forms will be addressed in detail in the respective sections.

Figure 8 illustrates the bond relaxation of a regular oscillator dimer and the O:H–O bond oscillator pair coupled by the O:⇔:O inter-lone-pair repulsion [128]. At equilibrium, the ($d$, $E_b$) coordinate corresponds to the bond length and bond energy. Perturbation acts on the potential and dislocates the ($d$, $E_b$) to a new equilibrium [$d_b\Pi(1 + \varepsilon_J)$, $E_b(1 + \Sigma\Delta_J)$]. For instance, a compression or a tension perturbs the $U(R)$ in a manner given in Figure 8a. Compression stores energy into a bond by shortening and stiffening the specific bond while tension does it oppositely, along the path of $f(P)$.

The coupling of the intermolecular O:H and intramolecular H–O bond by the O:⇔:O repulsion aims to feature the energetics and dynamics of water ice, in terms of O:H–O segmental cooperative relaxability and specific-heat disparity [124]. A perturbation relaxes the O:H–O bond always by shortening one segments and lengthens the other, which initiates the anomalous behavior of water ice when responding to stimulus. Thus, the bond is relaxed in length and energy and all the detectable properties of the substance vary. Thermal activation not only fluctuates the vibrating oscillator but also elongates and softens the bond. The perturbation may modify the shape of the potential, but it is not that important or out of immediate concern.

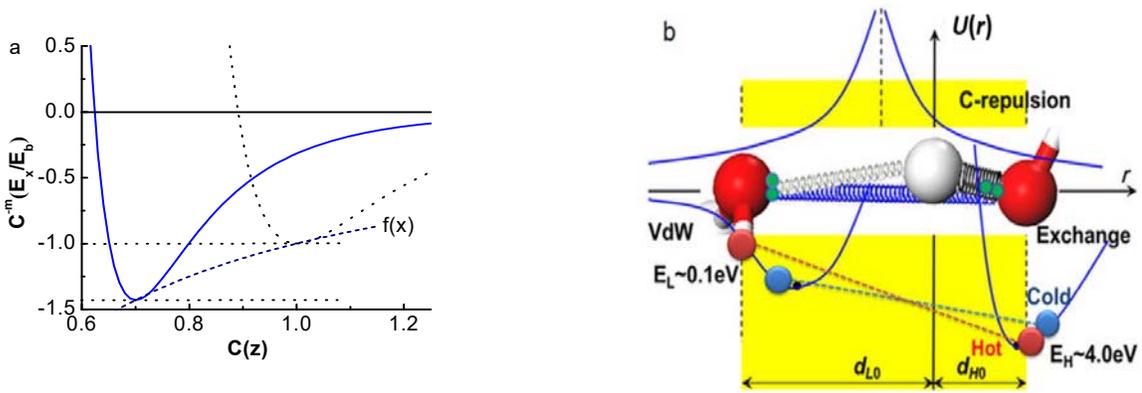

Figure 8. Relaxation of the regular dimer oscillator [7] and the O:H–O oscillator pair [128]. The bond relaxes in length and energy under a stimulus ($x = z, P, T, \varepsilon$, etc.) along the $f(x) = [d_b\Pi(1 + \varepsilon_J), E_b(1 + \Sigma\Delta_J)]$ path transiting the potential curve from one equilibrium to the other under stimulation. A perturbation relaxes the O:H–O bond always by shortening one segments and lengthens the other.

2.3.2. Phonon Frequency Relaxation

In place of the extrinsic processes of Raman scattering, one shall focus on the solution to the Hamiltonian of a vibration system. The solution is in the Fourier series with multiple frequencies being folds of the primary mode [121]. For instance, the 2D mode is twofold frequency that of the primary D mode of diamond. Therefore, one can clarify that the 2D mode does not arise from the extrinsic double resonant process of Raman scattering. A Hamiltonian perturbation such as the interlayer interaction, polarization, or the nonlinear effect only relaxes the folded frequencies to deviate from the references. In fact, the number-of-layer reduction shifts the D/2D peaks from 1367/2720 to 1344/2680 cm$^{-1}$, is within this right expectation.

The number-of-layer derived contrasting trends of the Raman frequency shift for graphene suggest that the origin of the G mode is completely different from that of the D/2D modes. On the other hand, an application of any perturbation will relax the involved bonds, resulting in the phonon frequencies change. Phonon band splitting is with expectation if the direction mismatch between the uniaxial strain and the C–C bond in the graphene.

The Raman frequency of a specific x mode is expressed, $\omega_x = \omega_{x0} + \Delta\omega_x$, where $\omega_{x0}$ is the dimer vibration frequency and the reference from which the $\Delta\omega_x$ proceeds. Incorporating the variables $x_i = z, \varepsilon, T, P$ into eq (2), one can formulate the relative Raman shift,

$$\frac{\omega(z,\varepsilon,P,T) - \omega(1,\varepsilon,P_0,T_0)}{\omega(z_b,0,P_0,T_0) - \omega(1,0,P_0,T_0)} = \frac{zd_b}{d(z,\varepsilon,P,T)} \left( \frac{E(z,\varepsilon,P,T)}{E_b} \right)^{1/2}$$

(20)

As the first-order approximation, the vibration frequency shifts $\Delta\omega_x(z,d_z,E_z,\mu)$ from the referential $\omega_x(1,d_b,E_b,\mu)$, in the following manner,

$$\Delta\omega_x(z,d_z,E_z,\mu) = \omega_x(z,d_z,E_z,\mu) - \omega_x(1,d_b,E_b,\mu)$$

$$= \Delta\omega = \sqrt{\left.\frac{d^2 u(r)}{\mu dr^2}\right|_{r=d_z}} \propto \frac{1}{d_z} \left( \frac{E_z}{\mu} \right)^{1/2} \times \begin{cases} 1 & (G, E_g) \\ z & (D, A_g) \end{cases}$$

(21)

Considering the coordination-resolved shift of different vibration modes, one has to take $z = 1$ and $z$ for the redshift. For instance [36, 131], for the D/2D modes of graphene and the $A_g$ mode for 2-D structures, $z \neq 1$ is involved, which causes the phonon frequency redshift due to the collective vibrations of the oscillators. For the graphene G mode, and the $E_{2g}$ mode of TiO$_2$, $z \equiv 1$, which secures the phonon frequency blueshift when the feature size is reduced because of the dominance of dimer oscillation.

2.4. Formulation of Multifield Perturbation

2.4.1. Atomic Undercoordination

2.4.1.1. BOLS-LBA approach

The BOLS notion [9] suggests that bonds between fewer-coordinated atoms shrink spontaneously. The charge and energy densities are higher and the local potential well becomes deeper associated with and quantum entrapment. The locally densely entrapped charge will in turn polarize the valence electrons of atoms at the open end of the crystal. Hence, the BOLS defines the atomic cohesive energy, binding energy density at the atomic site, which provides perturbation to the Hamiltonian of the entire system. The following expresses the BOLS notion:

$$\begin{cases} d_z = d_b C_z = 2\{1+\exp[(12-z)]/(8z)\}^{-1} & \text{(bond contraction)} \\ E_z = E_b C_z^{-m} & \text{(bond strengthening)} \end{cases}$$

(22)

where $z$ denotes an atom in the $z^{th}$ atomic layer and $b$ denotes the bulk as a standard reference. The $z$ spans from the outermost surface inward up to three atomic layers. No bond order loss occurs to the layer of $z > 3$. The contraction coefficient $C_z$ depends only on the effective $CN$ (or $z$) of the specific atom regardless of the nature of the bond, except for the segmented hydrogen bond (O–H:O) in water ice. Molecular undercoordination shortens the H–O covalent bond but lengthens the H:O nonbond because of the coupling between electron pairs on adjacent $O^{2-}$ [124].

The LBA represents the true Fourier transformation of the phonon spectroscopy. The spectroscope sorts the constituent bonds according to their vibrational manners and frequencies. The LBA is applicable to all situation and samples: crystalline, non-crystalline, and those with or without defects or additives. One can thus focus on the performance of the representative bond toward the bond-phonon-property cooperativity.

A typical example is the graphene and carbon nanotube (CNT) that follows the BOLS prediction. Using the diamond C-C bond length of 0.154 nm and 0.142 nm for a graphite, one may deduce the effective CN $z_g = 5.335$ for the bulk graphite from the bond contraction coefficient $C_z$. The effective CN for the C atom in the bulk diamond. The C-C bond energy is $E_b = 7.37/12 = 0.615$ eV/atom. For the monolayer graphene of $z = 3$, $E_3 = 1.039$ eV. The atomic cohesive energy is 3.11 eV/atom for the monolayer graphene.

Numerical match to the measured elastic modulus [132-134], melting temperature of the SWCNT [132, 135], and the C 1s energy shifts for the edge and interior of a graphene, bulk graphite and diamond [136] has revealed consistently that the C-C bond at the graphene edge contracts from 0.154 to 0.107 nm with a 30% bond contraction and a 152% bond energy gain [132, 133]. The bond contraction induced polarization dictates the width dependence of the band gap of GNR [137], and the Dirac-Fermi polariton generation [61]. The C-C bond between the 3-coordinated atoms in GNR contracts by 18.5% to 0.125 nm with a 68% increase of bond energy [133]. The Young's modulus of the SWCNT is 2.6 TPa compared to the bulk value of 865 GPa. The wall thickness is 0.142 nm rather than the layer spacing of 0.34 nm for the SWCNT. Breaking a C-C bond between 2-coordinated carbon atoms requires 7.50 eV/bond that is 32% higher than the amount (5.67 eV/bond) required for breaking a bond between 3-coordinated atoms in a monolayer graphene [138].

2.4.1.2.    Atomic-site and Crystal-size Resolved Shift

Any perturbation $x$ mediates a detectable property $Q$ at an atomic site or for a cluster of given size and shape in the following core-shell manners [9]:

$$\frac{\Delta Q(x)}{Q(x_0)} = \begin{cases} \frac{\Delta q(x)}{q(x_0)} & (atomic-site) \\ \sum_{j \leq 3} \gamma_j \frac{\Delta q(x)}{q(x_0)} & (size-shape) \end{cases}$$

$$\gamma_j = \frac{V_j}{V} \cong \frac{N_j}{N} = \frac{\tau C_j}{K};$$

$$z_1 = 4(1 - 0.75/K), \ z_2 = z_1 + 2; \ z_3 = z_2 + 4 \ (spherical\ dot)$$

(23)

The $q$ being the density of the $Q$ is a function of bond length $d$, bond energy $E$, and bond order $z$. $\gamma_j$ is the number or volume ratio between the $j$th atomic layer and the entire specimen with subscript $j$ counting from the outermost atomic layer inward up to three. $\tau$ = 1, 2, 3 is the dimensionality for the thin plate, a cylindrical rod, and a spherical dot. The $K$ features the size of the nanocrystal, which is the number of atoms lined along the feature size. For bonds between atoms at sites of point defects, monatomic chains, monolayer atomic sheets or monolayer skins, no weighted superposition is considered.

The nanostructure core-shell configuration and the LBA approach [139], yields the following relation with $z_{ib} = z_i/z_b$ for the size-reduction induced Raman red ($z > 1$) and blueshift ($z = 1$),

$$\omega(K) - \omega(1) = [\omega(\infty) - \omega(1)](1 + \Delta_R)$$
$$or, \frac{\omega(K) - \omega(\infty)}{\omega(\infty) - \omega(1)} = \Delta_R < 0$$
$$\Delta_R = \sum_{i \leq 3} \gamma_i \left( \frac{\omega_i}{\omega_b} - 1 \right) = \begin{cases} \sum_{i \leq 3} \gamma_i \left( z_{ib} C_i^{-(m/2+1)} - 1 \right) & (z = z) \\ \sum_{i \leq 3} \gamma_i \left( C_i^{-(m/2+1)} - 1 \right) & (z = 1) \end{cases}$$

(24)

where $\omega_{xi}$ and $\omega_{x0}$ correspond to the phonon frequency at the *i*th surface atomic shell and inside the bulk. The $\omega_x(1)$ for an isolated dimer is the reference from which the frequency shift takes place upon nanocrystal and bulk formation.

The vibration of oscillator formed between nanosolid and host matrix or between grains stems the LFR mode in the THz regime. The relative motion of the individual atoms in a complex unit cell determines the optical modes.

2.4.2. Thermal Excitation: Debye Thermal Decay
2.4.2.1.    Lattice Debye Thermal Expansion

In place of the Grüneisen notion description, one may consider the thermal expansion coefficient $\alpha(t)$ for the LBA as follows [140]. From the definition, one can transit $L = L_0 \left( 1 + \int_0^T \alpha(t) dt \right)$ to

$$\alpha(t) \cong \frac{dL}{L_0 dt} = \frac{1}{L_0} \left( \frac{\partial L}{\partial u} \right) \frac{du}{dt} \propto -\frac{\eta_1(t)}{L_0 F(r)} = A(r) \eta_1(t)$$

(25)

where $(\partial L / \partial u) = -F^{-1}$ is the inverse of the gradient of the interatomic potential $u(r \approx d)$ nearby equilibrium. The $du/dt$ is the specific heat of Debye approximation. $A(r) = (-L_0 F(r))^{-1}$ approaches the invers of binding energy near equilibrium. Compared to the Grüneisen's volumetric thermal expansion coefficient (TEC) [102], the $\alpha(t) = (VB_T)^{-1}\gamma\eta_1(t)$, $\gamma/(VB_T) = [L_0 F(r)]^{-1} \cong$ constant is in the dimension of energy.

The thermal expansion coefficient $\alpha(T > \theta_D)$ is about $10^{-(6\sim7)}$ K$^{-1}$. The smaller expansion coefficient for nanoparticles [98-100, 141, 142] indicates an increase of the potential gradient, or stronger bond near the equilibrium - a narrowed shape of the interatomic potential for the contracted bond [7]. Thermal expansion is harder for the contracted bonds of a nanograin than those in the bulk standard. Approximating $A(r \approx d)$ to be a constant, the $\alpha(t)$ follows closely the single-bond Debye specific heat, $\eta(t)$. The current approach covers the full-range of $T$-dependent $\alpha(T)$ agrees exceedingly well with the experimental data for AlN, Si$_3$N$_4$, and GaN, see Figure 9 and Table 2 [140].

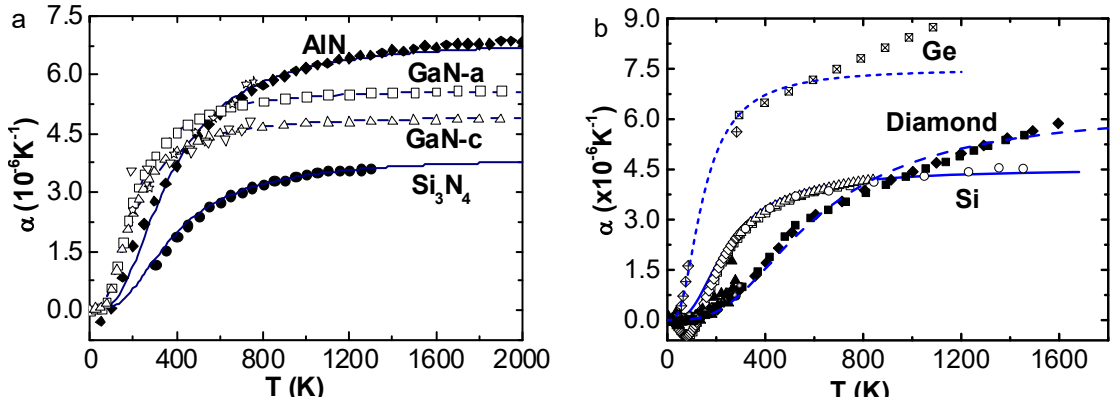

Figure 9 LBA reproduction (solid lines [140]) of the measured (symbols) temperature-resolved TECs for (a) AlN [143], Si$_3$N$_4$ [143], and GaN [142]), (b) Si [144, 145], Ge [146], and Diamond [141, 147-149]. Table 2 lists information derived from the fitting. (Reprinted with permission from [140]).

Table 2 Reproduction of the TECs and lattice parameters derived parameters, see Figure 9 [140].

|  | Refs [150, 151] | $\alpha(t)$ | | $l(t)$ | | | Mean | |
|---|---|---|---|---|---|---|---|---|
|  | $\theta_D$ (K) | $\theta_D$ (K) | $A(r)$ | $\theta_D$ (K) | $A(r)$ | $l_0$ (Å) | $\theta_D$ (K) | $A(r)$ |
| Si | 647 | 1000 | 0.579 | 1100 | 0.579 | 5.429 | 1050 | 0.579 |
| Ge | 360 | 600 | 0.966 | 500 | 1.035 | 5.650 | 550 | 1.001 |
| C | 1860 | 2500 | 0.811 | 2150 | 0.792 | 3.566 | 2325 | 0.802 |
| AlN | 1150 | 1500 | 0.888 | 1500 | 0.946(b) 0.881(c) | 3.110(b) 4.977(c) | 1500 | 0.882 |
| Si$_3$N$_4$ | 1150 | 1600 | 0.502 | 1400 | 0.888 | 7.734 | 1500 | 0.695 |
| GaN | 600 | 850 | 0.637 | 800 | 0.637(b) | 3.189(b) | 825 | 0.631 |

0.618(c)   5.183(c)

[a] The reference Debye temperature is documented in ref [100] for a-axis [b] and c-axis [c].

However, the negative TECs is beyond the scope of the present expression. Materials possessing negative TECs are often ascribed as the negative Grüneisen parameters of the transverse acoustic phonons. Generally, most materials expand upon being heated. Expansional cooling expansion such as graphite [152], graphene oxide [153], $ZrWO_3$ [154, 155], and compounds composed of N, F, and O [156] do exist because of the involvement of multiple-type interactions and their correlations like water ice [157]. Therefore, a superposition of the segmental specific heats and considers the coupling of the inter- and intra-molecular interactions would be necessary to reproduce the negative TECs [158].

2.4.2.2.   Debye Thermal Decay: Debye Temperature and Cohesive Energy

The specific heat $\eta_1(t)$ and its thermal integration is the conventionally termed internal energy, $U(T/\theta_D)$. When $x \ll 1$, $1+x$ approximation $\exp(x)$ is applied to describe the thermal bond expansion. The integration of the Debye specific heat $\eta(t)$ from 0 K to $T$ yields the $T$-induced bond weakening, $\Delta E_T$,

$$\Delta E_T = \int_0^T \eta(t)dt = \frac{\int_0^T C_v(t)dt}{z} = \int_0^T [\int_0^{\theta_D/T} \frac{9R}{z}\left(\frac{T}{\theta_D}\right)^3 \frac{y^4 e^y}{(e^y-1)^2} dy]dt$$
$$= \frac{9RT}{z}\left(\frac{T}{\theta_D}\right)^3 \int_0^{\theta_D/T} \frac{y^3 e^y}{e^y-1}dy$$

(26)

Where $R$ is the ideal gas constant, $\theta_D$ is the Debye temperature, and $C_v$ the specific heat. The reduced form of temperature $y = \theta_D/T$. The $\eta(t)$ is the specific heat per bond, which follows the specific heat of Debye approximation and closes to a constant value of $3R/z$ at high temperature. Figure 10 illustrates the Debye temperature dependence of the remnant energy of a bond subjecting to heating. The $U(T/\theta_D)$ is the loss of energy by thermalization toward bond dissociation and $1- U(T/\theta_D)/E_b(0)$ is the remnant fraction of the bond energy.

The Raman frequency transits gradually from the nonlinear to the linear form at higher temperature – Debye thermal decay. The slow drop of the Raman shift at considerable low temperatures because of the small $\int_0^T \eta(t)dt$ values.

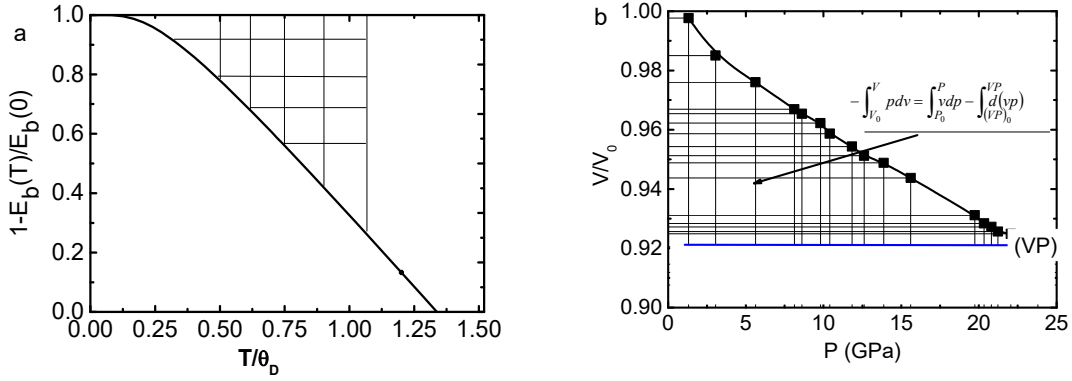

Figure 10. The average-bond bond oscillating dynamics. (a) Thermal energy loss (shaded area) corresponds to the difference between unity and the specific heat integration and (b) the mechanical deformation energy gain (shaded area) to the classical form of free energy [26], $-\int_{V_0}^{V} pdv = \int_{P_0}^{P} vdp - \int_{(VP)_0}^{VP} d(vp)$

2.4.3. Mechanical Compression: Elasticity and Energy Density

Likewise, the compressive distortion energy density gain, $\Delta E_P$, equals,[7]

$$\begin{cases} \Delta E_P = -\int_{V_0}^{V} p(v)dv = -V_0 \int_0^P p(x)\frac{dx}{dp}dp = V_0 P^2 \left[\frac{1}{2}\beta - \frac{2}{3}\beta' P\right] \\ x(P) = V/V_0 = 1 - \beta P + \beta' P^2; \; \frac{dx}{dp} = -\beta + 2\beta' P \end{cases}$$

or

$P(x) = 1.5B_0(x^{-7/3}-x^{-5/3})[1+0.75(B'_0-4)(x^{-2/3}-1)]$ (Birch-Mürnaghan)

(27)

The $V_0$ is the referential volume at the ambient pressure and zero temperature. The $x(P)$ is another form being equivalent to the Birch-Mürnaghan (BM) equation. The nonlinear compressibility $\beta$ and the bulk modulus $B_0$ meet the $\beta B_0 \cong 1$ relation in dimensionality [150].

Figure 10b illustrates the P-V profile and its integration. In the P-V profile, only the gridded area $-\int_{V_0}^{V} pdV \cong -\int_{P_0}^{P} p\frac{dV}{dp}dp$ contributes to the density of energy for the whole body [97]. In dealing with the single average bond, the bond length $d(z, t, p)$ replaces the atomic volume $V(z, t, p)$ and a force $f$ replaces the pressure. The $\beta(p)$ keeps a constant for the elastic deformation and then the integration $\int_0^P \beta(p)dp = \beta P$ holds [159, 160].

For a given sized specimen, one can formulate the $T$ and $P$ resolved $B$ and $\Delta\omega$ by joining Eqs (26) and (27),

$$\left.\begin{array}{c}\frac{B(T,P)}{B(0,0)}\\ \frac{\Delta\omega(T,P)}{\Delta\omega(0,0)}\end{array}\right\} \cong \left\{\begin{array}{c}\left(1+\frac{\Delta E_P - \Delta E_T}{E_0}\right)\exp\left[3\left(-\int_0^T \alpha(t)dt + \int_0^P [\beta - \beta' p]dp\right)\right]\\ \left(1+\frac{\Delta E_P - \Delta E_T}{E_0}\right)^{1/2}\exp\left(-\int_0^T \alpha(t)dt + \int_0^P [\beta - \beta' p]dp\right)\end{array}\right.$$

### 2.4.4. Uniaxial Stretch: Single Bond Force Constant

The data of tensile strain for the monolayer graphene [161] and $MoS_2$ [162] enabled the derived information of the average-bond force constant and the relative direction between the strain and a specific bond in the specimen. The strain-effect is given as [163],

$$\frac{\omega(z,\varepsilon)-\omega(1,0)}{\omega(z,0)-\omega(1,0)} = \frac{d(0)}{d(\varepsilon)}\left(\frac{E(\varepsilon)}{E(0)}\right)^{1/2} = \frac{\left(1-d^2\int_0^\varepsilon \kappa\varepsilon d\varepsilon/E_{z1}\right)^{1/2}}{1+\varepsilon}$$

$$\cong \frac{\left[1-\kappa'(\lambda\varepsilon')^2\right]^{1/2}}{1+\lambda\varepsilon'}$$

With $\kappa' = \kappa d_z^2/(2E_z) =$ constant

(18)

To feature the orientation mismatch between a bond and the uni-axial strain, one can introduce a strain coefficient $\lambda$ bounded by $0 \leq \lambda \leq 1$. The $\varepsilon = \lambda\varepsilon'$ is for a bond that is not along the applied strain. The effective force constant $\kappa = 2E_z\kappa'd_z^{-2}$ being constant at limited strain. In the experiment, one can hardly measure the force constant $\kappa_0$ for a certain bond but the average strain of the entire specimen. The $\lambda$ characterizes the actual strain of the orientated bonds in a basic unit from the complex of multiple bonds in different orientations. The strain can be along ($\lambda = 1$) or perpendicular ($\lambda = 0$) or random ($0 < \lambda < 1$) to a certain bond in the specimen. The bond along the strain is subject to a maximal extension and its phonon frequency shifts most; the bond perpendicular to the strain is subject to zero extension and keeps its phonon frequency. It is therefore not surprising that the uniaxial strain can split the phonon band into two components and the component separation varies with the angle between the bond and the uniaxial strain [36].

## 2.5. From spectroscopy to Spectrometrics

A phonon spectral peak, a vibration mode or a phonon band, probed by either Raman scattering or infrared absorption, features the Fourier transformation of all bonds vibrating in the similar frequency, irrespective of their locations or orientations or structural phases. Fourier transformation gathers the vibration modes according to the bond stiffness and their population in terms of the intrinsic oscillators' force constants. The force constant corresponds to the curvature of the resultant potentials that the vibrating dimer is subject at equilibrium. Multifield activation, electron entrapment and polarization, and light radiation contribute to the crystal potentials whose curvatures determine intrinsically the lattice oscillation dynamics.

Amplifying the phonon spectroscopy to spectrometrics aims to capturing the conventionally-unexpected information on the bonding dynamics by avoiding spectral peak decomposition or empirical simulation using hypothetic models. Compared with the electron spectrometrics that applies to conductive or semiconductive substance under high vacuum, the phonon spectrometrics extends to liquidius and insulating species under multifield perturbation. The phonon spectrometrics directly captures bond relaxation information but the electron spectrometrics captures the effect of bond relaxation on the electronic dynamics in various energy levels or bands [6].

When an external perturbation is applied, the interatomic potential will change, and the characteristic phonon frequency will evolve or transit from its first state to a new one in terms of abundance, stiffness, and fluctuation order, which gives profound information on the bond relaxation that triggers the electronic structure and properties of the substance. External perturbations include atomic irregular-coordination, mechanical activation, thermal excitation, charge injection by doping and aqueous solvation, electrification and magnification, radiation, etc. Correlating the phonon relaxation to the applied perturbation yields the ever-unexpected information.

Besides direct calculation of the ω(x) to match the measured frequency shift of a substance under x perturbation, the other convenient method is to find the fraction of abundance transition and frequency shift using the differential phonon spectrometrics (DPS) strategy [112, 164]. The DPS strategy distills only the phonon abundance and bond stiffness transition from the reference to conditioned states. This strategy monitors the phonon and bond relaxation both statically and dynamically. The DPS peak integral gives the fraction of bonds, or number of phonons transiting from the standard reference to the conditioned states under the change of the degree of freedom such as the solute concentration C in aqueous solutions and crystal size D for nanostructures,

$$f_x(C) = \int_{\omega_m}^{\omega_M} \left[ \frac{I_{solution}(C,\omega)}{\int_{\omega_m}^{\omega_M} I_{solution}(C,\omega)d\omega} - \frac{I_{H_2O}(0,\omega)}{\int_{\omega_m}^{\omega_M} I_{H_2O}(0,\omega)d\omega} \right] d\omega.$$

$$\omega_{COG} = \frac{\int_{\omega_m}^{\omega_M} \omega I(\omega)d\omega}{\int_{\omega_m}^{\omega_M} I(\omega)d\omega}$$

The division, $f(C)/C$, is the vicinal number of bonds per solute in the hydration volumes, which features the hydration volume and its local electric field. The center-of-gravity (COG) of the frequency $\omega_{COG}$ is obtained by integrating the population $I(\omega_x)$ from $\omega_{xm} = 0$ to $\omega_{xM} = 0$ [165].

2.6. Summary

From the perspective of Fourier transformation and LBA consideration, one can resolve the multifield bond oscillation dynamics and direct measure the $\Delta\omega(z, d, E, \mu)$ without involving the Grüneisen constants or the extrinsic multiple phonon resonant scattering or the optical phonon degeneration. Reproduction of the excited $\Delta\omega$ by perturbations such as bond-order-imperfection, compression, tension, thermal activation, and charge injection by solvation turns out the ever-unexpected information on the bond length and energy, single-bond force-constant, binding energy density, mode cohesive energy, elastic modulus, Debye temperature, etc.. One can thus reconcile the perturbation-relaxation-property cooperativity of a substance. This set of theoretical and experimental approaches provide not only efficient means for the local, dynamics, and quantitative information on the multifield bond oscillation dynamics but also comprehension of the physics behind observations.

## 3. Layered Structures

Highlights

- Atomic-undercoordination induced phonon relaxation imprints the bond order-length-energy.
- Compression stiffened phonons yield the binding energy density and elastic modulus.
- Phonon Debye thermal decay turns out Debye temperature and atomic cohesive energy.
- Stretching softened and separated phonon band leads to the single bond force constant.

3.1. Wonders of the 2D Structures

Two-dimensional (2D) substance such as black phosphorus (BP), graphene nanoribbon (GNR), and $MX_2$ (M = Mo, W; X = S, Se) have emerged as an amazing group of materials with high tunability of chemical and physical properties that do not demonstrate by their bulk parents [166-168]. For example, the few-layered BP shows high carrier mobility, up to $10^3$ cm$^2$V$^{-1}$s$^{-1}$, and high current on/off ratios, up to $10^5$ Hz at room temperature. Unrolling the single-walled carbon nanotubes (SWCNTs) triggers many intriguing properties that cannot be seen from the CNT or a large graphene sheet [169, 170]. In addition to the observed edge Dirac fermion states [171, 172] with ultrahigh electrical and thermal conductivity [173], and unexpected magnetization [174, 175], the bandgap of a GNR expands monotonically with the inverse of ribbon width [170, 176, 177].

One can turn the indirect bandgap of the bulk $MoS_2$ from 1.2 eV to the direct bandgap of 1.8 eV by simply reducing the number of layers from infinitely large into single [178, 179]. The indirect bandgap also degenerates into a direct for the few-layered $MoSe_2$ [180]. Bandgap transition from indirect to direct depends not only on the number of layers, but also on the material and the operation conditions [181]. Kelvin and conductive atomic force microscopy (AFM) revealed that the surface potential well goes deeper linearly with thickness from the bulk value of - 7.2 mV to −427 mV when the bulk $MoS_2$ turns into monolayer [60]. These entities make the 2D semiconductors an appealing type of materials for applications as flexible, miniaturized, and wearable electronic devices and energy management such as photovoltaic cells [50, 182], field effect transistors [56, 183, 184], light-emitting diodes [185, 186], and photodetectors [51, 187, 188].

3.1.1. Orbital Hybridization and Structure Configuration

Figure **11** illustrates the bond configuration for the monolayer BP, GNR, and $MX_2$. The C, P, S and Se atoms undergo the sp$^3$- orbital hybridization with three (for C $2s^2p^2$ and P $3s^2p^3$), and two (for S $3s^2p^4$ and Se $4s^2p^4$) bonds to their neighbors. Each C atom bonds covalently to its three neighbors in a plane and leaves one unpaired electron for the π bonding. Each P atom has three tetrahedrally directed bonds to its three neighbors and one dangling lone pair. In the same group of oxygen, S and Se share the same $s^2p^4$ electronic configuration in the 3$^{rd}$ and 4$^{th}$ outermost electronic shells. The S and Se atoms form each a tetrahedron with two bonds to the sandwiched M atoms and two electron lone pairs

exposing outwardly of the sandwiched layer. The extent of the orbital hybridization for the P, S and Se is less obvious than nitrogen and oxygen because of the orbital order occupancy of the valance electrons but their bond configuration does show the same trend [4].

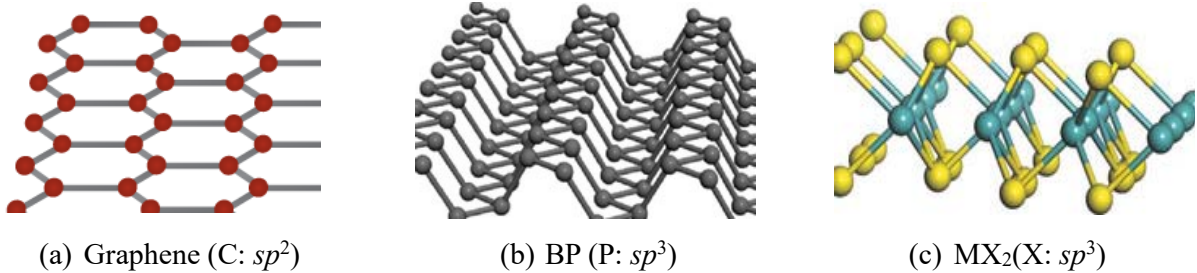

(a) Graphene (C: $sp^2$)　　　(b) BP (P: $sp^3$)　　　(c) MX$_2$ (X: $sp^3$)

Figure 11. Bond configuration for the GNR, BP, and MX$_2$. The C, P, S and Se atoms undergo the sp orbital hybridization with three (C $2s^2p^2$; P $3s^2p^3$), and two (S $3s^2p^4$, Se $4s^2p^4$) bonds to their neighbors, which generate (a) one unpaired $\pi$ electron and (b) one and (c) two electron lone pairs. The main layer consists of (a) one, (b) two and (c) three sublayers interacting with the next layer through van der Waals forces.

Figure 12 illustrates the atomic dislocations in the Raman-active and the IR-active vibration modes [35-39]:

1) The transverse optical phonon (TO) and the longitudinal optical (LO) Raman modes shift in frequency with the number-of-layer of the 2D materials.
2) The $E_{2g}$ mode for the WX$_2$ and TiO$_2$ and the G mode for graphene undergo blueshift while the $A_{1g}$ mode and the D mode shift to lower frequency as the feature size is reduced.

The main layers of these materials are composed of (a) one, (b) two and (c) three sublayers interacting with their next main layers through van der Waals interaction, which makes the mechanical filtration of these layered structures possible. For the MX$_2$ honeycomb-like structure, they can be taken as a positively charged plane of the transition-metal M$^{4+}$ cations sandwiched between two planes of negatively charged chalcogen X$^{2-}$ anions[189]. Each of the sandwiched M$^{4+}$ cation bonds to its four neighboring X$^{2-}$ anions. The MX$_2$ layer performs like a giant atom with the positive M$^{4+}$ core and the negative X$^{2-}$ shells. The MX$_2$ main layers interact one another through the short-range X$^{2-}$:⇔: X$^{2-}$ repulsion (the same as the O$^{2-}$:⇔:O$^{2-}$ super hydrogen-bond in basic solutions [190]) and the slightly longer $M^{4+} \sim X^{2-}$ Coulomb attraction. Such a set of interactions stabilizes the layered structures.

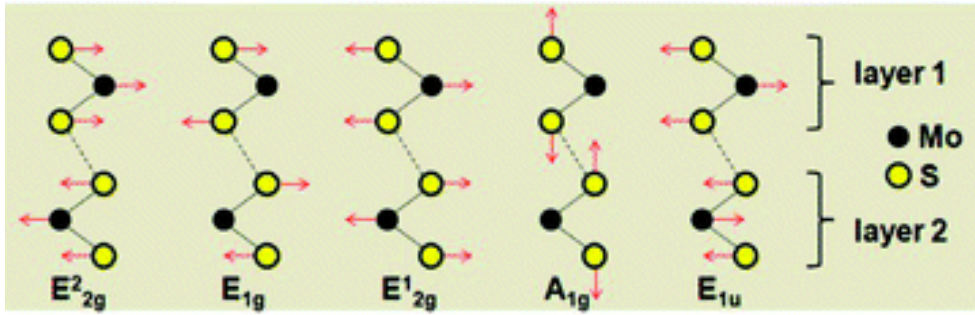

Figure 12. Atomic displacements of the four Raman-active modes and one IR-active mode ($E_{1u}$) in the unit cell of the bulk $MoS_2$ crystal as viewed along the [1000] direction (Reprinted with copyright permission from [39])

Furthermore, ribbons of these layered stuff have zigzag- and armchair-edge. Most importantly, bonds between undercoordinated atoms in the layered structures are shorter and stronger and the bonds at edges are even shorter and stronger than they are in the bulk trunks. The bond contraction and the associated core electron quantum entrapment and valence electron polarization govern the performance of these undercoordinated systems [6, 9].

3.1.2. Phonon Frequency Tunability

A huge database has been established toward the phonon spectroscopy/microscopy of the layered 2D structures and their nanoribbons subjecting to various stimuli. The stimuli include the number-of-layer (n) [21, 43, 49, 191, 192], uniaxial strain ($\varepsilon$) [63, 193, 194], mechanical compression (P) [65], thermal excitation (T) [111, 195], defect density and location [196], substrate interaction [197, 198], hydrogenation or passivation [199, 200], impurity or doping [201], incident photon polarization [202], edge conditioning [203, 204], etc.

Graphene shows two Raman active modes as do its allotropes. The G mode at ~1580 cm$^{-1}$ was assumed to result from the sp$^2$ network planer vibration [205]. The 2D mode at ~2680 cm$^{-1}$ was understood as a second-order scattering of double resonant process [206]. In the presence of defects or edge atoms, the defect-induced D band appears at frequencies around 1345 cm$^{-1}$ with intensity being varied with edge conditions compared with the D band of bulk diamond featured at 1331 cm$^{-1}$ [203, 204]. The monolayer BP shows the $A^1_g$, $A^2_g$ and $B_{2g}$ and the MX$_2$ shows the $A_{1g}$ and $E^1_{2g}$ Raman active modes.

The phonon behavior of the layered BP, GNR, and MX$_2$ [60, 207, 208] follows the same trend but by different amounts under the same excitation. The D/2D bands of graphene [20, 21] and the $A_{1g}$ mode of the MoS$_2$ and the WS$_2$ [39] undergo a redshift when the number-of-layer is reduced. For instance, under light radiation of 514.5 nm wave length, the D/2D modes shift up from 1367/2720 to 1344/2680 cm$^{-1}$ when the graphite turns to be the monolayer graphene [43]. In contrast, when the n is reduced from twenty to one a blueshift happens to the G band that shifts up from 1582 to 1587 cm$^{-1}$ [21, 43]. The $B_{2g}$ and $A_{2g}$ modes for BP[209, 210] and the $E^1_{2g}$ mode for MoS$_2$ and WS$_2$ [39] undergo a blueshift as well when their number-of-layer is reduced. When the n is increased from a few to multiple, the Raman peaks turn from the dominance of the monolayer component to the dominance of the bulk component [21]. It is even amazing that the D band reflectivity is one order geater at the armchair edge compared to the zigzag edge of the GNR [203, 204].

Introducing the uniaxial strain to the 2D substances by bending the layered structures on the elastic substrates can lower the work function to improve their carrier mobilities and raise their photoelectron emissivity [51, 187, 211]. The uniaxial tensile strain softens and splits the phonon bands of graphene [63] and MX$_2$ and narrows the bandgap of MX$_2$ [212]. The amount of the phonon band splitting varies with not only the strain but also its relative strain-bond direction of the layered structures [194, 213]. Under the compressive strain, the Raman shift extrapolates the trend of tension in an opposite direction [161].

Mechanical compression depresses the V-P profile but raises all the phonon frequencies [65, 214] [215, 216]. Contrastingly, heating [45] softens the Raman phonon to lower frequencies. Heating from ultra-low temperature to the well-above room-temperature softens the $A^1_g$, $A^2_g$ and $B_{2g}$ modes of BP [215, 216] and the $A_{1g}$ and $E^1_{2g}$ phonons of WS$_2$ and MoS$_2$ as well in the Debye thermal decay fashion [167, 207]. Heating from 0 to 300 K also reduces the excitonic transition energy of the WSe$_2$ in a similar manner [217]. Both the bandgap and phonon frequency initially drop at low temperature slowly and then linearly at high temperatures, being the form of the difference between a constant and the Debye specific heat integration [7]. The effects of mechanical compression and thermal excitation hold universal to both the layered and the bulk materials.

Current understanding of the multifield effect on the phonon frequency shift of the layered substance is focused on two perspectives. One is the continuum bond oscillating dynamics correlating the phonon frequency evolution as a plain function of the stimuli, like the Gibbs free energy. The other is the geometry of the Brillouin zone and electronic contribution to the contributing to the Raman photon scattering and phonon decaying dynamics. The effect of T on the Raman shift is ascribed as optical phonon decaying into multiple acoustic components [25, 80, 107-109] and the pressure effect is described in terms of Grüneisen notion – the slope of the Raman shift with respect to pressure, $\partial\omega/\partial P$ or $\partial(Ln\omega)/\partial(LnP)$ [63]. A "two-phonon double-resonance scattering" is suggested to govern the strain effect on phonon band splitting and softening [213]. The number-of-layer effect on the Raman shift follows hypothetically to the inverse number-of-layer $n^{-\gamma}$ with γ being a tunable parameter [35, 79].

The afore-mentioned empirical approaches could fit to the measurements independently, theoretically despite multiple freely adjustable parameters involved. However, consistent insight into the "external stimulus – bond relaxation – phonon frequency– macroscopic property" correlation and finding means for detection remain yet to be explored. For instance, mechanisms govern the opposite size trends of the phonon frequency shifts under the number-of-layer reduction and the edge-discriminated photon reflectivity is beyond the scope of available premises.

One may be concerned with what dictates the chemical and physical properties of the layered structures. In fact, bond formation and relaxation and the associated energetics, localization, entrapment, and polarization of electrons mediate the macroscopic performance of substance accordingly [4]. Compared with the coordination-resolved electron spectrometrics [6] that probes the local bond length and energy and the electronic dynamics under high vacuum, the Raman spectrometry features are very sensitive to various stimuli in amiable manners, which offers rich information on the local bonding dynamics under external excitation.

Advancing the phonon spectroscopy to spectrometrics aims at quantitative information that the spectroscopy can hardly offer. The phonon and electron spectrometrics complement each other providing comprehensive information on the performance of bonds and electrons of a system under perturbation. The key challenge is how to correlate and formulate the spectral feature to the bonding identities and the way of responding to excitation.

The aim of this section is to show that a combination of the BOLS-LBA [7, 9] and the phonon spectrometrics could reconcile the (n, ε, T, P) effect on the Raman shifts with clarification, formulation, and quantification of the bond and phonon relaxation dynamics.

## 3.2. Numerical Reproduction of Phonon Relaxation

### 3.2.1. Number-of-layer Dependence

#### 3.2.1.1. Formulation

For the z dependence of the phonon frequency shift follows,

$$D_L(z) = \frac{\Delta\omega(z)}{\Delta\omega(z_b)} = \frac{\omega(z)-\omega(1)}{\omega(z_b)-\omega(1)} = \frac{d_b}{d_z}\left(\frac{E_z}{E_b}\right)^{1/2} \begin{cases} \frac{z}{z_b} & (z-depression) \\ 1 & (z-elevation) \end{cases}$$

where,

$$\frac{d_b}{d_z}\left(\frac{E_z}{E_b}\right) = \left(\frac{C_z}{C_b}\right)^{-(1+m/2)}$$

(28)

This formulation clarifies that the direct z >1 involvement derives the redshift and the other blueshift of phonons. Numerical matching to the experimental result yields the ω(1) and m with ω($z_b$) and ω(z) as input for calculations, as demonstrated in the following subsections.

#### 3.2.1.2. Graphene and Black Phosphorus

Bond-order and the pattern of *sp*-orbital hybridization variations fascinate carbon allotropes including diamond, graphite, graphene, nanotube (CNT), $C_{60}$, nanobud (CNB), and GNRs with properties that are amazingly different. As an insulator, Diamond is transparent to light of all wavelengths but Graphite is an opaque conductor; the former has the ideally $sp^3$-orbital hybridized structure but the has the nonbonding unpaired π-bond electrons arising from the $sp^2$-orbital hybridization. Both SWCNT and monolayer graphene possess high electrical conductivity, ductility, tensile strength, and thermal conductivity. However, the involvement of the two-coordinated edge atoms endows GNR to perform quite differently from the SWCNT or the infinitely large monolayer graphene [61, 137, 138, 218]. Raman phonon frequencies relax under perturbations such as the allotropic coordination environment, pressure, and temperature [96, 219].

The database as a function of the number-of-layer [20, 21] of graphene and BP [210] enabled the verification of the framework for thickness dependence. In numerical calculations for graphene, the known effective bond length $d_g$ = 0.142 nm and $z_g$ = 5.335 for the bulk graphite, $z$ = 3 for the single layer graphene, and m = 2.56 for carbon were taken as input. Errors in measurements render only the accuracy of the derivatives but not the nature and the trends of observations.

Taking $z_g$ = 5.335 for the bulk graphite as a reference, one can obtain the reference ω(1) and the z-dependent frequency ω(z) for the vibration [219]. With the known D/2D peak shifting from 1367/2720 to 1344/2680 cm$^{-1}$ and the G mode shifting from 1582 to 1587 cm$^{-1}$ when the $z_g$ of graphite turns to be the monolayer (z = 3) graphene [20, 21, 43, 220], one can calibrate the relative shift of the vibration bands. For the monolayer graphene (z = 3, m = 2.56),

$$C_x(z, z_g) = \frac{\omega_x(z) - \omega_x(1)}{\omega_x(z_g) - \omega_x(1)} = \left(\frac{C_z}{C_{z_g}}\right)^{-(m/2+1)} \begin{cases} \frac{z}{z_g} & (D, 2D) \\ 1 & (G) \end{cases}$$

$$= \left(\frac{0.8147}{0.9220}\right)^{-2.28} \times \begin{cases} \frac{3.0}{5.335} = 0.7456 & (D, 2D; z = 3) \\ 1 = 1.3260 & (G; z = 3) \end{cases}$$

And the reference frequency,

$$\omega_x(1) = \frac{\omega_x(z) - \omega_x(z_g)C_x(z, z_g)}{1 - C_x(z, z_g)} = \frac{\omega_x(3) - \omega_x(z_g)C_x(3, z_g)}{1 - C_x(3, z_g)} = \begin{cases} 1276.8 & (D) \\ 1566.7 & (G) \\ 2562.6 & (2D) \end{cases} (cm^{-1})$$

The z dependent phonon frequency for graphene,

$$\omega(z) = \omega(1) + [\omega(z_g) - \omega(1)] \times D_L(z)$$
$$= \begin{cases} 1276.8 + 90.2 \times D_D(z) & (D) \\ 2563.6 + 157.4 \times D_{2D}(z) & (2D) \\ 1566.7 + 16.0 \times D_G(z) & (G) \end{cases}$$

(29)

Figure 13 shows the BOLS-LBA derivatives of the z-resolved Raman shifts of the G mode and the D/2D modes [21, 63, 161]. Likewise, inset b is the z-dependence of the BP $A^1_g$ mode [210]. The DPS for the layered BP revealed that reduction of the number-of-layer from multiple to single transits the $A_{2g}$ phonon from 466 to 470 cm$^{-1}$ and the $B_{2g}$ from 438 to 441 cm$^{-1}$. The n- resolved phonon positive shift indicates that both the $A_{2g}$ and $B_{2g}$ modes arise from oscillation of an invariant number of bonds [221]. The $\omega(1)$ for the BP was determined as $A_g^1$(360.20 cm$^{-1}$), $B_{2g}$(435.00 cm$^{-1}$), and $A_g^2$(462.30 cm$^{-1}$).

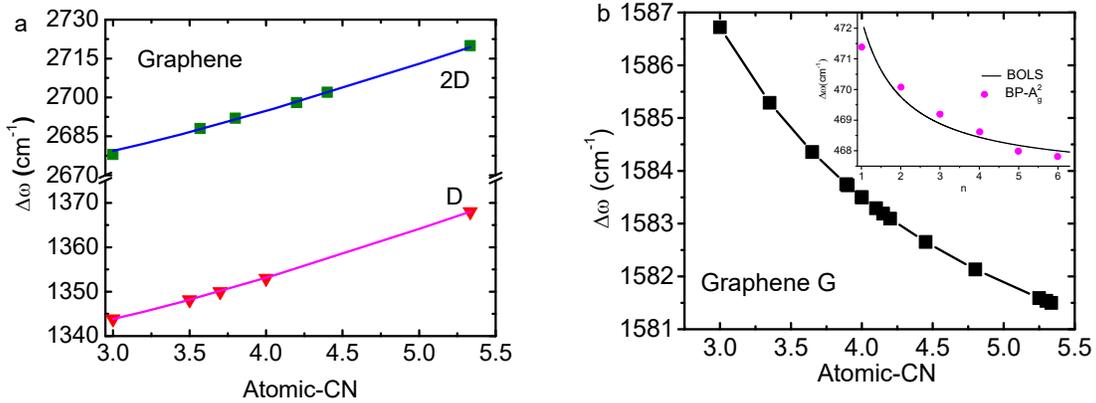

Figure 13 BOLS-LBA duplication of the z-resolved Raman shift for (a) the D/2D bands and (b) the G band for the layered graphene [20, 21, 43] and $A_g$ mode for the layered BP (inset f) [210] with derivative of information in Table 3. The scattered data counts the number of layers. (Reprinted with copyright permission from [221])

3.2.1.3.    (W, Mo)(S, Se)$_2$

The DPS in Figure 14 resolves the n-induced phonon frequency relaxation [222]. Phonons transit from the bulk component (DPS valleys) to the undercoordinated skin component (DPS peaks) of the layered structures. The frequency shifts of the two vibration modes suggests that the collective oscillation of bonds to the z-neighbors govern the $A_{1g}$ redshift while the single bond vibration drives the $E^1_{2g}$ blueshift, which are the same to the D/2D mode and the G mode of graphene, respectively.

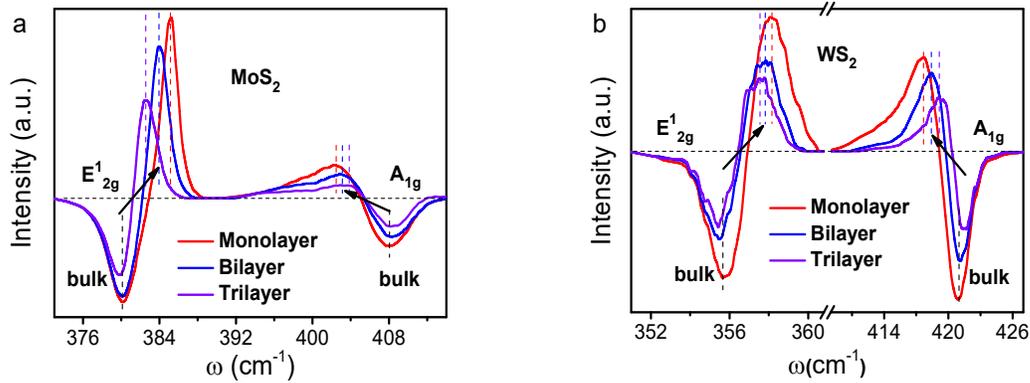

Figure 14. Number-of-layer resolved Raman DPS for (a) MoS$_2$ [39, 223] and (b) WS$_2$ [224]. Phonons transit from the bulk component (DPS valleys at 380 and 408 cm$^{-1}$ for the E$_{2g}$ and the A$_{1g}$ of MoS$_2$ and at 356 and 421 cm$^{-1}$ for the WS$_2$) to the skin (n-resolved DPS upward-shift peak for the E$_{2g}$ and downward for the A$_{1g}$) of the layered structures. The collective oscillation of z-neighbors govern the A$_{1g}$ redshift while the single bond vibration drives the E$^1_{2g}$ blueshift (Reprinted with copyright permission from [222])

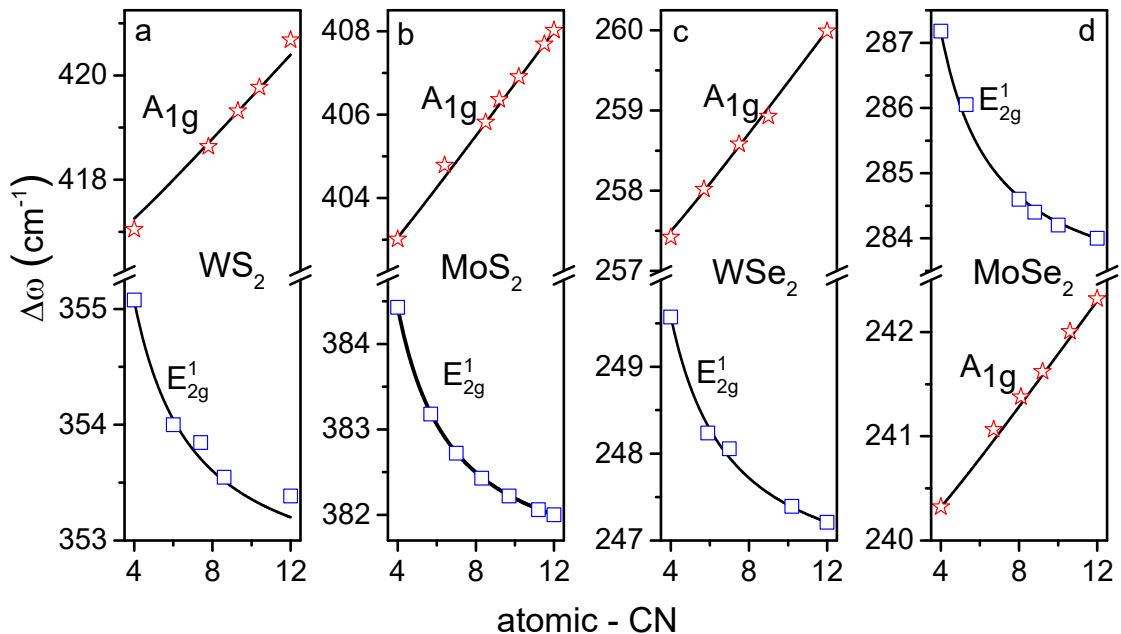

Figure 15 BOLS-LBA reproduction of the z-resolved Raman shift for (a-d) (Mo, W)(S$_2$, Se$_2$) with derivative of information in Table 3. The scattered data corresponds to the number of layers. (Reprinted with copyright permission from [221])

With the known $\omega(n = \infty)$ for the bulk and $\omega(n = 1)$ as input parameters, one can derive information show in Table 3 for the layered $MX_2$ by repeating the same calculation iteration of graphene [222]. Figure 15 shows the BOLS-LBA matching to the phonon frequency shift of the layered $MX_2$ with the derive m and $\omega_L(1)$ tabulated in Table 3. Theoretical reproduction of the n-dependence of Raman shift verifies that the dimer bond interaction dictates the $E^1_{2g}$ mode blueshift but the collective interaction of an atom with its z-neighbors governs the $A_{1g}$ mode redshift.

3.2.1.4.     Atomic-CN versus the Number-of-layer

The scattered data in Figure 13 and Figure 15 correspond to different number of layers while the lateral axis is the effective atomic CN, which gives rise to the z - n correlation as shown in Figure 16. When the n is greater than 6, the z saturates at the bulk value of 5.335 ($z = 2.55 + 0.45n$) for graphite and the bulk value of 12 ($z = 2.4 + 1.6n$) for the layered $MX_2$. Agreement between theoretical predictions and measured Raman shifts and the z-n transformation function prove the essentiality of the BOLS framework for the vibration dynamics in these 2D structures.

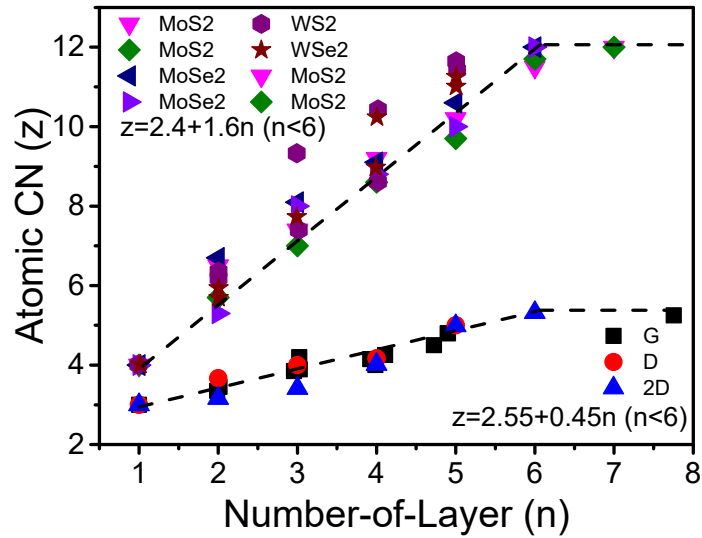

Figure 16. The atomic CN correlates to the number-of-layer linearly. For n > 6, the z saturates to the bulk values for graphite ($z = 5.335$) and for the fcc standard ($z = 12$) (Reprinted with copyright permission from [36, 222]).

### 3.2.2. Strain-induced Phonon Softening and Band Splitting

The data of tensile strain for the monolayer graphene [161] and MoS$_2$ [162] enabled the derived information of the average-bond force constant and the relative direction between the strain and a specific bond in the specimen. The strain-effect is given as [163],

$$\frac{\omega(z,\varepsilon)-\omega(1,0)}{\omega(z,0)-\omega(1,0)} = \frac{d(0)}{d(\varepsilon)}\left(\frac{E(\varepsilon)}{E(0)}\right)^{1/2} = \frac{\left(1-d^2\int_0^\varepsilon \kappa\varepsilon d\varepsilon / E_{z1}\right)^{1/2}}{1+\varepsilon}$$

$$\cong \frac{\left[1-\kappa'(\lambda\varepsilon')^2\right]^{1/2}}{1+\lambda\varepsilon'}$$

With $\kappa' = \kappa d_z^2/(2E_z) = $ constant

(30)

To feature the orientation mismatch between a bond and the uni-axial strain, one can introduce a strain coefficient λ bounded by $0 \leq \lambda \leq 1$. The $\varepsilon = \lambda\varepsilon'$ is for a bond that is not along the applied strain. The effective force constant $\kappa = 2E_z\kappa'd_z^{-2}$ keeps constant at limited strain. One can not measure the force constant $\kappa_0$ for a certain bond but the average of the entire specimen. The λ corresponds to the actual strains of the orientated bonds in a basic unit. The strain can be along (λ = 1) or perpendicular (λ = 0) or random (0 <λ <1) to a certain bond in the specimen. The bond along the strain is subject to a maximal extension and its phonon frequency shifts most; the bond perpendicular to the strain is subject to zero extension and keeps its phonon frequency, as illustrated in Figure 17a inset. The mechanical strain splits thus the phonon band and the amount of splitting depends functionally on the angle between the a specific bond and the uniaxial strain [36].

Combining eqs (29) and (30) yields the joint z and ε effect on the Raman shifts for GNR:

$$\omega(z,\varepsilon) = \omega(1,0) + \left[\omega(z_b,0)-\omega(1,0)\right]D_L(z) \times \frac{\left[1-\kappa'(\lambda\varepsilon')^2\right]^{1/2}}{1+\lambda\varepsilon'}$$

$$= \begin{Bmatrix} 1276.8 \\ 2562.6 \\ 1566.7 \end{Bmatrix} + D_L(z) \times \frac{\left[1-\kappa'(\lambda\varepsilon')^2\right]^{1/2}}{1+\lambda\varepsilon'} \times \begin{Bmatrix} 90.2 & (D) \\ 157.4 & (2D) \\ 16.0 & (G) \end{Bmatrix}(cm^{-1})$$

(31)

Figure 17 shows reproduction of the tensile strain induced red shifting and band splitting for (a) the 2D mode of graphene [213] and the $E_{2g}$ mode for $MoS_2$.[162]. Reproduction of the measurements turned out the reduced force constant, $\kappa' = \kappa d_z^2/(2E_z) = 0.30$, corresponding to $\kappa = 6.283$ N/m for graphene (z = 3) for graphene. The compressive [63, 161] strain only extrapolates the curve to opposite direction of the tensile strain of graphene [161]. Inset a illustrates the geometric relation between the uni-axial tensile strain and the C–C bonds denoted 1, 2, and 3. The $\lambda = 1.0$ for the lower branch indicates that strain is along a certain C-C bond of the GNR and the M-S bond of the $MoS_2$. The upper branches of $\lambda = 0.31$ and 0.12 are for the other bond of graphene and $MoS_2$, respectively.

For the GNR instance, the strain is along bond 2, $\theta = 0°$, $\lambda = 1$. The strain $\varepsilon_1 = \varepsilon_3 = \lambda\varepsilon_2 < \varepsilon_2$, the $\varepsilon_2$ is the maximum; at $\theta = 30°$, the strain is perpendicular to bond 3, $\varepsilon_1 = \varepsilon_2 > \varepsilon_3 \sim 0$. The hexagonal bond geometry allows one to focus on the strain only in the angle ranging of 0° and 30° with respect to a certain bond. Accordingly, there should be a branch without frequency shift as it is subject to $\varepsilon_3 \sim 0$ at $\theta = 30°$. The $\lambda = 0.31$ for GNR and 0.12 for $MoS_2$ and the $\lambda = 1.0$ for the lower branches in both cases mean that one C-C bond and one Mo-S bond projected to the surface are along the strain. The bonds 1 and 3 are lengthened by 31% with respect to bond 2 as shown inset a. The other Mo-S bond projection is elongated by 12%. The Mo-S bonds projecting along the layer. As the $\lambda$ changes with the angle between the bond and strain, any scale of band splitting and frequency shift due to strain can be resolved.

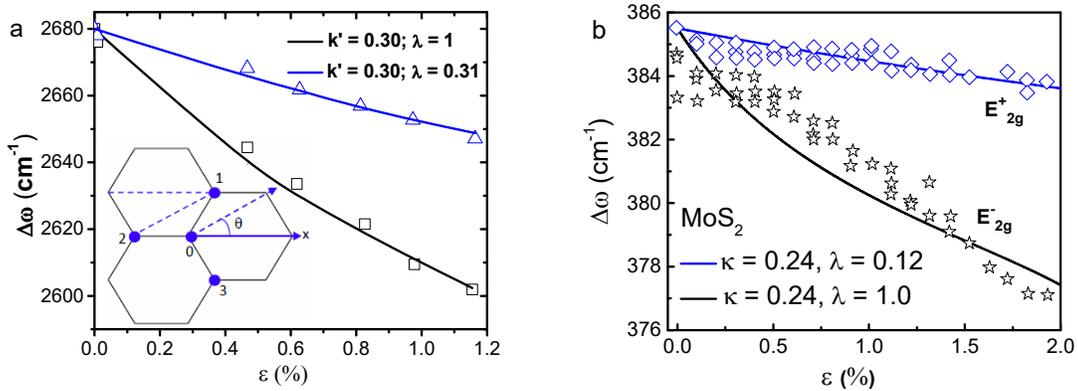

Figure 17 BOLS-LBA reproduction of strain effect on the (a) 2D mode of graphene [161] and (b) the $E_{2g}$ mode of $MoS_2$ [162]. Inset (a) illustrates the geometric relation between the $C_{3v}$ bonds to the $C_{2v}$ uni-axial strain. One extreme situation at $\theta = 0°$(along a C-C bond), $\varepsilon_1 = \varepsilon_3 = \lambda\varepsilon_2 < \varepsilon_2$ and the other at

θ = 30° (perpendicular to a C-C bond), $\varepsilon_1 = \varepsilon_2 > \varepsilon_3 \sim 0$. (Reprinted with copyright permission from [225])

One can readily estimate the force constant $k_0$ for the single C–C bond in the monolayer graphene from the C$_{3v}$ bond configuration symmetry and the deduced force constant $\kappa = 6.283$ N/m. The resultant $\kappa_{13} = 2\kappa_0$ for the parallel bonds labeled 1 and 3. This resultant bond connects to bond 2 in series and therefore the resultant force constant of the three bonds is $\kappa_{123} = 2\kappa_0/3$. Hence, the C–C bond force constant $\kappa_0 \approx 9.424$ N/m. Likewise, the force constant κ per M–X bond is derived as $\kappa_0 \approx 2.56$ N/m.

### 3.2.3. Mechanical Compression and Thermal Excitation
#### 3.2.3.1. Formulation

Using the $1+x \cong \exp(x)$ at $x \ll 1$ approximation for thermal expansion, one can formulate the thermal and pressure effects as follows (y = P, T) [96, 226],

$$\frac{\omega(z,y)-\omega(1,y_0)}{\omega(z,y_0)-\omega(1,y_0)} = \frac{d(y_0)}{d(y)}\left(\frac{E(y)}{E(y_0)}\right)^{1/2} \cong \begin{cases} (1-\Delta_T)^{1/2} \times \exp\left(-\int_{T_0}^T \alpha(t)dt\right) \\ (1+\Delta_P)^{1/2} \times \exp\left(+\int_{P_0}^P \beta(p)dp\right) \end{cases}$$

(32)

The thermally- and mechanically-induced energy perturbations $\Delta_T$ and $\Delta_P$ follow the relations [7],

$$\begin{cases} \Delta_T = \int_{T_0}^T \frac{\eta(t)dt}{E_z} = \int_{T_0}^T \frac{C_v(t/\theta_D)dt}{zE_z} \\ C_v(\tau,T) = \tau R\left(\frac{T}{\theta_D}\right)^\tau \int_0^{\theta_D/T} \frac{x^{\tau+1}e^x}{(e^x-1)^2}dx \end{cases}$$

And,

$$\Delta_P = -\int_{V_0}^V \frac{p(v)dv}{E_z} = -\frac{V_0}{E_z}\int_1^X p(x)dx$$

$$= \frac{\int_{P_0}^P v(p)dp - \int_{(VP)_0}^{VP} d(pv)}{E_{den}}$$

With,

$$\begin{cases} p(x) = \dfrac{3B_0}{2}\left(x^{-7/3} - x^{-5/3}\right) \times \left[1 + 3\dfrac{(B_0' - 4)(x^{-2/3} - 1)}{4}\right] & (B-M) \\ x(p) = 1 - \beta P + \beta' P^2 & (Polynomial) \end{cases}$$

(33)

The $\Delta_T$ is the integral of the Debye specific heat division by the bond energy. At temperature T higher than $\theta_D$, the two-dimensional specific heat $C_v$ converges a constant of $\tau R$ with R being the idea gas constant. The Debye temperature $\theta_D$ and the atomic cohesive energy $E_{coh} = zE_z$ are adjustable parameters used in calculating the $\Delta_T$. The $\Delta_P$ is derived from the integral of the Birch-Mürnaghan state-of-equation [227, 228]. The $V_0$ denotes the volume of a bond under the ambient pressure.

The compressibility β, β′ and binding energy density $E_{den} = E_z/V_0$ are tunable parameters variables used in calculating $\Delta_P$. The x(P) is an alternative of the equation-of-state. Matching the Birch-Mürnaghan equation to the measured x(P) curve, one can derive the β, β′, the bulk modulus $B_0$ ($\beta B_0 \cong 1$) and its first-order differentiation $B'_0$. Substituting the integrals (33) into (32), one can derivatives of the $\theta_D$, α, and $E_{den}$ and the compressibility derived from the x(P) relation with the known $E_{coh}$ by reproducing the P- and T-resoved Raman frequency shifts.

### 3.2.3.2. MX₂ Debye Thermal Decay

Figure 18 shows the reproduction of the thermally softened Raman shifts for the $MX_2$ with derived $E_{coh}$, and $\theta_D$ given in Table 3. The $\omega_x(1)$ and $\Delta\omega_x$ can also be obtained from matching the number-of-layer effect on the phonon frequency shift. Theoretical match is realized by adjusting the $\theta_D$ and the $E_{coh}$ without needing any hypothetic parameters [229], as the Raman shift features the $(E/d^2)^{1/2}$ that relaxes following functionally the integration of Debye specific heat. The small $\int_0^T \eta dt$ value at very low temperatures determines the slow decay of the Raman shift as the $\eta(t)$ is proportional to $T^\tau$ for the τ-dimensional system. Therefore, the $\theta_D$ determines the width of the shoulder, the $1/E_{coh}$ and the thermal expansion coefficient determine the slope of the curve at high temperatures.

Results derived from the matching to measurements favor the BOLS-LBA notion of the bond thermal relaxation. The atomic cohesive energy for the $MX_2$ follows the $E_{coh-MoS2}(1.35) < E_{coh-WS2}(2.15) < E_{coh-MoSe2}(2.30) < E_{coh-WSe2}(3.21)$ order. The $MoS_2$ shows a higher rate of bond energy attenuation and thermal expansion than those of others, because of its intrinsically weaker cohesive energy. Besides, the Debye temperature $\theta_D$ and the offset $\Delta\omega_x$ of frequency follow the same rank of cohesive energy. These observations suggest that the bond nature determines the relaxation dynamics under thermal excitation.

The cohesive energy is correlated to the electronegativity difference between the M and X constituent atoms. The electronegativity is valued at W(1.7), Mo(1.8), S(2.5), and Se(2.4) [230]. Differences in electronegativity for the compounds represent the polarity of chemical bonds, which follows the order: $MoSe_2(0.6) < WSe_2(0.7) = MoS_2(0.7) < WS_2(0.8)$. However, The order of atomic cohesive energy and the order of the electronegativity are different.

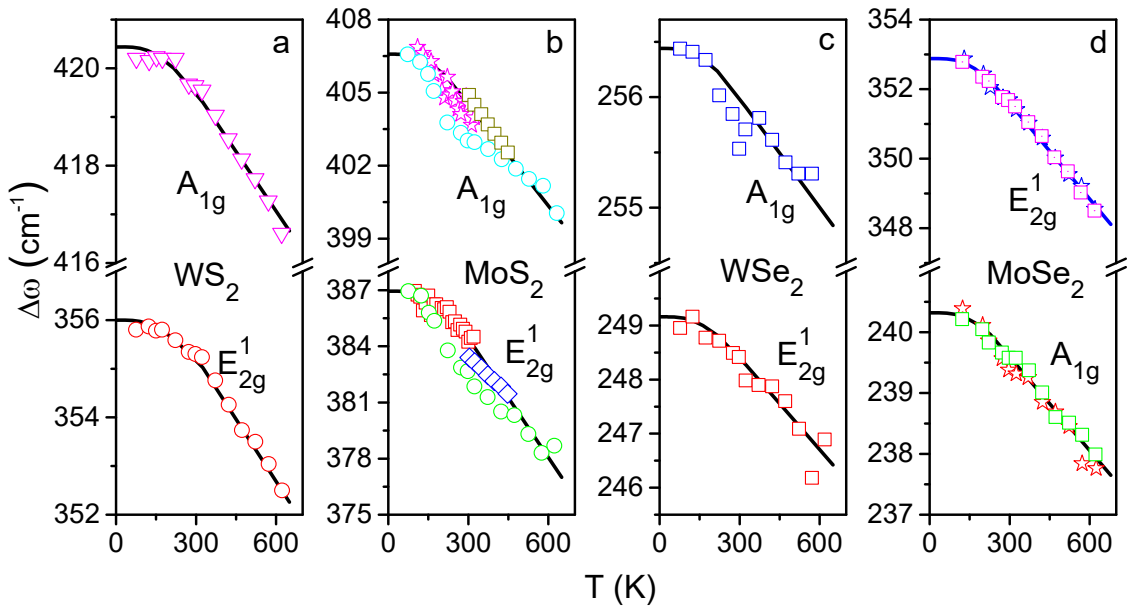

Figure 18. BOLS-LBA theoretical reproduction (solid lines) of the measured (scattered) Debye thermal decay of the $A_{1g}$ and $E^1_{2g}$ modes for (a) $WS_2$ [231], (b) $MoS_2$ [232-234], (c) $WSe_2$ [207] and (d) $MoSe_2$ [207]). (Reprinted with copyright permission from [229])

3.2.3.3.     BP and (W, Mo)$S_2$

Figure 19 shows the BOLS-LBA reproduction of the measured pressure and temperature dependent Raman shifts for BP [215, 235, 236]. The reproduction of the P-stiffened and T-softened Raman shifts results in the compressibility β, energy density $E_{den}$, elastic modulus and the Debye temperature featured in Table 3 [221]. The energy density $E_{den}$ = 9.46 eV nm$^{-3}$ and the compressibility $\beta$ = 0.32 GPa$^{-1}$. The $\theta_D$ = 466 K and the $E_m$ = 2.11 eV.

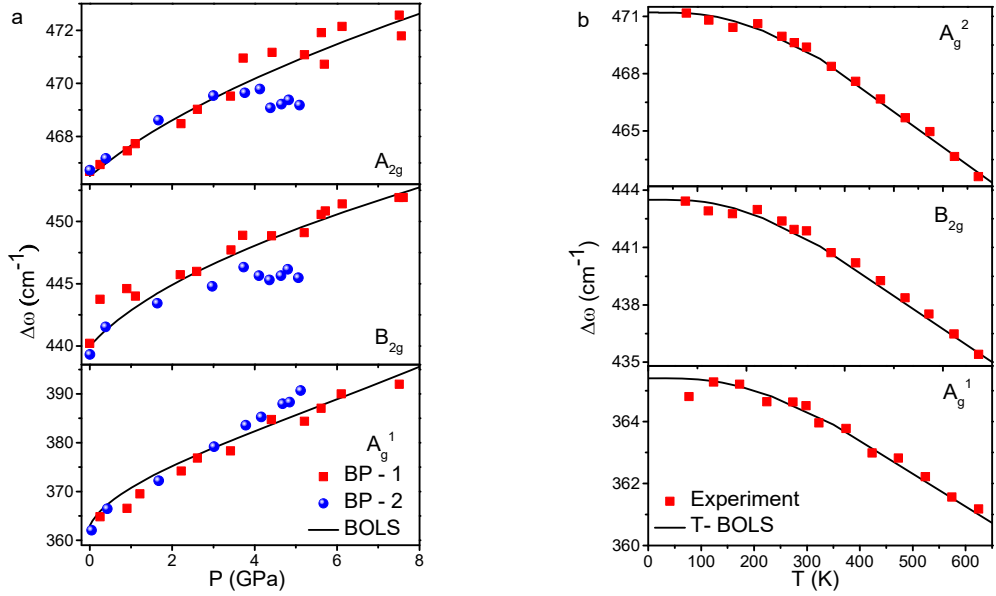

Figure 19. Theoretical reproduction of the (a) P- and (b) T-dependent $A_g^1$, $B_{2g}$, and $A_g^2$ mode for the few-layered BP [215, 235, 236] (Reprinted with copyright permission from [221])

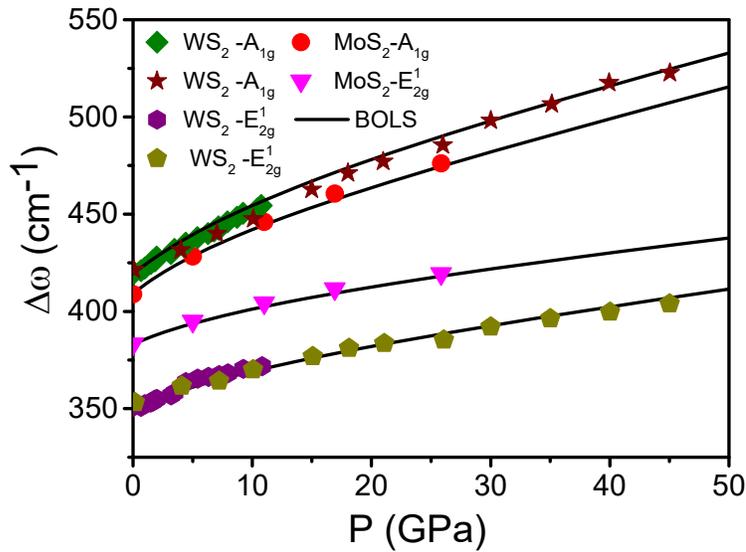

Figure 20 BOLS-LBA reproduction of the (a) compression-resolved phonon frequency shift of WS$_2$ [237] and MoS$_2$ [238].

The database as a function of temperature [239] and pressure [65] for graphene and the pressure for (W, Mo)S$_2$ [238] allows for information from the theoretical matching to measurements. Figure 20 shows BOLS-LBA reproduction of the pressure-resolved phonon relaxation for (W, Mo)S$_2$ with derived information summarized in Table 3.

3.2.3.4.     Carbon Allotropes

With the derived $\omega_x(1)$ and the measured $\omega_x(z_g)$, one can reproduce the thermal decay and P-resolved Raman shift of carbon allotropes, which derived the binding energy density, compressibility, Debye temperature, and mode cohesive energy, see Table 4. Matching to the measured Raman 2D mode frequency thermal decay for graphene [45] CNB [240], C$_{60}$ [241] in Figure 21a turns out that $\theta_D$ = 540 K, with the given E$_c$ of 3.11 eV/atom [36]. The $\theta_D$ is about one third of the T$_m$ =1605 K for the SWCNT [132]. At T ~ $\theta_D$/3, the Raman frequency shifts gradually from the nonlinear form to a linear when the temperature rises. The small $\int_0^T \eta dt$ values and the $\eta(t) \propto T^\tau$ relationship slower the Raman decay for the $\tau$-dimensional system at very low temperatures [239].

Figure 21a also shows the compression effect on the 2D mode of graphite [198, 242], SWCNT [243, 244], and diamond [70]. The $\theta_D$ determines the width of the shoulder and the thermal expansion and the inverse mode-cohesive energy E$_{mod-coh}$ determine the slope of the linear part at high temperatures. With the known relation E$_{m-coh}$(z) = C$_z^{-m}$E$_{m-coh}$(bulk), the bulk mode-cohesive energy and the bond nature index m = 2.56 for carbon, one can estimate the effective CN of the CNB and C$_{60}$, as shown in Table 4.

Different from the atomic cohesive energy, the mode cohesive energy varies from phase to phase for the same phonon band. For example, the atomic cohesive energy is 7.37 eV but G mode cohesive energy is 0.594 eV for a diamond. The former corresponds to energy of atomic evaporation of the specific crystal while the latter to energy activating the specific vibration mode. From the derivatives, the C$_{60}$ has fewer effective CN than the graphene because of the curvature strain. The C$_{60}$ shows the highest bond energy, atomic cohesive energy, and mode cohesive energy among all allotropes. All allotropes show much lower Debye temperatures than that of diamond, 2230 K, due to the undercoordination effect.

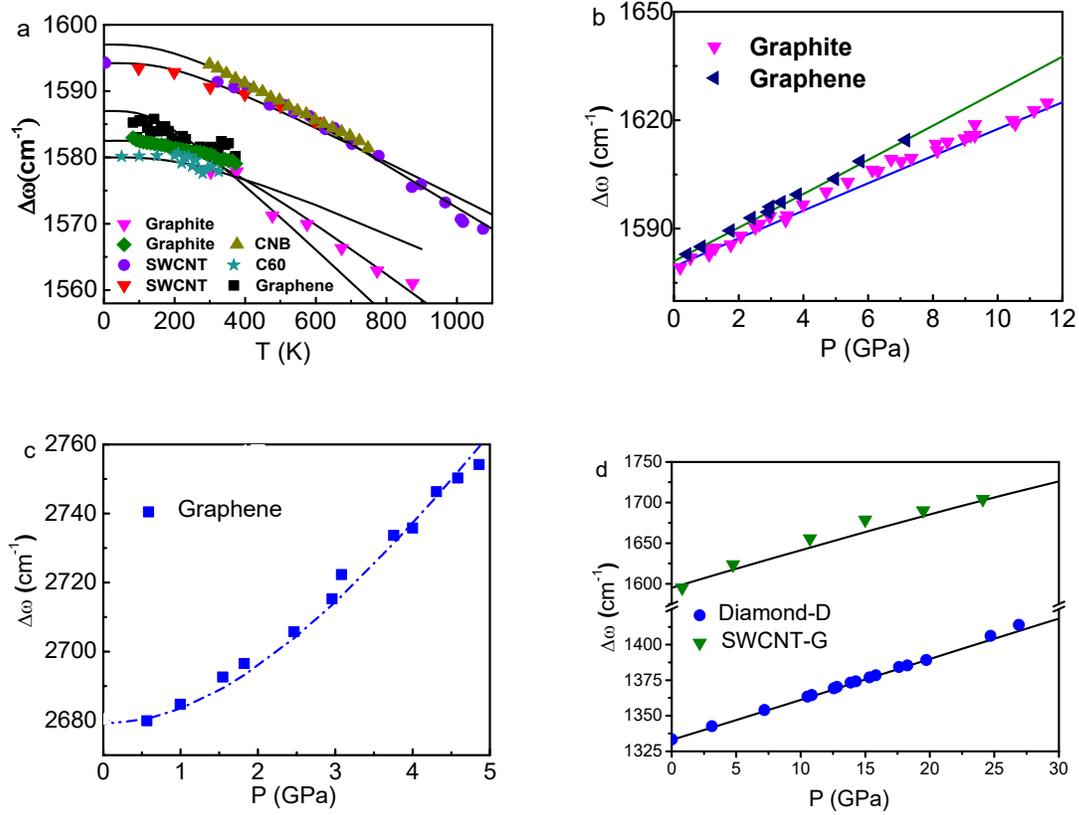

Figure 21. Thermal decay of Raman shifts at the ambient pressure for (a) diamond [70], CNB [240] and $C_{60}$ [245], graphene [45], graphite [242, 246] and SWCNT [243, 244]. (b) *P*-enhanced Raman shifts of diamond [247], SWCNT [248], graphite [249], graphene [65], and (c) graphene [65] at low pressures and (d) (a) diamond [247], SWCNT [248], measured at room temperature with derived information summarized in Table 3 [36, 192] (Reprinted with permission from [36, 192]).

Matching to the measured Raman 2D mode thermal decay for graphene in Figure 20b turns out that $\theta_D = 540$ K, with the given atomic cohesive energy of 3.11 eV/atom. The match to the measured P-enhanced Raman shift in Figure 20b yields the energy density of 320 eV/nm$^3$, the compressibility of $\beta = 1.145 \times 10^{-3}$ (GPa$^{-1}$) and its first derivative $\beta' = 7.63 \times 10^{-5}$ (GPa$^{-2}$).

Figure 21b-d show the BOLS-LBA reproduction of the P-enhanced Raman shifts carbon allotropes. Consistency to measurements yields the binding energy density ($E_{den}$) and the compressibility, as listed in Table 4. The $E_{den}$ of the graphene and SWCNT's are both higher than that of the diamond, agreein with XPS measurement [7]. Since the elastic modulus is proportional to the $E_{den}$, the elastic modulus of graphene and SWCNT are higher and they are hardly compressible in comparison to diamond.

Provided with a P-enhancement of the Raman shift, one can readily derive the elastic modulus for $C_{60}$ and CNB, as well.

### 3.2.4. Edge Discriminative Raman Reflectivity

Scanning tunneling microscopy [250] observations and quantum computations [61] confirmed that the Dirac-Fermi polarons preferentially generate at the zigzag-edge of monolayer graphene [251] and at sites surrounding graphite surface atomic vacancies [252]. The uniform √3d longer atomic distance along the edge endows the isolation and polarization of the dangling electrons by the locally and deeply entrapped core and bonding electrons. The dangling bond electrons at the armchair-edge and the reconstructed zigzag-edge of graphene (5 and 7 atomic rings), however, tend to form quasi-triple-bond between the edge atoms of shorter d distance. The polarized unpaired dangling bond electrons scatter the incident radiation substantially and hence lowers the Raman reflectivity of the D band at the zigzag edge in comparison to that at the arm-chaired edge [203, 204].

Table 3 Quantitative information derived from reproducing the (n, ε, T, P) dependent Raman shift of the layered BP, GNR, and $MX_2$ [36, 221, 225, 229].

| Stimuli | quantity | | graphene | BP | $MoS_2$ | $WS_2$ | $MoSe_2$ | $WSe_2$ |
|---|---|---|---|---|---|---|---|---|
| n | $\omega(1)(cm^{-1})$ | | $\begin{cases}1276.8 & (D)\\ 2562.6 & (2D)\\ 1566.7 & (G)\end{cases}$ | $\begin{cases}360.20 & (A_{1g})\\ 435.00 & (B_{2g})\\ 462.30 & (A_{2g})\end{cases}$ | 399.65($A_{1g}$) 377.03 ($E^1_{2g}$) | 416.61($A_{1g}$) 352.26 ($E^1_{2g}$) | 237.65($A_{1g}$) 265.12 ($E^1_{2g}$) | 254.84($A_{1g}$) 246.42 ($E^1_{2g}$) |
| | Bond nature index m | | 2.56 | 4.60 | 4.68 | 2.42 | 2.11 | 1.72 |
| | Bond length (nm) | Monolayer | 0.125 | 0.224/0.223[253] | 0.249/0.241[254] | 0.243/0.242[254] | 0.250/0.254[254] | 0.264/0.255[254] |
| | | bulk | 0.154 (Diamond) | | 0.255 | 0.284 | 0.276 | 0.286 | 0.301 |
| | | Monolayer | 1.04 | 0.527 | 0.338 | 0.538 | 0.575 | 0.613 |

|   |   |   |   |   |   |   |   |
|---|---|---|---|---|---|---|---|
|   | Bond energy (eV) | bulk | 0.615[255] (Diamond) | 0.286[255] | 0.181 | 0.390 | 0.435 | 0.487 |
| ε | Force constant κ (Nm$^{-1}$) |   | 6.28 | - | 2.56 | - | - | - |
| T | Debye θ$_D$ (K) |   | 540 | 466/400 [256] | 250 | 530 | 193 | 170 |
|   | Atomic E$_{coh}$ (eV/atom) |   | 3.11 | 2.11 | 1.35 | 2.15 | 2.30 | 2.45 |
|   | Thermal expansion α(10$^{-6}$K$^{-1}$) |   | 9.00 | 22.0[257] | 1.90 /8.65 | 10.10/- | 7.24/12.9 | 6.80/10.6 (α$_a$/α$_c$) |
| P | Energy density E$_{den}$ (eV/nm$^3$) |   | 320 | 9.46 | 21.9 | 29.6 | - | - |
|   | ββ′(10$^{-3}$GPa$^{-1}$/GPa$^{-2}$) |   | 1.15/0.0763 | 0.32/2.20 [258] | 13.30/0.96 | 25.60/0.88 | - | - |
|   | B$_0$/B′$_0$(GPa/-) |   | 690/5 (704/1)[259] | 84.10/4.69[260] | 47.7/10.6 | 63.00/6.50[237, 261] | - | - |

Table 4 Quantitative information derived from reproducing the measured T- and P- resolved Raman shift for graphene, SWCNT, C$_{60}$, CNB, graphite and diamond [192].

| Parameters |   | SWCNT | Graphite | C$_{60}$ | CNB | Diamond |
|---|---|---|---|---|---|---|
|   |   | G mode | | | | D mode |
| P | CN | 3.0 | 5.335 | 2.465 | 5.605 | 12 |
|   | A (ω-P slope) | 0.024[243] | 0.028[242] | 0.0165[241] | 0.0273[240] | 0.033[70] |
|   | α(10$^{-6}$ K$^{-1}$) | 8 | 8[262] | 1 | 5 | 0.8 [263] |
|   | $B_0/B_0^{'}(GPa/-)$ | 865/5.0 | 39/10.0[259] | — | — | 446/3.6[264] |

|  | β/β′($10^{-3}$GPa$^{-1}$/GPa$^{-2}$) | 1.156/0.0770 | 18.440/0.4427 | — | — | 2.120/0.0035 |
|---|---|---|---|---|---|---|
|  | $E_{den}$ (eV/nm$^3$) | 456.32 | 347.68 | — | — | 249.59 |
| T | d(nm) | 0.125 | 0.142 | 0.118 | 0.143 | 0.154 |
|  | $E_b$ (eV) | 1.038 | 0.756 | 1.22 | 0.743 | 0.614[150] |
|  | $E_{mode-coh}$ (eV) | 0.817 | 0.700 | 1.188 | 0.718 | 0.594 |
|  | $θ_D$(K) | 600 | 1000 | 650 | 550 | 2230[265] |

### 3.3. Summary

The notion of multifield bond oscillation has reconciled the following thrusts

1) Connection the continuum approaches to the quantum theory by focusing on the perturbed bond relaxation to the phonon frequency.
2) Reproduction of the spectrometrically measured phonon frequency shift.
3) comprehension of the Multifield driven bond-phonon-property correlation of a substance;
4) Quantitative information on bonding dynamics derived from the combination of the multifield bond oscillation framework and its driven phonon spectrometrics.

Being able to connect the continuum to quantum premises, the combination of the BOLS-LBA notion and the enabled Raman spectroscopy has enabled the quantitative information about the perturbation-relaxation-property that is beyond the scope of other available approaches. The practice of the average-bond bond oscillating dynamics has reconciled (z, ε, P, T) effects on the Raman shifts of BP, GNR, and MX$_2$ with not only clarification of the intrinsic phonon relaxation origin but also information on the bonding dynamics and their properties to the 2D species. Consistency between theory expectations and Raman measurements clarified and quantified the following:

1) The number-of-layer reduction induced phonon frequency redshift results from the undercoordination induced bond contraction and bond strength gain. The collective vibration of bonds between a certain atom and its z neighbors dictates phonons undergoing redshift while the dimer interaction dictates the phonons undergoing blueshift occurred when the number-of-layer is reduced. Matching the number-of-layer reduction induced phonon frequency shift offers the local bond length, bond energy, band nature index, referential wavenumber $\omega(1)$ and its bulk shift, and the effective coordination numbers for the layered 2D structures.

2) Bond stretching by the directional strain resolves the phonon red-shifting and phonon band splitting. Reproduction of the strain effect on phonon relaxation gives rise to the average-bond force constant and the relative direction between the bond and the strain.

3) Bond thermal expansion softens all phonons. One can derive the mode cohesive energy $E_{coh}$ and Debye temperature $\theta_D$ from reproducing the Debye thermal decay of phonon frequencies.

4) Mechanical compression stiffens all phonons. Theoretical matching to the effect of compression derives the binding energy density $E_{den}$, compressibility $\beta$, bulk modulus $B_0$ and their first derivatives $\beta'$ and $B'_0$.

Findings proved the essentiality of the BOLS-LBA notion that has empowered the Raman spectroscopy substantially in gaining coordination-resolved, quantitative information about the multifield dynamics of the representative bond and the properties of a crystal. Most strikingly, freely hypothetic adjustable parameters are avoided in the entire process of data analysis. The attainments may inspire a new yet practical way of thinking about the stimulus-bond-property correlation and approaches dealing with substance from the perspective of average-bond approach with interdisciplinary nature, offering references for functional materials design and synthesis. Extending the exercises to covering more stimuli and more properties would be even more challenging, fascinating and rewarding.

# 4. Sized Crystals

Highlight

- DPS resolves the skin thickness of the core-shell structured nanocrystals and liquid droplets.
- Intergrain coupling emits THz waves with frequency proportional to the inverse of grain size.
- Dimer and collective oscillators originate optical phonon frequency shift contrastingly.
- Phonon frequency and elasticity hold the same trends of compression and Debye thermal decay.

## 4.1. Skin Thickness of the Core-shelled Structures

Using the DPS, one can confirm the core-shell structure and determine the skin thickness of solid nanocrystals and water nanodroplets. The DPS was conducted to distill the skin bond vibration frequencies from their bulk mixture by differencing the bulk reference spectra from those of the sized samples. The DPS not only determines the skin-shell thickness but also distinguishes the performance of bonds and electrons in the skin shells in terms of length and stiffness.

Figure 22 insets show the peak-area normalized Raman spectra collected from the sized $CeO_2$ nanocrystals [40] and from the $H_2O$(95%) + $D_2O$(5%) droplets [266] under the ambient conditions. The measurements were focused on the D-O mode (centered at 2500 $cm^{-1}$) for the liquid and the mode centered at 464 $cm^{-1}$ for the $CeO_2$. The spectra peak area normalization to one unit aims to minimizing the experimental artifacts. Insets c and d illustrate the core-shell structure and its potential well, which shows the BOLS effect on bond contraction and potential well depression in the undercoordinated skin region up to three atomic layers [267].

The DPS thus resolves the phonons transiting their population from the bulk (valley) to the skin (peak). The DPS blueshift from the valley to the peak represents the skin bond stiffness gain and the redshift is associated with the undercoordination-induced subjective polarization of the surface electrons that screen and split the local potentials [6]. The $CeO_2$ skin covering sheets show both redshift and blueshift dominated by the bond contraction (blueshift) and the electron polarization (redshift). However, water droplets show only the D-O bond contraction featured at 2500 $cm^{-1}$ as the polarization softens the O:H phonons in the THz regime, from 200 to 75 $cm^{-1}$ (ref [129]). The polarization on the high-frequency D-O bond is indistinguishable.

An integration of the DPS peak is the number/volume fraction of bonds transiting from the core to the skin covering shell of the nanostructures. For a spherical structure of D across, the $f(D) = \Delta V/V = 3\Delta R/R = 6\Delta R/D$, which gives rise to the shell thickness $\Delta R$ of 0.5 nm for the $CeO_2$ nanocrystals and 0.09 nm for the $H_2O$ droplets. The 0.5 nm is two atomic diameters and 0.09 nm is the length of the H–O dangling bond that is 10% shorter than the H–O bond of 0.10 nm length in the bulk water [124], which is beyond the scope of available methods. The direct measure of the skin-shell thickness proves for the universal core-shelled configured liquid and solid nanostructures. The unusual performance of bonds and electrons in the skin-shells and the varied skin/volume ratio dictate the size dependency of nanostructures [267].

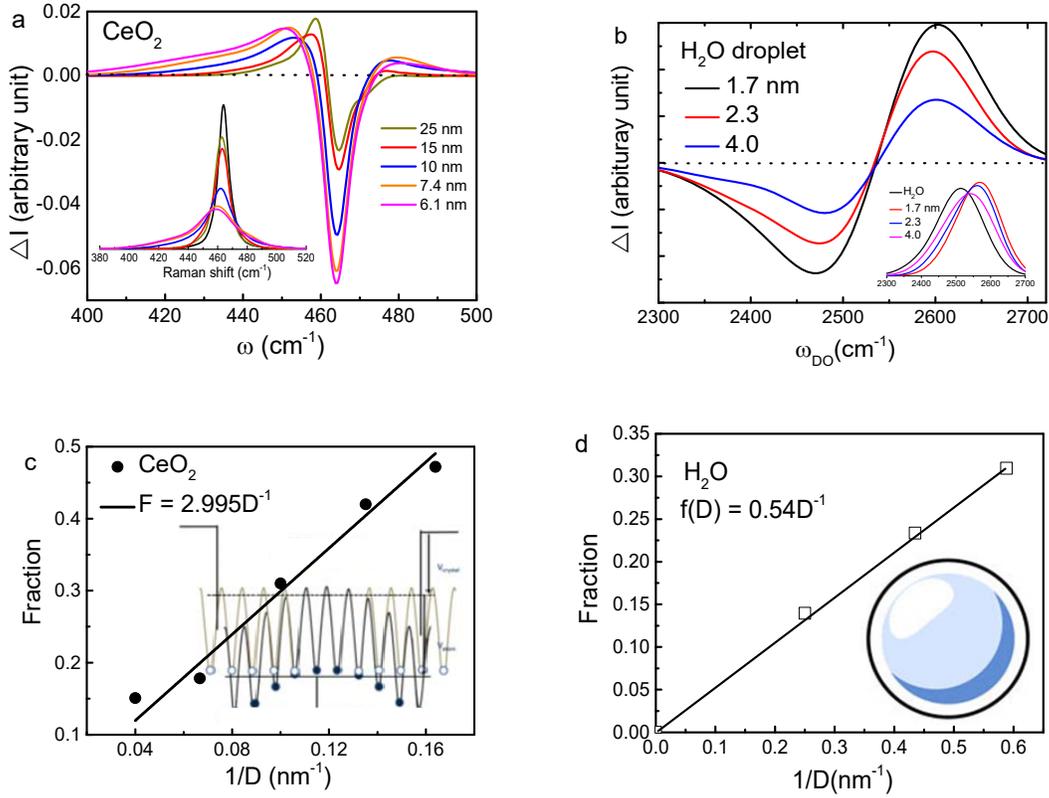

Figure 22. DPS verification of the core-shell configuration for the sized (a) CeO2 nanocrystals and (b) H2O droplets with insets a and b showing the spectral area normalized peaks (original spectral data are sourced from [40, 266]). Insets c and d illustrate the core-shell structure and its BOLS defined potential well depression in the skin covering shells. The fraction coefficient f(D) = $\Delta V/V = 3\Delta R/R = 6\Delta R/D$ defines the skin-shell thickness (b) $\Delta R$ = 0.5 nm for CeO2 nanocrystals and (d) $\Delta R$ = 0.09 nm of the H–O dangling bond length for H2O droplets.

4.2. Nanocrystals: Size and Thermal Effects

4.2.1. Raman Shift of the Core-shelled Crystals

Based on the proven core-shell configuration of nanostructures, one can derive the size-induced phonon relaxation in a spherical nanostructure of the dimensionless form of size K that is the number of atoms of d across lined along the radius of a spherical dot. Measurements show that the Raman shift $\omega(K) - \omega(\infty)$ changes with the inverse of crystal size AK-1. The A is the slope. A > 0 or < 0 corresponds to the phonon frequency blue or the redshift.

At the ith atomic layer, the effective atomic CN follows the relation: z1 = 4(1-0.75/K), z2 = z1 +2, z3 = z2+ 4, and z4 = 12. The i counts from the outermost layer inward. Incorporating the BOLS notion into the spectrometric measurements of the core-shelled nanostructures, see section 2.4.1.2, yields,

$$\omega(K) - \omega(\infty) = \begin{cases} \dfrac{A}{K} & \text{(Measurement)} \\ \Delta_R(\omega(\infty) - \omega(1)) & \text{(Theory)} \end{cases}$$

$$\omega(K) - \omega(\infty) = \begin{cases} \dfrac{-A}{K} & \text{(Measurement)} \\ \Delta_R(\omega(\infty) - \omega(1)) & \text{(Theory)} \end{cases}$$

with

$$\Delta_R = \sum_{i \leq 3} \gamma_i \left( \dfrac{\omega_i}{\omega_b} - 1 \right) = \begin{cases} \sum_{i \leq 3} \gamma_i \left( z_{ib} C_i^{-(m/2+1)} - 1 \right) & (z = z) \\ \sum_{i \leq 3} \gamma_i \left( C_i^{-(m/2+1)} - 1 \right) & (z = 1) \end{cases}$$

(5)

Hence, the frequency shift from the reference dimer vibration frequency $\omega(1)$ to the bulk value, $\omega(\infty) - \omega(1) = -A/(\Delta_R K)$, is a constant as $\Delta_R \propto K^{-1}$. The $\gamma_i = \Delta V/V = \tau C_i/K$ is the skin/volume ratio and m is the bond nature index correlating to the bond length and energy. One can obtain the reference frequency $\omega(1)$ given the slope A of the experimental profile. The sum over the skin shell represents that the skin bond relaxation dictates the intrinsic size dependency of the phonon frequency shift for nanostructures.

### 4.2.2. Skin dominated Size Dependency
#### 4.2.2.1. Group IV Nanocrystals

With the core-shell configuration and the known bulk frequency shifts as references, one can reproduce the size dependence, or z-resolved Raman shift of nanostructures. Figure 23 shows the size K-resolved Raman shift for the group IV nanostructures. Except for the layered graphene G mode that undergoes blueshift with the inverse of thickness 1/K, the rest vibration modes for the group IV elemental nanocrystals follow the linear $\Delta\omega$-1/K dependence. The linear 1/K dependence can be expressed empirically in terms of the extended Grüneisen constant, $\partial\omega/\partial(1/K)$, though the physics behind the trends is tremendously propounding – undercoordination induced skin bond contraction originates and the skin/volume ratio governs the magnitude of the phonon frequency shift.

The CN-reduction derived phonon frequency redshift occurs to the covalent system of Ge, Si, diamond, and the D and 2D modes of graphite, which evidence that the phonons of the covalently bonded systems are dominated by the collective oscillation of the associated oscillators. However, the G mode dimer oscillation exists in the anisotropic graphite and graphene. Theoretical reproduction of the measured Raman shift derived the bond nature index, the reference $\omega(1)$ as Table 5 listed. The bond length and energy follow the BOLS regulation.

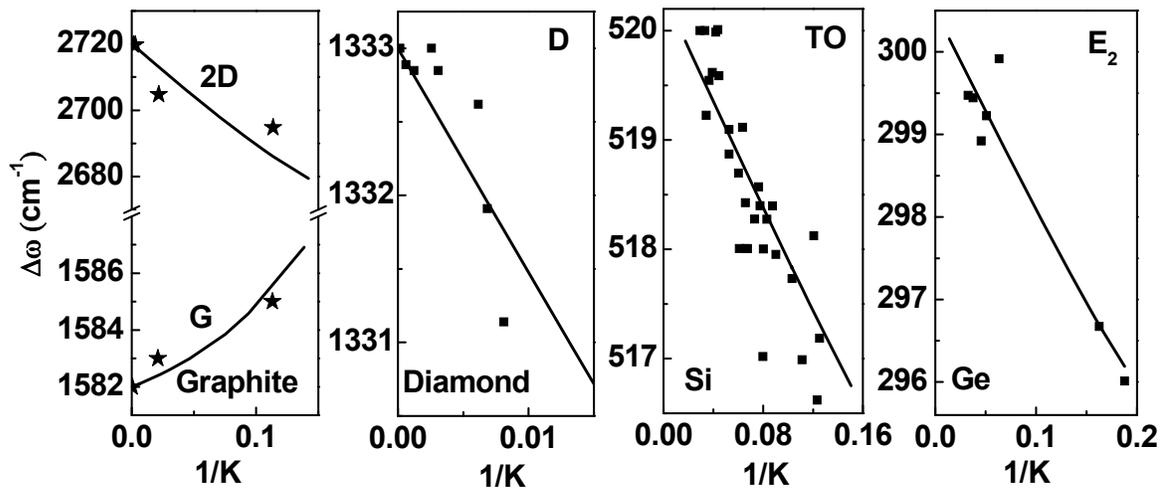

Figure 23. BOLS reproduction of the K-resolved Raman shift for (a) Graphite [268] and Graphene [20, 21, 43, 220], (b) Diamond [269, 270], (c) Si [271], and (d) Ge [272, 273] with derivative of information in Table 1.

4.2.2.2.   TiO$_2$

The anatase TiO$_2$ has six Raman active modes of 3E$_g$ at 144, 196, and 639 cm$^{-1}$, 2B$_{1g}$ at 397 and 519 cm$^{-1}$, and 1A$_{1g}$ at 513 cm$^{-1}$. The rutile TiO$_2$ has four Raman odes of A$_{1g}$ at 612 cm$^{-1}$, B$_{1g}$ at 143 cm$^{-1}$, E$_g$ at 447 cm$^{-1}$, and B$_{2g}$ at 826 cm$^{-1}$. The A$_{1g}$ mode of rutile phase at 612 cm$^{-1}$ undergoes a redshift while the E$_g$ mode of anatase phase at 144 cm$^{-1}$ is subject to a blueshift as the crystal size turns to be smaller [274-277]. The size trends of the phonon relaxation and the Yong's modulus Y(K) follow the linear dependence of the inverse size of the crystals but at slightly different slopes, both of which are governed by the skin-shell bond length and energy relaxation.

One can determine the $m$ value; the $\omega(1)$ and its bulk shift $\omega(\infty) - \omega(1)$ from matching the BOLS prediction to the measured/calculated size-reduction stiffened $Y(K)$ and the softened/stiffened Raman modes. The size-reduction induced Raman $E_g$ mode blueshift is governed by the same mechanism of the graphene G mode arising from dimer oscillator.

Theoretical reproduction of the measured size trend, in Figure 24, for the (a) $Y(K)$[278], and the $\Delta\omega(K)$ (b) of the $E_g$ mode of anatase phase (144 cm$^{-1}$) [277], and (c) the $A_{1g}$ mode of rutile phase(612 cm$^{-1}$) [274] for the TiO$_2$ under the ambient conditions. Theoretical calculation optimized the bond nature index $m = 5.34$ [33]. The $A_{1g}$(612 cm$^{-1}$) and the $E_g$(144 cm$^{-1}$) shift contrastingly with the crystal size. Decoding the size effect on the Raman optical shift yields the $\omega(1) = 610.25$ cm$^{-1}$ for TiO$_2$ dimers and their bulk shifts of 1.75 cm$^{-1}$ for the $A_{1g}$ mode, and $\omega(1) = 118.35$ cm$^{-1}$ and their bulk shifts of 25.65 cm$^{-1}$ for the $E_g$ mode of anatase phase.

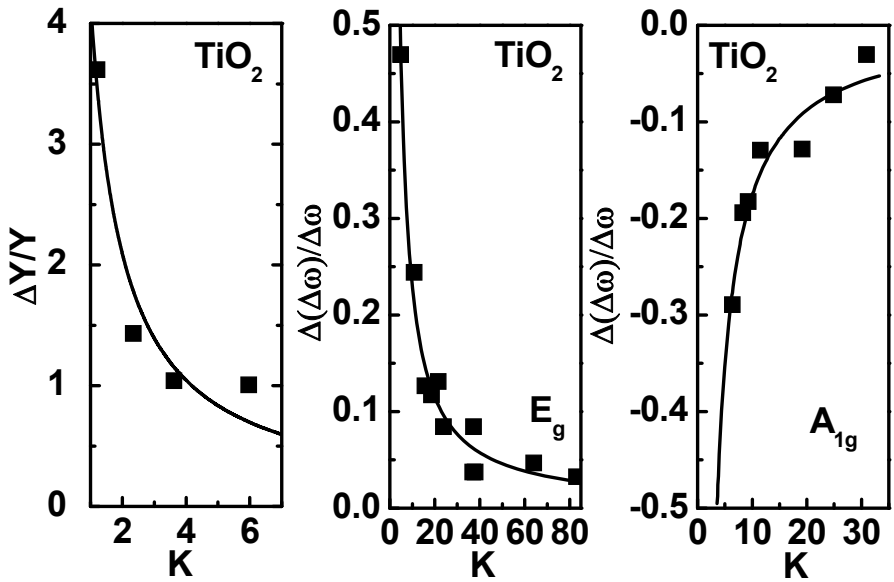

Figure 24. Theoretical reproduction (solid curves) of the experimental (scattered data) K-dependence of the (a) elastic modulus [278], (b) the $E_g$ mode at 144 cm$^{-1}$ and (c) the $A_g$ mode at 612 cm$^{-1}$ [274, 277].

4.2.2.3.    CeO$_2$, SnO$_2$, ZnO, CdS, CdSe, and Bi$_2$Se$_3$

The LO phonons of CeO$_2$ [40, 87] SnO$_2$ [279], and InP [280], ZnO [281, 282], CdS$_{0.65}$Se$_{0.35}$ [283] and CdSe [95, 284] show consistently the size-reduction induced redshift. The LO frequency goes lower when the CdS film is thinner than 80 nm and the CdSe nanodot is smaller than 9.6 nm [22]. When the CdSe crystal size is reduced to 3.8 nm, the peak frequency shifts by about 3 cm$^{-1}$ with respect to the bulk [23]. The Raman-active modes of Bi$_2$Se$_3$ nanoplatelets show a few wave numbers redshift as the thickness is decreased in the vicinity of $\sim$ 15 nm [19], similar to that of the D and 2D modes in the number-of-layer resolved graphene [20, 21].

Repeating the same iteration of numerical reproduction of the measurements, one able to determine the $\omega(1)$ or the $\omega(\infty) - \omega(1)$ and the bond nature index. Figure 25 shows typically the size softened optical phonons for compound semiconductors. Results justify the BOLS validity with derived information on the corresponding nanograin dimer vibration frequency of reference.

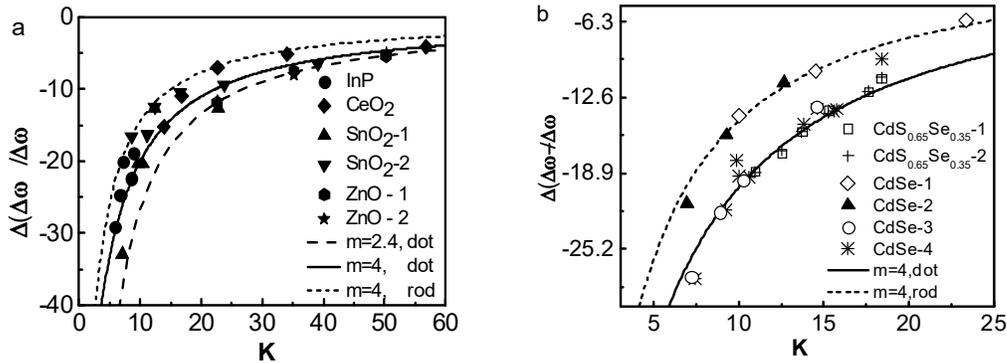

Figure 25. BOLS reproduction of the measured (scattered datum) size-resolved LO phonons of (a) InP [280], CeO$_2$ [40, 87] SnO$_2$ [279], ZnO [281, 282], and (b) CdS$_{0.65}$Se$_{0.35}$ [283] and CdSe [95, 284] nanoparticles with derived information given in Table 5.

Table 5. Nanograin dimer vibration frequencies and their redshift with size increases.

| Material | m | Mode | $\omega(\infty)$(cm$^{-1}$) | $\omega(1)$ (cm$^{-1}$) |
|---|---|---|---|---|
| Graphene / | | 2D | 2565.1; | 2562.6; |
| Graphite | 2.56 | G | 1587.0 | 1566.7 |
| Diamond | | D | 1333.3 | 1276.8 |

| Material | | Mode | | |
|---|---|---|---|---|
| Si | 4.88 | TO | 520.0 | 502.3 |
| Ge | 4.88 | E$_2$ | 302 | 290.6 |
| TiO$_2$ | 5.34 | A$_{1g}$ | 612 | 610.25 |
| | | E$_g$ | 639 | 600.00 |
| | | E$_g$ | 144 | 118.35 |
| CeO$_2$ | | LO | 464.5 | 415.1 |
| SnO$_2$ | | A$_{1g}$ | 637.5 | 613.8 |
| InP | | LO | 347 | 333.5 |
| ZnO | | E$_2$ | 441.5 | 380 |
| CdS$_{0.65}$Se$_{0.35}$ | 4.0 | LO$_1$ CdSe-like | 203.4 | 158.8 |
| | | LO$_2$ CdS-like | 303 | 257.7 |
| CdSe | | LO | 210 | 195.2 |
| CdS | | LO | 106.23 | 106.57 |
| Bi$_2$Se$_3$ | | A$^2_{1g}$ | 72.55 | 40.57 |

4.2.3. Intergrain Interaction Derived THz Phonons

The LFR phonon frequency shifts linearly with the inverse size $K$. The $K$ increases toward infinity, the LFR disappears, $\omega(\infty) = 0$. This fact implies simply that the intergrain interactions originate both the LFR mode and its LFR blueshift due to size reduction. The nanoparticle-substrate and intergarain interactions determine the slope of the ω-1/$K$ profile. Different from the Raman mode redshift arising from collective vibration in the surface region of the nanograin, the LFR blueshift is predominated by intergrain interactions. Linearization of the LFR size trend yields information about the interparticle interaction, as slisted in Table 6 [285].

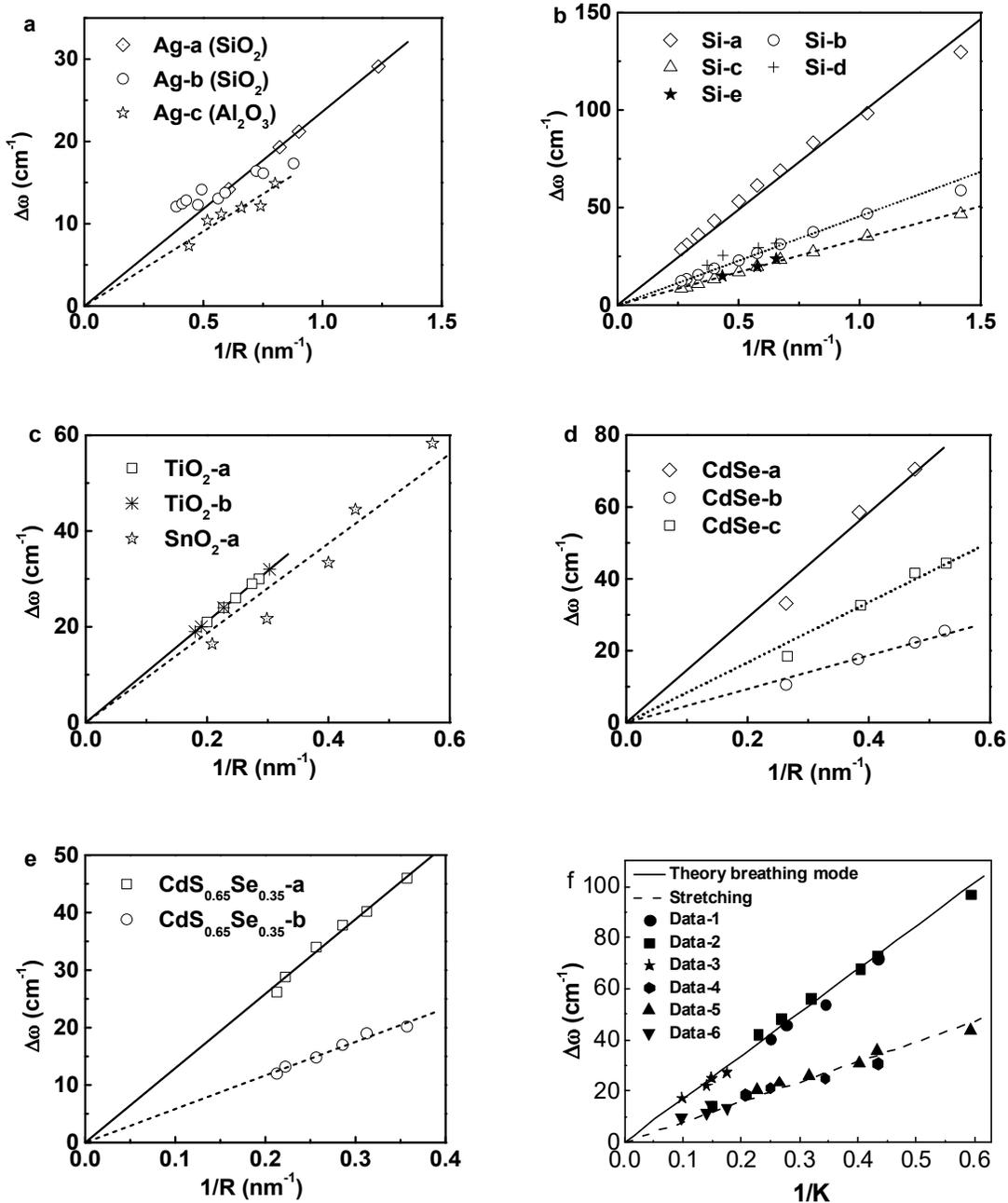

Figure 26. Size-reduction derived LFR mode for (a) Si [35, 38, 79, 88, 271, 286], (b) Ag-a [287], Ag-b [83] Ag-c [82], (c) $TiO_2$ [288] and $SnO_2$ [87], (d) CdSe-a, CdSe-b and CdSe-c [289], (e) $CdS_{0.65}Se_{0.35}$-a (embedded in glass) [283], (f) ZnO [290-292] nanograins (Reprinted with permission from [9]).

Table 6 Linearization of the LFR modes for nanograins.

| Sample | A (slope) | Sample | A (slope) |
|---|---|---|---|
| Ag-(a, b) | 23.6 ± 0.72 | Ag-c | 18.2 ± 0.56 |
| CdSe-1-a | 146.1 ± 6.27 | CdSe-1-b | 83.8 ± 2.84 |
| CdSe-1-c | 46.7 ± 1.39 | Si | 97.77; 45.57; 33.78 |
| SnO-a | 93.5 ± 5.43 | TiO-(a, b) | 105.5 ± 0.13 |
| CdS-a | 129.4 ± 1.18 | CdS-b | 58.4 ± 0.76 |

4.2.4. Joint Size and Thermal Effect

Figure 27 shows the theoretical reproduction of the joint crystal size and thermal effect on the phonon frequency shift of CdSe nanocrystals [293, 294]. The reproduction shows consistently that the crystal dimensionality reduction from bulk to a rod or a dot at the nanometer scale shorten the length and enhances the energy of bonds involved in the collective oscillation, showing the redshift. However, the characteristic phonon frequency follows the Debye thermal decay that provides information of the Debye temperature and the atomic cohesive energy. The inverse of the slope at higher temperatures is the inverse of $E_{coh}$, which is also the Grüneisen parameter, $\partial\omega/\partial T$.

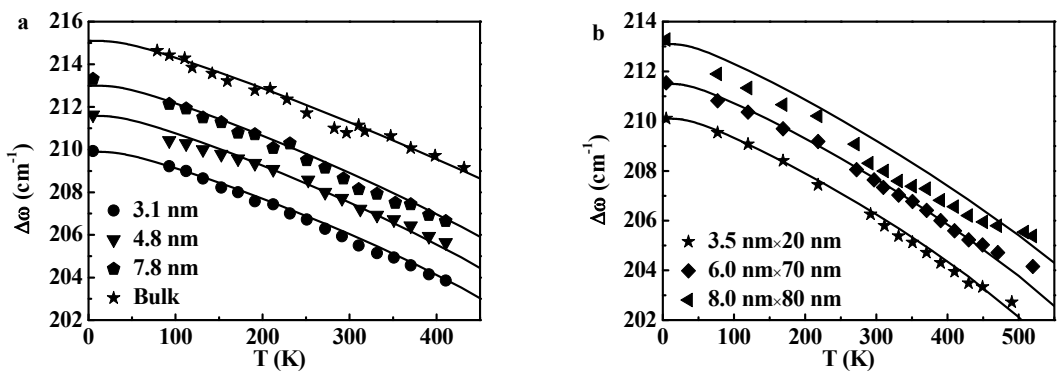

Figure 27. Numerical reproduction (solid curves) of experimental (scattered data)[293] Raman thermal decay of (a) CdSe nanoparticles and (b) cylindrical nanorods of different radius and lengths derived with information given in Table 7 (reprinted with permission from [294]).

Table 7 Information yielded from theoretical matching to the Raman thermal decay of for sized CdSe nanoparticles and nanorods [294].

|      | Input |         |       | Output  |         |            |            |                    |
|------|-------|---------|-------|---------|---------|------------|------------|--------------------|
|      | R (nm) | $A(d\omega/dT)$ | $<z>$ | $d_z$(nm) | $E_z$ (eV) | $E_{coh}$ (eV) | $\theta_D$ (K) | $\omega(z)$(cm$^{-1}$) |
| Bulk | ∞     | 0.0142  | 12    | 0.2940  | 0.153   | 1.84       | 450        | 215.0              |
|      | 7.8   | 0.0162  | 9.14  | 0.2883  | 0.169   | 1.57       | 300        | 213.0              |
| Dot  | 4.8   | 0.0168  | 7.36  | 0.2824  | 0.204   | 1.47       | -          | 211.6              |
|      | 3.1   | 0.0170  | 4.81  | 0.2666  | 0.250   | 1.40       | -          | 209.9              |
|      | 8.0   |         | 10.14 | 0.2906  | 0.160   | 1.65       | -          | 213.1              |
| Rod  | 6.0   | 0.0157  | 9.52  | 0.2892  | 0.164   | 1.54       | -          | 211.7              |
|      | 3.5   | 0.0174  | 7.76  | 0.2840  | 0.176   | 1.42       | -          | 210.1              |

4.2.5. Vibration Amplitude and Frequency

According to Einstein's relation, $\mu(c\omega x)^2/2z = k_B T$, the atomic vibrational amplitude and frequency of an oscillator in a surface ($z = 4$) follows,

$$\frac{\omega_1}{\omega_b} = z_{ib} C_1^{-(m/2+1)} = \begin{cases} 0.88^{-3.44}/3 = 0.517 & (Si) \\ 0.88^{-3/2}/3 = 0.404 & (Metal) \end{cases}$$

$$\frac{x_1}{x_b} = (z_1/z_b)^{1/2} \omega_b/\omega_1 = (z_b/z_1)^{1/2} C_1^{(m/2+1)}$$

$$= \begin{cases} \sqrt{3} \times 0.88^{3.44} = 1.09 & (Si) \\ \sqrt{3} \times 0.88^{3/2} = 1.43 & (Metal) \end{cases}$$

(34)

The surface atomic vibrational amplitude is indeed greater [295, 296] than it is in the bulk while its vibration frequency becomes lower. The vibration frequency and magnitude sensitive not only to the *m* value but also the coordination environment, and the surface curvature of a spherical dot.

4.3. Bulk Crystals: Compression and Thermal Excitation

4.3.1. Group IV Semiconductors

Reproduction of the Debye thermal decay of the Raman shift and Young's modulus for group IV elemental crystals, shown in Figure 28, yields information on the atomic cohesive energy $E_{coh}(0)$ and Debye temperature given in Table 8. Deviation between the derived $E_{coh}(0)$ for the $\omega(T)/\omega(0)$ and $Y(T)/Y(0)$ exists for diamond and Si but the $E_{coh}(0)$ for Ge is consistent [297].

**Error! Not a valid filename.**

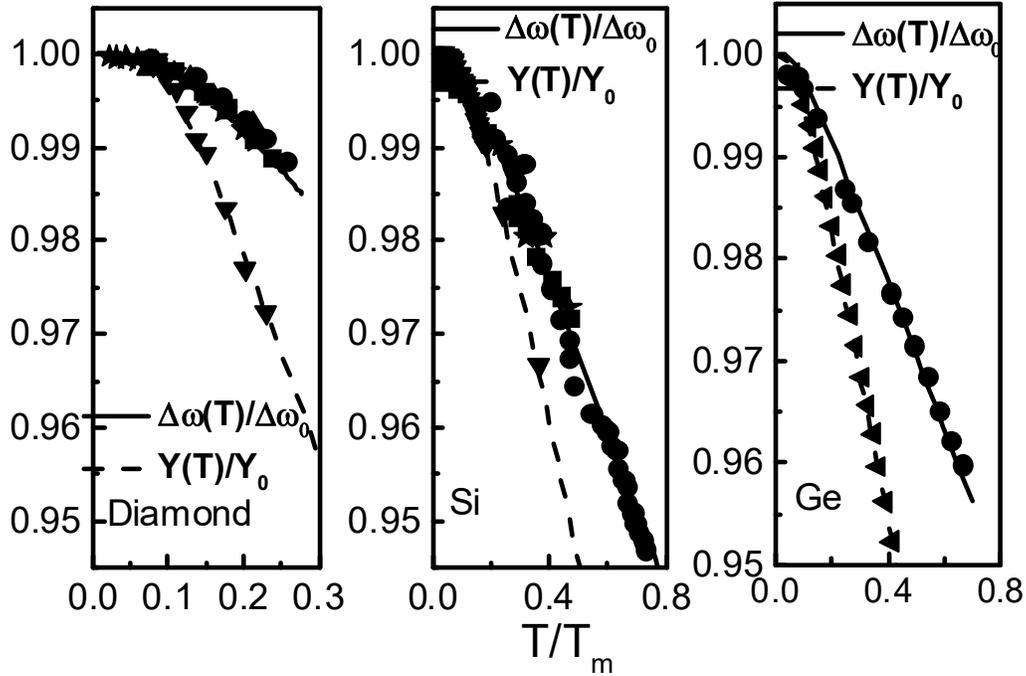

Figure 28. Theoretical reproduction of the thermal decay of the Raman shift and the Young's modulus of (a) Diamond [67, 70, 298-300], (b) Si [301, 302], and (c) Ge [108, 109, 301, 303] yields information of $E_{coh}(0)$ tabulated in Table 8 (Reprinted with permission from [297]).

Table 8 Information on the $T_m$, the Debye temperature, $\theta_D$, thermal expansion coefficient, $\alpha(T)$, and the atomic cohesive energy $E_{coh}(0)$ for Si, Ge, and Diamond [297].

|  | $\alpha(10^{-6}K^{-1})$ [140] | $T_m(K)$ | $\theta_D(K)$ | $E_{coh}(0)$ (eV) | | | |
|---|---|---|---|---|---|---|---|
|  |  |  |  | Raman | Y | Mean | Ref.[150] |
| Si | 4.5 | 1647 | 647 | 2.83 | 4.33 | 3.58 | 4.03 |
| Ge | 7.5 | 1210 | 360 | 2.52 | 2.65 | 2.58 | 3.85 |
| Diamond | 55.6 | 3820 | 1860 | 6.64 | 5.71 | 6.18 | 7.37 |

4.3.2. Group III-Nitrides

Figure 29 features the TP-BOLS reproduction of the P and T dependent Raman shift for group III-V semiconductors. Yielded information is summarized in Table 9 on the reference frequencies $\omega(1)$, mode cohesive energy ($E_{coh}$). The Debye temperature, melting temperature and thermal expansion coefficient were used as input parameters.

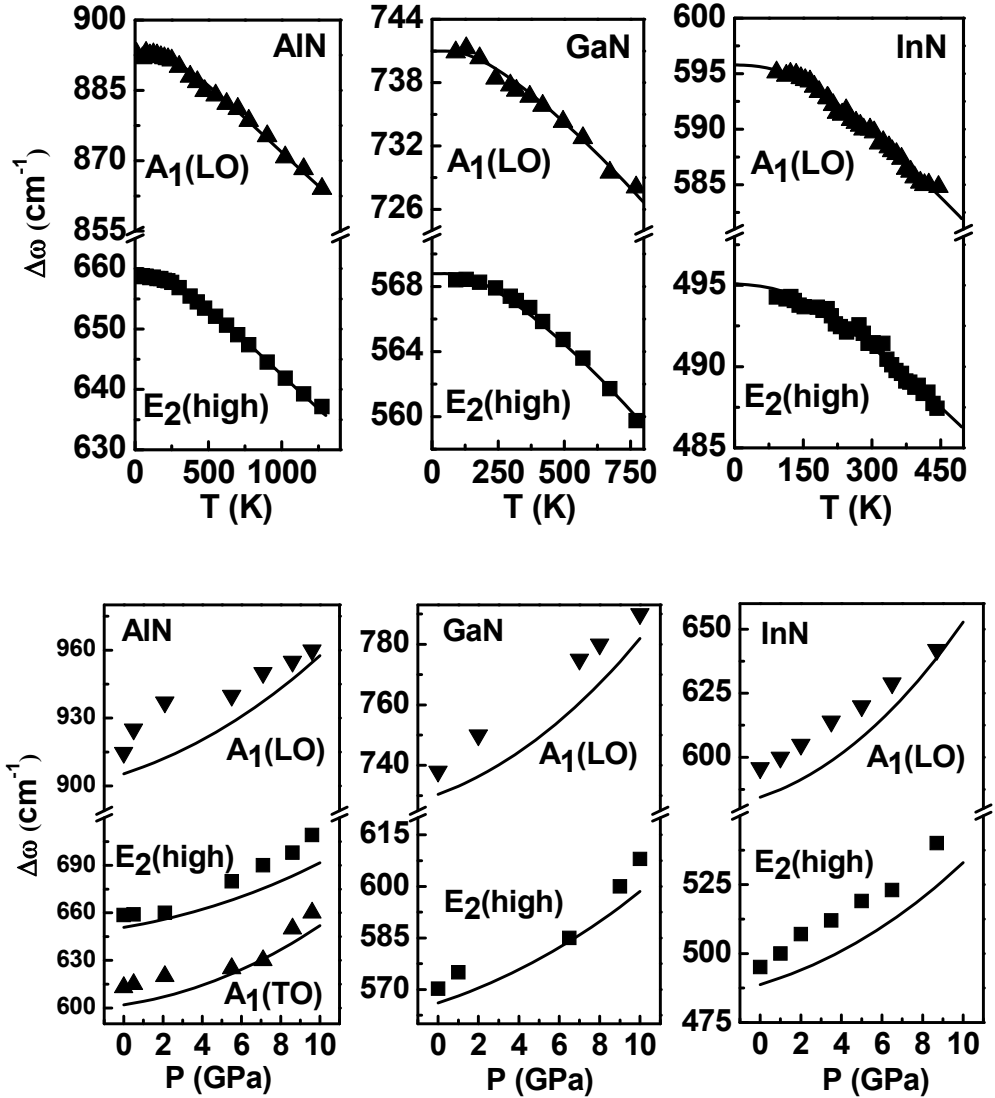

Figure 29. Simulation of the Raman shift for (a) AlN [304-307] (b) GaN [308-311], and (c) InN [312, 313]. Table 9 summarizes the extracted phonon frequency $\omega(1)$ and *mode* cohesive energies $E_{coh}(0)$ (Reprinted with permission from [226]).

Table 9 The intrinsic Raman frequency $\omega(1)$ and the *mode* cohesive energy, $E_{coh}(0)$, of various Raman active modes [97, 226]. The known Debye temperature ($\theta_D$), melting temperature ($T_m$), and the thermal expansion coefficient $\alpha(T)$ were used as input.

| | $T_m$ (K) | $\theta_D$ (K) | $\alpha$ [140] (Ref) | Raman mode | $\omega(1)$ (cm$^{-1}$) | $E_{coh}(0)$ (eV) | $\sigma$ (10$^{-4}$) |
|---|---|---|---|---|---|---|---|
| AlN | 3273 | 1150 | [34] | E$_2$(high) | 658.6 | 1.13 | 0.94 |
| | | | | A$_1$(LO) | 892.6 | 1.21 | 1.54 |
| | | | | A$_1$(TO) | 613 | 0.71 | 1.74 |
| | | | | E$_1$(LO) | 914.7 | 1.31 | 1.52 |
| | | | | E$_1$(TO) | 671.6 | 1.19 | 1.04 |
| GaN | 2773 | 600 | [142] | E$_2$(high) | 570.2 | 1.44 | 0.57 |
| | | | | A$_1$(LO) | 738 | 0.97 | 1.20 |
| | | | | A$_1$(TO) | 534 | 1.26 | 1.28 |
| | | | | E$_1$(LO) | 745 | 0.95 | 0.68 |
| | | | | E$_1$(TO) | 561.2 | 1.59 | 1.12 |
| InN | 1373 | 600 | [141] | E$_2$(high) | 495.1 | 0.76 | 1.15 |
| | | | | A$_1$(LO) | 595.8 | 0.50 | 0.92 |

4.3.3. TiO$_2$ and ZnO

TP-BOLS matching of the documented size [274, 278], temperature [27, 314], and pressure [27] dependence of the $B$ and $\Delta\omega$ [315] for TiO$_2$ and ZnO allows the verification of the BOLS notion for the multifield effect. Table 10 summarizes information yielded from the matching to the experiments. Generally, the frequency of the TO phonon undergoes a redshift when the solid size reduced. All the vibration modes undergo thermal Debye decay and mechanical compressive stiffening. s [276, 314, 316]. However, for the rutile TiO$_2$ A$_{1g}$ mode at 612 cm$^{-1}$ undergoes a redshift while the anatase E$_g$(144 cm$^{-1}$) mode undergoes a blueshift when the solid size is reduced [274-277].

Theoretical matching to the measured and calculated size [274, 278], temperature [27, 314], and pressure [27] dependence of the $B$ and $\Delta\omega$ [315] for TiO$_2$ at room temperature allows one to verify the theoretical strategies to extract information as summarized in Table 10. Numerical reproduction of the TiO$_2$ phonon shift yields the bond nature index m = 5.34 [33]. The single dimer vibration (z = 1) creates the TiO$_2$ E$_g$ mode of size-reduction induced blueshift.

In the $T$-dependent curves, the shoulder is related to the $\theta_D$ and the slope at higher temperature depends on the atomic cohesive energy, $E_{coh} = zE_z$. Matching the two sets of $B(T)$ and the $\Delta\omega(T)$ data will improve the reliability of the derivatives. The theoretical match of the measured $T$ dependent $B(T)$[27] and the $\Delta\omega(T)$[314] of the E$_g$(639 cm$^{-1}$) mode for anatase phase TiO$_2$ in

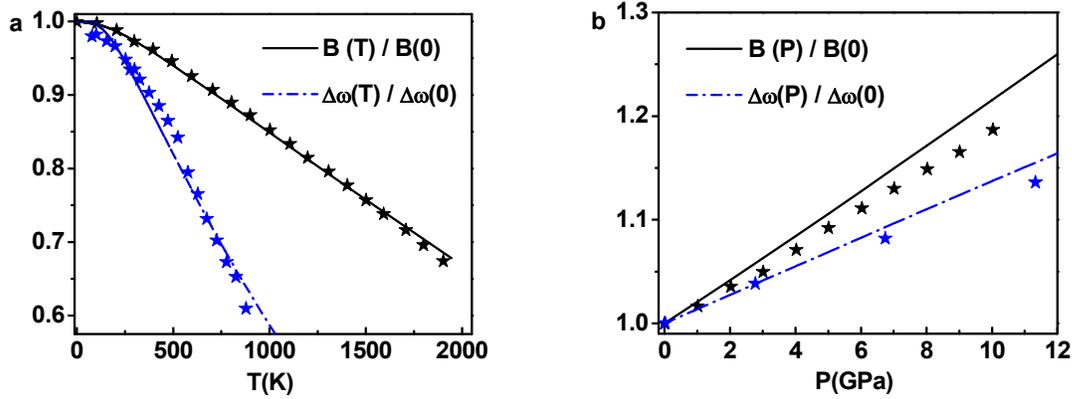

**Figure 30** turns out the $\theta_D$ of 768 K and the $\omega(1)$ = 600 cm$^{-1}$. The derived $\theta_D$ is in good agreement with the literately documented 778 K [317]. The cohesive energy $E_c$ is estimated as 1.56 eV.

Theoretical reproduction of the experimental $x$-$P$ curve [227, 228, 315] yielded the compressibility, $\beta$ = 6.84×10$^{-3}$ GPa$^{-1}$ and bulk modulus, $B_0$ = 143 GPa, and their first derivatives, $\beta'$ = $-1.21\pm0.05\times10^{-4}$ GPa$^{-2}$ and $B'_0$ = 8.86 [131]. The energy density 0.182 eV/Å$^3$. Based on the $[\Delta\omega]^2/[Yd]\equiv 1$ relation, one can derive the $\Delta\omega(P)$, which agrees with the $P$-dependent $B(P)$[27] and $\Delta\omega(P)$[276] for the anatase TiO$_2$ E$_g$(639 cm$^{-1}$) mode.

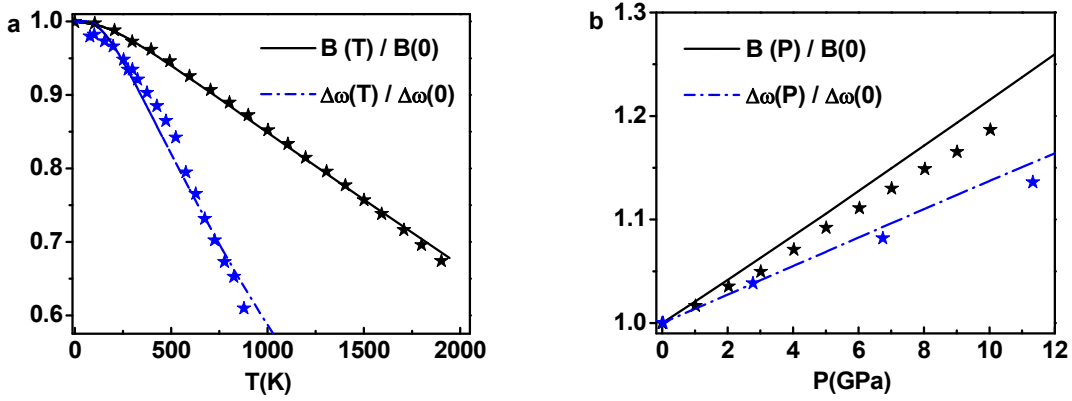

Figure 30. (a) temperature [314] and (b) pressure [27, 276] resolved bulk modulus and the $A_{1g}$ mode shift [131, 274] for $TiO_2$. Table 10 summarizes the derivatives on the mode cohesive energy and Debye temperature (Reprinted with permission from [131]).

ZnO with wurtzite structure belongs to the $C_{6v}^4$ symmetry group. At the Γ point of the Brillouin zone, optical phonons have the following irreducible representation: $\Gamma_{opt} = A_1 + 2B_1 + E_1 + 2E_2$, where the $A_1$ and the $E_1$ modes are polar and can be split into transverse (TO) and longitudinal optical (LO) phonons, with all being the Raman and infrared active. The non-polar $E_2$ modes are the Raman active, while the $B_1$ modes are the Raman inactive [80, 318, 319]. These Raman modes contain important physical information. For example, the $E_2$ (high) mode represents the bending modes of O-O atoms [101].

With the known Debye temperature $\theta_D$ = 310 K, bulk modulus $B$ =160 GPa and $B_0'$ = 4.4 [320], the frequency ω(1) [80, 319, 321], and the thermal expansion coefficient $\alpha(t)$ [322] as input, one can find the mode cohesive energy by matching the theory to the experimental results. Theoretical match of the measured $T$ dependent $B(T)$[27] and the $\Delta\omega(T)$ [314] of the $E_g$(639 cm$^{-1}$) mode for anatase phase $TiO_2$ in Figure **30**a leads to the ω(1) = 600 cm$^{-1}$ and the $\theta_D$ of 768 K, which is in good agreement with the reported value of 778 K [317]. The cohesive energy $E_{coh}$ is derived as 1.56 eV. At $T \leq \theta_D/3$, the relative $B$ and $\Delta\omega$ turns from nonlinear to linear when the temperature is increased. The slow decrease of the $B$ and $\Delta\omega$ at very low temperatures arises from the small $\int_0^T \eta(t)dt$ values as the specific heat $\eta(t)$ is proportional to $T^3$ at very low temperatures.

Figure 31a and b show the Debye thermal decay of the (a) $Y(T)$ and $\omega(T)$, (b) band gap $E_g(T)$ and (c) the pressure dependent $Y(P)$ and $\omega(P)$ modes of $E_1$(LO, 595 cm$^{-1}$), $E_2$(high, 441.5 cm$^{-1}$), $E_1$(TO, 410 cm$^{-1}$), $A_1$(TO, 379 cm$^{-1}$) and $B_1$(LO, 302 cm$^{-1}$) phonon modes for ZnO crystals [282, 319, 323]. Numerical agreement yielded the referential $\omega(1)$ for the modes of $E_1$(LO, 510 cm$^{-1}$), $E_2$(high, 380 cm$^{-1}$), $E_1$(TO, 355 cm$^{-1}$), $A_1$(TO, 330 cm$^{-1}$) and $B_1$(LO, 271 cm$^{-1}$). Bond thermal and compressive relaxation determine the thermal decay and the P-enhancement of these quantities depending bond length and energy.

Figure 31c and d presents theoretical match with the measured pressure-dependent Raman shift of $E_1$(LO, 595 cm$^{-1}$), $E_2$(high, 441.5 cm$^{-1}$), $E_1$(TO, 410 cm$^{-1}$), $A_1$(TO, 379 cm$^{-1}$) and $B_1$(LO, 302 cm$^{-1}$) phonon modes for ZnO at room temperature [319, 323]. Agreement between predictions and experimental observations allows us to determine the $\omega(1)$ of $E_1$(LO, 510 cm$^{-1}$), $E_2$(high, 380 cm$^{-1}$), $E_1$(TO, 355 cm$^{-1}$), $A_1$(TO, 330 cm$^{-1}$) and $B_1$(LO, 271 cm$^{-1}$) modes. The change of the bond energy is dependent on the ambient temperature and pressure. Therefore, the competition between the thermal expansion the pressure-induced compression determines the blueshift of Raman peaks.

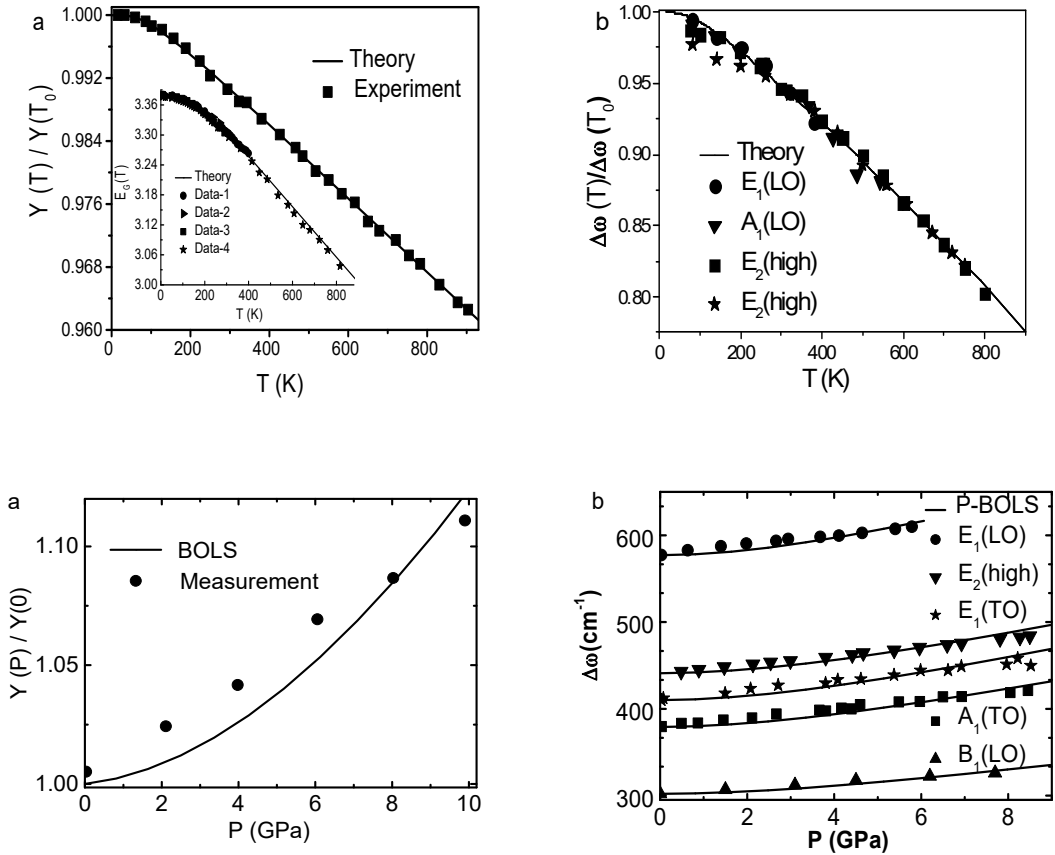

Figure 31 Theoretical (solid lines) reproduction of the measured (scattered datum) Debye thermal decay of the (a) Young's modulus [324] band gap $E_G$ (inset) [324-328] and (b) the Raman phonons of the optical modes ($E_1(LO)$, $A_1(LO)$ and $E_2(high)$)[80, 321] at the atmospheric pressure for ZnO with confirmation of $\theta_D$ = 310 K and $E_b$ = 0.75 eV per bond. (c) Reproduction of the pressure hardening of the elastic modulus [329] and (d) optical phonon frequencies ($E_1(LO)$, $E_2(high)$, $E_1(TO)$, $A_1(TO)$ and, $B_1(LO)$) yielded the binding energy density ($E_d$) of 0.097 eV/Å$^3$ for ZnO [319, 323] (Reprinted with permission from [282]).

Table 10. Information yielded from reproduction of the size, pressure, and temperature resolved bulk modulus and the Raman shift for TiO$_2$ [131] and ZnO bulk [131, 139, 319].

| Stimulus | Quantity | TiO$_2$[27, 314, 324] | ZnO[139, 319] |
|---|---|---|---|
| T | Cohesive energy $E_c = z_b E_0$(eV/atom) | 1.56 | 0.75 |
| | Debye temperature, $\theta_D$ (K) | 768 (778[317]) | 310 |
| P | Bulk modulus, $B_0/B'_0$(GPa/-) | 143/8.86 (167/-[330]) | 160/4.4 |
| | Compressibility, $\beta/\beta'$ (GPa$^{-1}$/GPa$^{-2}$) | 6.84×10$^{-3}$ (1.21±0.05)×10$^{-4}$ | 6.55×10$^{-3}$ 1.25×10$^{-4}$ |
| | Energy density, $E_{den}$ (eV/ Å$^3$) | 0.182 | 0.097 |

4.3.4. Other Compounds

Figure 32 shows the T-BOLS reproduction of the Raman thermal decay for the group II-VI semiconductors. Table **11** summarizes information of the Debye temperature ($\theta_D$) and atomic cohesive energy ($E_{coh}$) in comparison to the documented $\theta_D$. Figure 33a and b show the theoretical match to the measured elasticity thermal decay of KCl, MgO, Al$_2$O$_3$, Ma$_2$SO$_4$ crystals. Figure 33 c and d shows the theoretical match of the reported T dependence of $l(T)$ for BaXO$_3$ cubic perovskites [28]. The agreement between the theory and the reported temperature dependence of $l(T)$ of cubic perovskites justifies the validity of the Debye thermal expansion, as given in Table **11**.

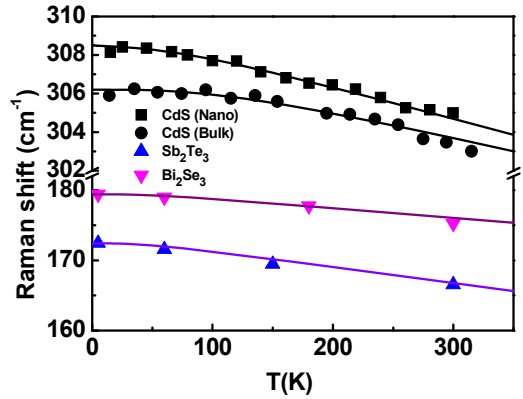

Figure 32. Raman thermal decay for CdS and CdSe [331] and Bi$_2$Se$_3$ [19, 22, 23, 81] and Sb$_2$Te$_3$ [19, 22, 23, 81] with derived information as given in Table 8.

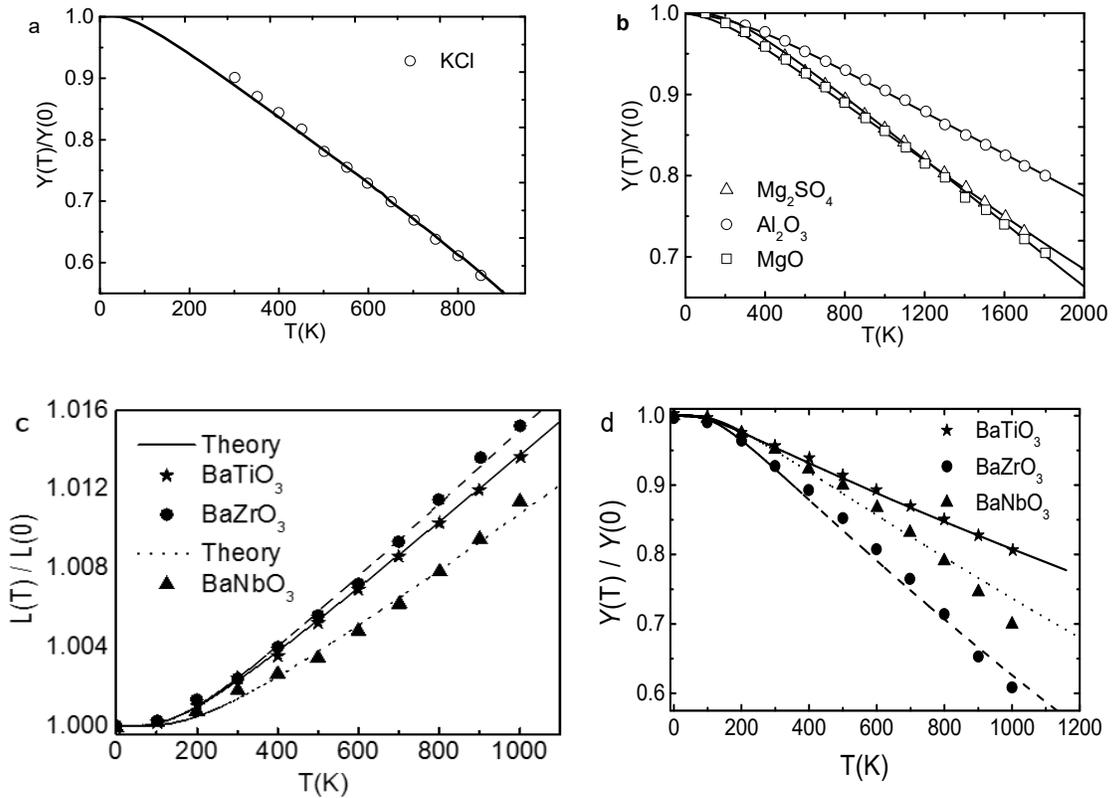

Figure 33. Theoretical reproduction (continued lines) of the measured (scattered data) Debye thermal decay of the Young's modulus and lattice constant for various specimens [27, 28, 314]. Table **11** lists the yielded mean atomic cohesive energy.

Table 11. Atomic cohesive energy $E_b(0)$ yielded from fitting to the Debye thermal decay of the Young's modulus. The $T_m$ and $\theta_D$ for bulk are input parameters [131, 294, 297, 332].

| Sample | $T_m$ (K) | $\theta_D$ (K) | $E_{coh}(0)$ (eV/atom) |
|---|---|---|---|
| KCl [105] | 1044 | 214 | 0.57 |
| MgSO$_4$[105] | 1397 | 711 | 2.80 |
| Al$_2$O$_3$ [104] | 2303 | 986 | 3.90 |
| MgO [105] | 3100 | 885 | 1.29 |
| BaTiO$_3$ [28] | 1862 | 580 | 2.50 |
| BaZrO$_3$ [28] | 2873 | 680 | 0.74 |
| BaNbO$_3$ [28] | - | 740 | 1.10 |
| CdS (bulk) | 1678[333] | 450/460 [334] | 2.13 |
| CdS (2 nm) | 979.5[335] | 300/300[334] | 1.72 |
| Sb$_2$Te$_3$ | 629[336] | 165/162[337] | 1.09 |
| Bi$_2$Se$_3$ | 710[338] | 185/182[339] | 1.24 |

4.4. Summary

The BOLS-LBA theory and the multifield phonon spectrometrics has enabled reconciliation of the crystal size, pressure, and temperature effect on the phonon and elasticity relaxation of the core-shelled ZnO, TiO$_2$, CeO$_2$, CdS, CdSe, and Bi$_2$Se$_3$ nanostructures and the group IV, III-nitride, II-oxide bulk crystals with derivative of the conventionally-unexpected information. Reproduction of the size trend of phonon frequency shift has derived the bond nature index, bond length and energy, effective atomic coordination number, intergrain interaction strength, dimer referential vibration frequency. Theoretical match to Debye thermal decay derives cohesive energy and Debye temperature. Reproduction of pressure effect leads to binding energy density and elasticity. The phonon and elasticity relaxation under perturbation results merely bond length and energy relaxation, which has little to do with the multiple phonon resonant scatter or the optical phonon degeneration into multiple acoustic phonons.

# 5. Water and aqueous solutions

Highlight

- Hydration of ions, lone pairs, protons, dipoles mediates the O:H–O network and solution properties.
- DPS distills the phonon abundance-order-stiffness transition of O:H–O bonds under perturbation.
- Ionic screened polarization and interanion repulsion stem the Hofmeister hydration volume.
- H↔H and O:⇔:O repulsion, ionic polarization, and solute bond shrinkage feature Lewis solutions.

## 5.1. Water and Aqueous Solutions

Water ice is ubiquitously important to agriculture, climate, environment, quality-of-life, and sustainability. O:H dissociation for water harvesting and H–O bond dissociation for $H_2$ generating are strategies to conquering resource crisis; hydration and solvation laid foundations to ionic rejection in water desalination and protein dissolution; aqueous drug-cell and water-protein interfaces are of great importance to microbiology, disease curing, DNA regulating and signaling. Therefore, grasp with factors dictating the performance of the undercoordinated $H_2O$ molecules and their electrons is pivotal to engineering the surface water and controlling its reaction, transition, and transport dynamics as one wishes.

Water ice responds to perturbation such as compression, heating, electromagnetic radiation, and molecular undercoordination leading to anomalies such as ice friction, ice floating, regelation, supercooling/superheating, warm water cooling faster, etc. Acid, base and salt solvation makes the solution even amazing. The O:H–O bonds (or HB) in the hydration shells of solutes perform differently from they do in the ordinary water.

Much work has been focused on the motion manner and dynamics of the excessive hydrated charge (proton and line-pair) and molecules with debating mechanisms. One often considers the solute motion dynamics in terms of phonon or molecular lifetime in the hydrating states, drift motion diffusivity, by taking the $H_2O$ motif as the basic structural unit. Unfortunately, progress in understanding the solute capabilities of mediating the solution HB network and properties is slower than it should be because lacks knowledge of the solvent structure and the O:H–O bond attributes. Little attention has been paid to the full-frequency Raman spectra or the correlation between the frequency shifts of the H–O and the O:H characteristic peaks – O:H–O bond segmental phonon cooperativity. It is unclear how the $H^+$, $HO^-$, $Y^+$, and $X^-$ ions functionalize collaboratively the hydration network and properties of the Lewis-Hofmeister solutions.

The aim of this section is to show that a combination of the O:H–O bond cooperativity notion [124] and the differential phonon spectrometrics (DPS) [112, 164], verified the expectations of H↔H formation that disrupts the acidic solution network and the surface stress and the O:⇔:O super–HB formation that compresses and polarizes the basic solutions. The screened $X^-$ and $Y^+$ in all solutions serve as point polarizers to form the supersolid hydration shell [340]. The combination of the ionic polarization and the H↔H fragilization or the O:⇔:O compression, and the undercoordinated solute H–O bond contraction mediate the HB network and the performance of Lewis-Hofmeister solutions. This premise is much more profoundly revealing than focusing on the solute motion dynamics to the understanding of acid and basic solvation dynamics, solute-solvent interactions and solute extraordinary capabilities on HB transition.

Figure 34 shows the full–frequency Raman spectra for the 0.1 malar ratio, monovalent (a) HX acid [341], (b) YHO base [342], and (c) YX salt [343] solutions collected under the ambient conditions and (d) heated water [344]. The Raman spectrum covers the phonon bands of O:H stretching vibration at < 200 cm$^{-1}$, the ∠O:H–O bending band centered at 400 cm$^{-1}$, the ∠H–O–H bending band at 1600 cm$^{-1}$, and the H–O stretching band centered at 3200 cm$^{-1}$. The O:H stretching, molecular rotational and torsional vibrations are within the THz regime, one can hardly discriminate these contributions one from the other. The H–O stretching phonon band can be decomposed into the bulk (3200 cm$^{-1}$), the skin or the surface having a certain thickness (3450 cm$^{-1}$), and the surface dangling H–O bond or called free radical (3610 cm$^{-1}$) directing outwardly of the surface [124]. Likewise, the O:H stretching vibration phonon centered at 75 cm$^{-1}$ features the undercoordination–induced skin O:H elongation and polarization, the ~200 cm$^{-1}$ peak feature the O:H vibration for the four–coordinated molecules in the ordinary bulk water.

Focusing on the evolution of the characteristic peaks for the stretching vibrations of O:H centered at < 200 cm$^{-1}$ and the H–O at >3000 cm$^{-1}$ would suffice. Insets recap the reaction at solvation:

$$HX + nH_2O \Rightarrow X^- + H_2O^+(H \leftrightarrow H)OH + (n-1)H_2O \Rightarrow X^- + (n-5)H_2O + [H_{11}O_5]^+$$

$$YHO + nH_2O \Rightarrow Y^+ + H(O^-:\Leftrightarrow:O^{2-})H_2 + (n-1)H_2O \Rightarrow Y^+ + (n-5)H_2O + [H_9O_5]^-$$

$$YX + nH_2O \Rightarrow Y^+ + X^- + nH_2O$$

(35)

The full–frequency Raman spectra reveal that charge injection derives none vibration features unless the excessive feature at 3610 cm$^{-1}$ due solute HO$^-$ contraction. Charge injection and thermal excitation do relax the H–O bond and the O:H nonbond cooperatively.

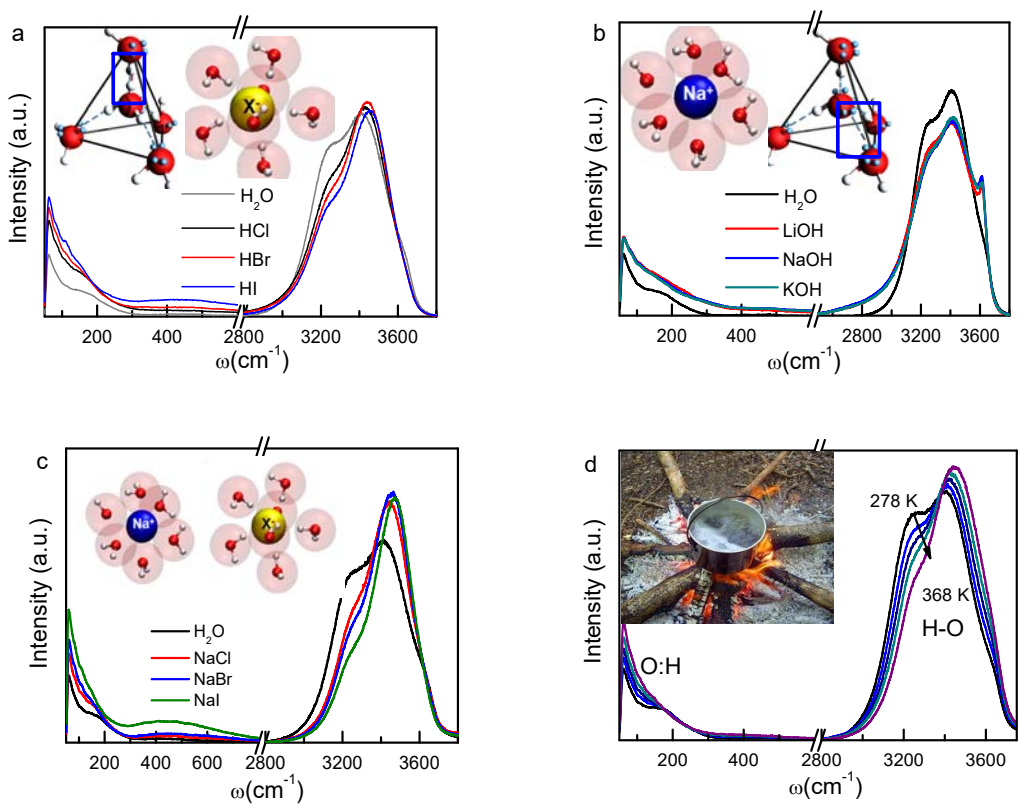

Figure 34. Full-frequency Raman spectroscopy for the room-temperature, 0.1 molar concentrated (a) HX/H$_2$O [341], (b) YOH/H$_2$O [342], and (c) NaX/H$_2$O [343] solutions compared with the spectra of (d) heated water [344]. Inset a illustrates the [H$_{11}$O$_5$]$^+$ unit cell with the framed H↔H point breaker and the X$^-$ point polarizer; Inset b shows the [H$_9$O$_5$]$^-$ with the framed O:⇔:O point compressor and polarizer and the Y$^+$ polarizer; inset c shows that the Y$^+$ and X$^-$ ions form each a hydration shell through polarization and the hydrating H$_2$O dipole screen shielding. (reprinted with permission from [341-344])

The vibration mode features the Fourier transformation of bonding vibrating in the similar frequencies regardless of their orientations and locations in real space. The peak shape is the population function and the peak maximum the highest popularity, the peak central $\omega_x$ represent the stiffness of the vibrating bonds as a function of bond length $d_x$ and bond energy $E_x$, $(\omega_x)^2 \propto E_x/d_x^2$. The peak integral is proportional to the number of bonds contributing to the phonon abundance of the vibration mode x. Thus, one can distill the fraction-stiffness transition of the O:H–O phonons by differentiating the spectrum from deionized water as a reference from those collected from solutions or water under perturbation such as heating and compressing upon all the spectral peak area being normalized [340]. This (DPS) strategy gives direct information on the O:H–O bonds transition from its vibration mode of ordinary water to the hydrating in terms of the number fraction (phonon abundance), stiffness (frequency shift), and fluctuation (line width) with much more ease and quantitative information than using the conventional spectral peak Gaussian-type decomposition.

## 5.2. Water and Ice: O:H−O Segmental Specific Heats
### 5.2.1. Conservation and Restriction

We proposed, verified and systematically demonstrated the impact of his unique 'hydrogen bond theory'. Firstly, we coupled the O:H van der Walls and the H–O intramolecular interactions using the O:⇔:O inter-lone-pair repulsion to feature the energetics and dynamics of water ice, in terms of hydrogen bond segmental cooperative relax-ability and specific-heat disparity. The specific-heat defines the structure phases at the ambient pressure, density anomalies, and the quasisolid (QS) phase of cooling expansion. Wen then examined the response of O:H–O bonds and electrons to mechanical and thermal excitation, electrostatic polarization, magnetization, molecular undercoordination, charge injection by acid, base, salt, and molecular solvation, and their programmed combinations. The examination was conducted by employing electron and phonon spectroscopy, surface stress and solution viscosity detection, quantum computations, Lagrangian-Laplacian oscillation dynamics, and Fourier fluid transport thermodynamics, in combination with ultrafast phonon spectroscopy and X-ray K-edge absorption analysis.

Figure 35a illustrates the $2H_2O$ unit cell (or $5H_2O$ motif) of $C_{3v}$ symmetry having four HBs bridging oxygen anions. It is essential to treat water as a crystalline–like structure with well–defined lattice geometry, strong correlation, and high fluctuation. For a specimen containing N number of $O^{2-}$ anions, there are 2N protons $H^+$ and 2N lone pairs ":" and the O:H−O bond configuration conserve regardless of structural phase [345] unless excessive $H^+$ or ":" is introduced [341, 342].

The motion of a H₂O molecule or the tunneling transport of a proton H⁺ is subject to restriction. If the central H₂O rotates 60° and above around the C$_{3v}$ symmetrical axis of the 2H₂O, there will be a H↔H anti–HB and an O:⇔:O super–HB formed, which is energetically forbidden. Because of the H−O bond energy of ~4.0 eV, translational tunneling of the H⁺ is also forbidden. Breaking the H−O bond in vapor phase needs 121.6 nm laser radiation [346], estimated 5.1 eV because the extremely low molecular coordination numbers.

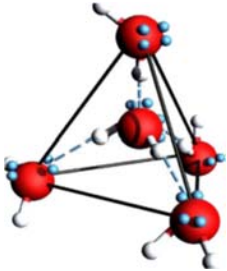

(a) 2H₂O unit cell  (b)  O:H−O bond specific heat  (c)  ρ(T) oscillation

Figure 35. (a) The 2H₂O primary unit cell contains four oriented O:H−O bonds defines liquid water as a crystal with molecular and proton motion restrictions [124]. (b) Superposition of the segmental specific heat η$_x$ defines the specific ratios η$_L$/η$_H$ [158] in (c) the five phases of density oscillation over the full range of temperature, for bulk water (T ≥ 273 K) and 1.4 nm water droplet (T < 273 K) [347]. The phase boundaries are subject to dispersion governed by Einstein relationship: Θ$_{Dx}$ ∝ ω$_x$ (x = L, H for the O:H and the H–O bond, respectively).

5.2.2. Specific Heat and Density Oscillation

It is necessary to decompose the specific heat of water ice into two components η$_x$(T/Θ$_{Dx}$) of Debye approximation [158]. The segmental specific heat meets two criteria. One is the Einstein relation ω$_x$ ∝ Θ$_{Dx}$ correlating the Debye temperature to the vibration frequency of the X segment, and the other is its thermal integration being proportional to bond energy E$_x$. The (ω$_x$, E$_x$) is (200 cm⁻¹, ~0.1 eV) for the O:H nonbond and (3200 cm⁻¹, ~4.0 eV) for the H–O bond. The Debye temperatures and the specific heat curves are subject to the ω$_x$ that varies with external perturbation. Figure 35b shows the superposition of the specific heats $\eta_x$, which defines five phases in Figure 35c showing density oscillation over the full temperature range [158].

The hydrogen bonding thermodynamics at a certain temperature is subject to the specific heat ratio, $\eta_L/\eta_H$. The segmental having a lower specific heat follows the regular thermal expansion but the other segment responds to thermal excitation oppositely because of the HB cooperativity by repulsion between electron pairs on adjacent $O^{2-}$.

5.2.3. O:H–O Length-stiffness Cooperative Relaxation

As the basic structural and energy exchange unit, the O:H−O bond integrates the intermolecular weaker O:H nonbond (or called van der Waals bond with ~0.1 eV energy) and the intramolecular stronger H−O polar–covalent bond (~4.0 eV), rather than either of the O:H or the H–O alone. The characteristics of the O:H−O bond is its asymmetrical and short–range interactions and coupled with the Coulomb repulsion between electron pairs on adjacent oxygen ions [124]. O:H−O bond segmental length and bond angle relaxation changes system energy, but structural fluctuation contributes little to the system energy on an average.

With the known H−O and O:H segmental length relaxation derived from $\rho(P)$ (1/V(P)) profile measured from compressed ice [348, 349] and the framework of tetrahedral-coordination for a water molecule [129], one can correlate the size $d_H$, separation $d_{OO}$, bond geometry, and mass density $\rho$ of molecules packed in water and ice in the following manner [350],

$$\begin{cases} d_{OO} = 2.6950\rho^{-1/3} & (Molecular\ separation) \\ \dfrac{d_L}{d_{L0}} = \dfrac{2}{1+exp[(d_H - d_{H0})/0.2428]}; & (d_{H0} = 1.0004\ and\ d_{L0} = 1.6946\ at\ 4\ °C) \end{cases}$$

(36)

The O:H nonbond and the H−O bond segmental disparity and the O−O coupling dislocate the $O^{2-}$ along the segmented O:H−O bond in the same direction but by different amounts under an external stimulus [158, 348, 351, 352]. The softer O:H nonbond always relaxes more than the stiffer H−O bond with respect to the $H^+$ as the coordination origin. The ∠O:H−O angle θ relaxation contributes to the geometry and mass density. The H−O relaxation absorbs or emits energy and the O:H relaxation dissipates energy caped with the 0.1 eV. The O:H−O bond segmental energy and specific heat disparity dictate the extraordinary adaptivity, cooperativity recoverability, sensitivity, properties of water and ice under external stimulus such as mechanical compression and thermal excitation.

Heating lengthens the O:H nonbond from 1.695 to 1.750 Å, but the H−O bond responds to heating contrastingly from 1.000 to 0.985 Å when heating from 273 to 377 K. The segmental disparity defines the manner of O:H−O thermal relaxation. The O:H having a lower specific heat follows the regular thermal expansion, which shortens the H−O bond by the O−O coupling [158]. Compression always shortens the O:H and lengthens the H−O regardless of the phase structures. The O:H nonbond contracts from 1.78 to 1.10 Å under compression up to 60 GPa when transits from the VII/VIII phase into phase X of identical O:H and H−O lengths. The segmental energy $E_x$ varies with a certain power of its inverse length $d_x$. The phonon spectroscopy probes the length and energy as the phonon frequency shift, or segmental stiffness, in the form of $(\Delta\omega_x)^2 \propto (E/d^2)_x$. A Lagrangian-Laplacian resolution to the oscillation dynamics of the O:H−O oscillator pair has transformed the measured segmental length and vibration frequency $(d_x, \omega_x)$ int to the respective force constant and binding energy $(k_x, E_x)$ and hence produces the potential paths of the O:H−O under continued perturbation [123, 128].

Table 12 summarizes the thermodynamics of O:H−O bond length and stiffness, mass density under the ambient pressure, and the molecular CN, polarization and pressure effect [158]. External perturbation changes the phase boundaries through the Einstein relation $\Delta\Theta_{Dx} \propto \Delta\omega_x(P, T, z, E,…)$.

Table 12. O:H–O segmental cooperative relaxation in length, vibration frequency, and surface stress with respect to $d_{L0}$ = 1.6946 Å, $d_{H0}$ = 1.0004 Å, $\omega_{H0}$ = 3200 cm$^{-1}$, $\omega_{L0}$ = 200 cm$^{-1}$, $\Theta_{DH}$ = 3200 K, $\Theta_{DL}$ = 198 K upon excitation by heating, compression, molecular undercoordination (skin, cluster, droplet, nanobubble).

| Phase | $(T_1-T_2)$ K | $\Delta T$ | $\Delta d_H$ | $\Delta d_L$ | $\Delta\omega_H$ | $\Delta\omega_L$ | Remark | Ref |
|---|---|---|---|---|---|---|---|---|
| Vapor ($\eta_L \cong 0$) | 377- | >0 | - | - | - | - | H$_2$O monomer | |
| Liquid ($\eta_L/\eta_H$ <1) | 277-377 | >0 | <0 | >0 | >0 | <0 | Liquid and solid thermal expansion | [158] |
| I$_c$ + I$_h$ ($\eta_L/\eta_H$ < 1) | 100-258 | | | | | | | |
| QS ($\eta_L/\eta_H$ > 1) | 258-277 | >0 | >0 | <0 | <0 | >0 | QS negative thermal expansion | |
| QS boundary ($\eta_L= \eta_H$) | 258; 277 | - | 0 | 0 | 0 | 0 | $\rho$ = 1.0; $\rho$ = 0.92 gcm$^{-3}$ | |
| XI ($\eta_L \cong \eta_H \cong 0$) | 0-100 | <0 | $\cong 0$ | | | | ∠O:H−O expands from 165 to 173 ° | |

| $\Delta z < 0; \Delta E \neq 0$ (polarization) | | | <0 | >0 | >0 | <0 | polarization; supersolidity | [340] [129] |
|---|---|---|---|---|---|---|---|---|
| $\Delta P > 0$ | | | >0 | <0 | <0 | >0 | $d_L$ and $d_H$ symmetrization | [351] |

Numerical reproduction of the Mpemba effect – hot water cools faster [353], evidences directly the essentiality of the 0.75 gcm$^{-3}$ mass density of the supersolid skin that promotes heat conduction outward the water of heat source. Exothermic reaction proceeds by bond elongation and dissociation while endothermic reaction proceeds by bond contraction and bond formation. The Mpemba effect integrates the O:H−O bond energy "storage-emission-conduction-dissipation" cycling dynamics. The energy storage is proportional to the H−O bond heating contraction and the rate of energy emission at cooling is proportional to its first storage. The skin higher thermal conductivity due to lower mass density benefits heat flow outward the solution, and the source-drain non-adiabatic dissipation ensures heat loss at cooling.

## 5.3. Lewis-Hofmeister Solutions
### 5.3.1. Conservation Rules Broken

Upon solvation, as given in eq (35), an HX acid molecule dissolves into an H$^+$ proton and an X$^-$ anion. The H$^+$ does not stay freely but bonds firmly to a H$_2$O molecule to form a tetrahedral H$_3$O$^+$ hydronium with one lone pair and three H$^+$, see Figure 34a. The H$_3$O$^+$ replaces a H$_2$O in the center of the 2H$_2$O unit cell while its four neighbors remain their orientations because of the O:H−O regulation of interactions with their rest neighbors [124, 157]. The alteration of the ":" with a H$^+$ at the H$_2$O→H$_3$O$^+$ transition breaks the 2N conservation with derivative of 2N+1 number of protons and 2N-1 lone pairs in the solution. The excessive 2N+1 – (2N-1) = 2 protons form uniquely an H$^+$↔H$^+$ anti–HB without any other choice. The H$_3$O$^+$ remains the tetrahedron configuration having three H−O bonds and one lone pair, which is similar to the situation of H$_{2n+1}$O$_{2n}^+$ cluster formation with n = 2 and 4 [354] but no shuttling between oxygen ions or freely hopping from one site to another.

Likewise, the YOH base is dissolved into a Y$^+$ cation and a hydroxide (HO$^-$ is HF-like tetrahedron with three lone pairs and one H$^+$). The HO$^-$ addition transits the 2N conservation into 2N+3 number of lone pairs ":" and 2N+1 number of protons. The excessively unbalanced 2N+3–(2N+1) = 2 lone pairs can only form the O:⇔:O point compressor between the central HO$^-$ and one of its H$_2$O neighbors that reserve their molecular orientations. This O:⇔:O compressor pertaining to each hydroxide has the same effect of but stronger mechanical compression on the network HB relaxation [351] because of the clustering of the two pairs of lone pairs and their weak binding energy to $O^{2-}$. The clustering of four electrons and weak bonding to $O^{2-}$ specify the point polarizer instead of a point breaker. $X^-$ and $Y^+$ in all solutions only polarize their neighboring $H_2O$ molecules to form hydration shells.

5.3.2. DPS of Lewis-Hofmeister Solutions

The DPS distils only phonons transiting into their hydration states as a peak above the x–axis, which equals the abundance loss of the ordinary HBs as a valley below the axis in the DPS spectrum. This process removes the spectral areas commonly–shared by the ordinary water and the high–order hydration shells. The DPS resolves the transition of the phonon stiffness (frequency shift) and abundance (peak area) by solvation. The fraction coefficient, $f_x(C)$, being the integral of the DPS peak, is the fraction of bonds, or the number of phonons transiting from water to the hydration states at a solute concentration C.

YOH solvation broadens the main peak shifting to lower frequencies [355-357]. The solute type and concentration resolved DPS shown in Figure 36 confirmed that YOH solvation indeed softens the $\omega_H$ phonon – O:⇔:O compression lengthens the neighboring H–O bond and softens its phonon to 3100 cm$^{-1}$ and below. Besides, a sharp peak appears at 3610 cm$^{-1}$ that is identical to the dangling H–O bond at water surface. The sharp peaks feature the less ordered H–O bond due the HO$^-$ The $\omega_x$ is less sensitive to the type of the alkali cations of the same concentration. These observations confirm that the O:⇔:O super–HB has the same but stronger effect of mechanical compression [358] and that the bond order-length-strength (BOLS) correlation [359] applies to aqueous solutions - bonds between undercoordinated atoms become shorter and stronger.

Observations justify that the O:⇔:O super–HB point compression (< 3100 cm$^{-1}$; > 220 cm$^{-1}$) has the same but much stronger effect of mechanically bulk compression (< 3300 cm$^{-1}$; > 200 cm$^{-1}$) at the critical pressure 1.33 GPa for the room-temperature water-ice transition [351], compared to the compression effect on water and ice. The $\Delta\omega_L$ for the YOH solutions at lower concentrations duplicates the $\Delta\omega_L$ feature of the mechanically compressed water, because of the compression and polarization [348]. Compression lengthens the solvent H–O bond and softens its phonon but relaxes the O:H nonbond contrastingly. The strong effect of compression overtones the effect of $Y^+$ polarization. The excessive peak at 3610 cm$^{-1}$ features the bond-order-deficiency induced H–O contraction of the due

HO⁻ solute, which is identical to the surface H–O dangling bond of 10% shorter, shorter and stiffer than the skin H–O bond featured at 3450 cm⁻¹ for ice and water, and bulk water at 3200 cm⁻¹ [360].

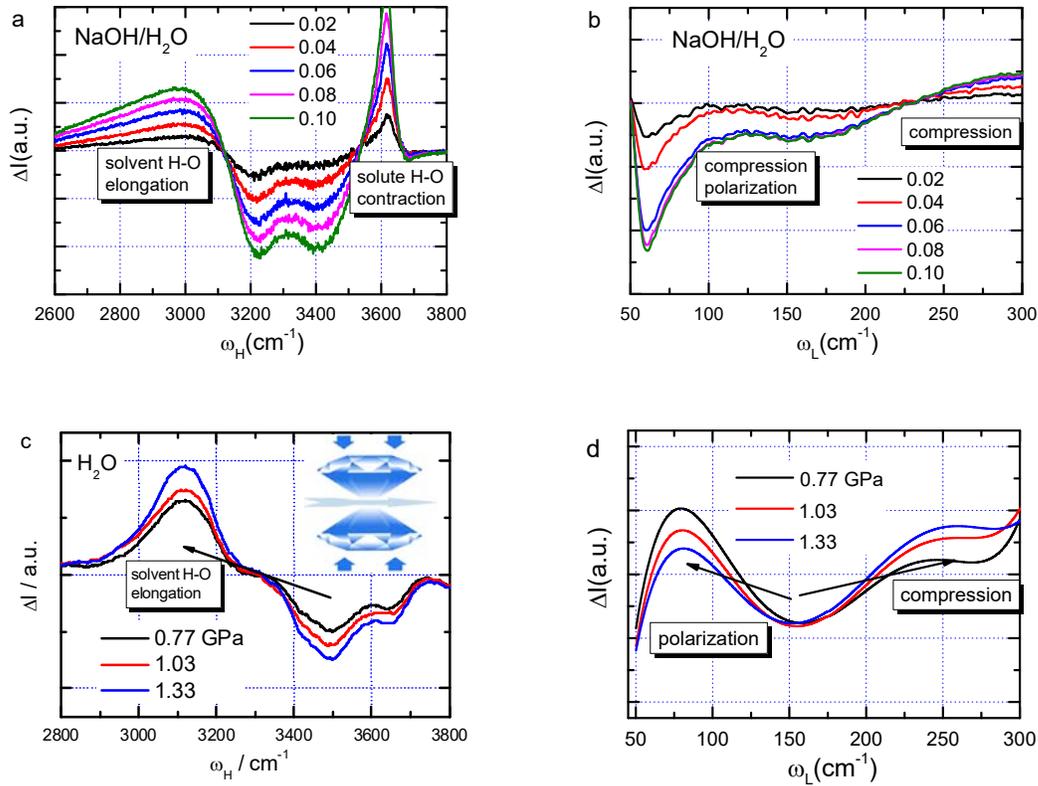

Figure 36. The $\omega_x$ DPS for the (a, b) concentrated NaOH/H$_2$O solutions [342] and the (c, d) mechanically compressed water [351] under the ambient conditions. O:⇔:O compression and polarization have the same but stronger effect of pressure on the $\omega_H$ softening and $\omega_L$ stiffening. The 3610 cm⁻¹ sharp feature arises from the undercoordinated solute H–O bond contraction that is the same to the dangling H–O radicals of 90 % length of its bulk value.

The broad hump at $\omega_H$ < 3100 cm⁻¹ shows the distant dispersion of the O:⇔:O compression forces acting on subsequent H$_2$O neighbors. The sharp feature at 3610 cm⁻¹ for the YOH indicates the strong localized nature of the solute H–O bond contraction. These DPS H–O vibration peaks clarify the origin for the two ultrafast processes in terms of phonon lifetime [356, 361]. The longer 200 ± 50 fs lifetime features the slower energy dissipation of higher-frequency solute H–O bond vibrating at 3610 cm⁻¹ and the other shorter time on 1–2 ps characterizes the elongated solvent H–O bond at lower-frequency of vibration <3100 cm⁻¹ upon HO⁻ solvation [190]. In the pump-probe ultrafast IR spectroscopy, the phonon population decay, or vibration energy dissipation, rate is proportional to the phonon frequency – the dynamics of higher frequency phonon is associated with a faster process.

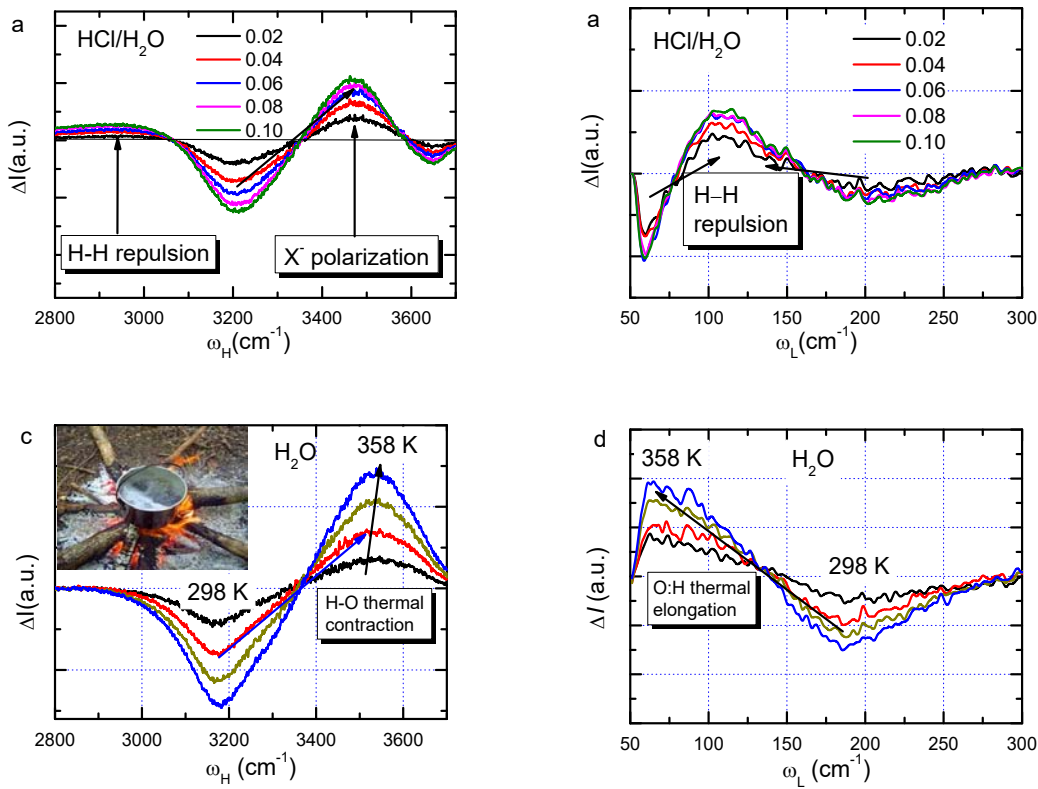

Figure 37. DPS $\omega_x$ for the (a, b) concentrated HCl/H$_2$O solutions [341], and the (c, d) heated water [344]. The blueshift in (a) arises from X$^-$ polarization and the humps below 3100 cm$^{-1}$ from H↔H repulsion. The joint effect of polarization and repulsion shifts the $\omega_L$ in (b) to 110 cm$^{-1}$. H–O thermal contraction shifts $\omega_H$ from 3200 to 3500 cm$^{-1}$ and the O:H elongation shifts the $\omega_L$ from 200 to 75 cm$^{-1}$ (reprinted with permission from [341])

Figure 37 compares the segmental $\omega_x$ DPS for the concentrated HX/H$_2$O solutions [341] and for the heated water [362]. The $\omega_L$ and the $\omega_H$ relax indeed cooperatively for all acidic solutions and the heated water. As the concentration increases from 0 to 0.1, X$^-$ polarization transits the $\omega_L$ from 180 to 75 cm$^{-1}$ and the $\omega_H$ from 3200 to 3480 cm$^{-1}$ (ref [363]). H↔H repulsion reverts the $\omega_L$ from 75 cm$^{-1}$ to 110 cm$^{-1}$. The DPS also resolves the effect of the H↔H repulsion on the H–O bond elongation featured as a small hump below 3050 cm$^{-1}$. The H↔H repulsion lengthens its neighboring H–O bonds, as being the case of mechanical compression and O:⇔:O point compression in basic solutions [351]. The fraction of H–O bonds elongation by H↔H repulsion decreases when the X⁻ turns from Cl⁻ to Br⁻ and I⁻ as the stronger polarizability of I⁻ annihilates the effect of H↔H repulsion. The small spectral valley at 3650 cm$^{-1}$ results from the preferential skin occupation of X⁻ that strengthens the local electric field. The stronger electric field stiffens the dangling H–O bond but the X⁻ screening weakens the signal of detection.

Acid solvation shares the same effect of segmental length and phonon stiffness relaxation and surface stress depression but different origins. X⁻ polarization and H−O thermal contraction stiffen the $\omega_H$, X⁻ polarization and O:H thermal expansion softens the $\omega_H$. H↔H fragilization and thermal fluctuation lowers the surface stress.

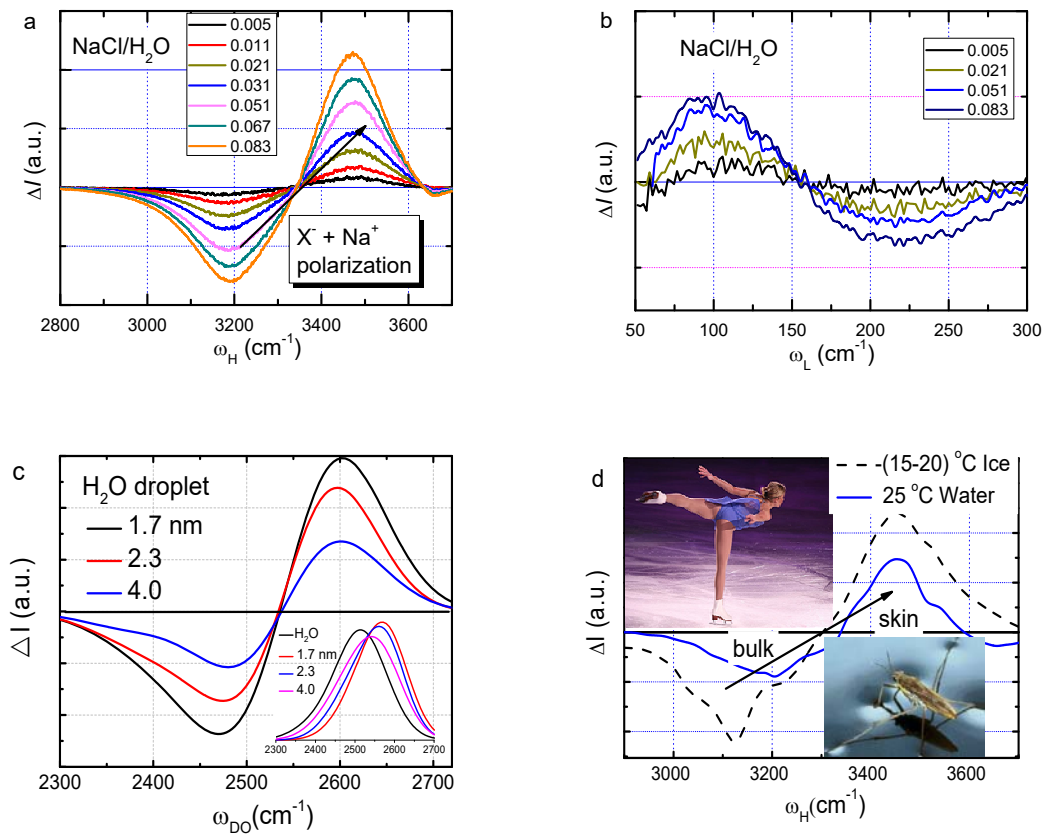

Figure 38. DPS $\omega_x$ for the concentrated (a, b) NaCl/H$_2$O solutions [343], (c) D-O phonon for the sized water (0.05D$_2$O + 0.95H$_2$O) droplets [266, 364], and the (d) $\omega_H$ for the skins of water and ice [360, 365]. Molecular undercoordination transits the $\omega_H$ from 3200 for water (at 25 °C) and 3150 cm$^{-1}$ for ice (-20 and -15 °C) to 3450 cm$^{-1}$ that is identical to both skins of water and ice showing anomalous toughness and slipperiness.

Figure 38 compares the $\omega_x$ DPS for the concentrated (a, b) NaCl/H$_2$O solutions [343], (c) D-O phonon for the sized (0.05D$_2$O + 0.95H$_2$O) water droplets [266, 364], and the (d) $\omega_H$ for the skins of water and ice [360, 365]. Ionic polarization and molecular undercoordination share the same effect on the H−O phonon blueshift, resulting the supersolid phase [364]. The supersolidity is characterized by the shorter and stiffer H−O bond and the longer and softer O:H nonbond, deeper O1s energy band, and longer photoelectron and phonon lifetime. The supersolid phase is less dense, viscoelastic, mechanically and thermally more stable. The O:H−O bond cooperative relaxation offsets boundaries of structural phases and raises the melting point and meanwhile lower the freezing temperature of water ice – known as supercooling and superheating. The softened O:H elasticity and adaptivity and the high repulsivity of the polarized supersolid skin takes the responsibility of slipperiness of ice and the tough skin of liquid water [360]. The high thermal diffusivity of the supersolid skin ensures the Mpemba paradox – warm water cooling faster [353].

5.3.3. Fraction of Bond Transition and Molecular Interactions

Integrating the $\omega_H$ DPS peaks for the HX, YHO, and YX yields the fraction coefficients $f_x(C)$ for the O:H−O bonds transition from its mode of ordinary water to the hydrating. The slope of the fraction coefficient, $df_x(C)/dC$, is proportional to the number of bonds per solute in the hydration shells, which characterizes the hydration shell size and its local electric field. A hydration shell may have one, two or more subshells, depending on the nature and size of the solute. The size and charge quantity determine its local electric field intensity that is subject to the screening by the local H$_2$O dipoles and the solute–solute interactions [366].

The $f_x(C)$ concentration trends show the following, see Figure 39 and Figure 40:

1) The $f_H(C) \equiv 0$ means that the H$^+$(H$_3$O$^+$) is incapable of polarizing its neighboring HBs but only breaking and slightly repulsing its neighbors [341].
2) The $f_Y(C) \propto C$ means the constant shell size of the small Y$^+$ cation (radius < 1.0 Å) without being interfered with by other solutes. The constant slope indicates that the number of bonds per solute is conserved in the hydration shell. The electric field of a small Y$^+$ cation is fully screened by the H$_2$O dipoles in its hydration shells; thus, no cation–solute interaction is involved in the YX or YHO solutions.

3) The $f_{OH}(C) \propto C$ (<3100, 3610 cm$^{-1}$) means that the numbers of the O:⇔:O compression–elongated solvent H–O bonds and the bond–order–deficiency shortened solute H–O bonds are proportional to the solute concentration. Bond order deficiency shortens and stiffens the bonds between undercoordinated atoms [124].

4) The $f_X(C) \propto 1-\exp(-C/C_0)$ toward saturation means the number of H$_2$O molecules in the hydration shells is insufficient to fully screen the X$^-$ (radius ~ 2.0 Å) solute local electric field because of the geometric limitation to molecules packed in the crystal–like water. This number inadequacy further evidences for the well–ordered crystal–like solvent. The solute can thus interact with their alike – only anion–anion repulsion exists in the X$^-$ – based solutions to weaken the local electric field of X$^-$. Therefore, the $f_X(C)$ increases approaching saturation, the hydration shells size turns to be smaller, which limits the solute capability of bond transition.

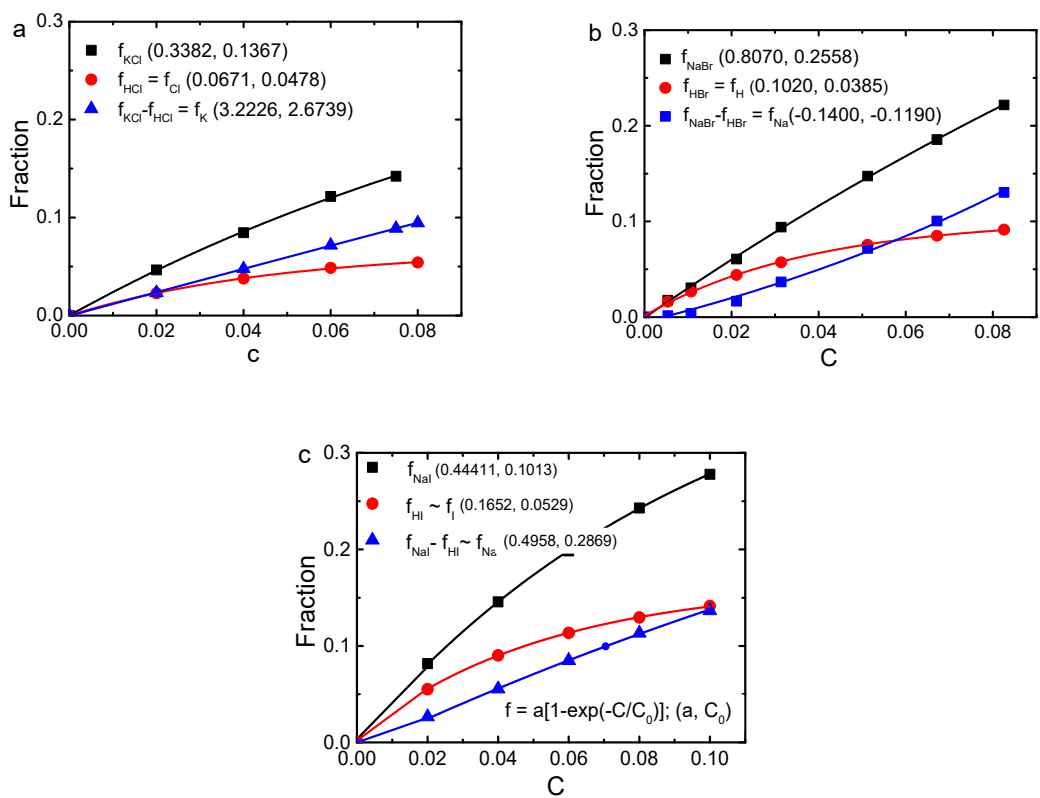

Figure 39. Fraction coefficients for the O:H–O bonds transition from the mode of ordinary water to hydrating by (a) NaCl, Cl$^-$, and Na$^+$, (b) NaBr, Br$^-$, and Na$^+$, and (c) NaI, I$^-$, and Na$^+$. The $f_{YX}(C) = f_Y(C) + f_X(C)$ given $f_H(C) \cong 0$ and $f_{HX}(C) \cong f_X(C)$ (reprinted with permission from [341, 343])

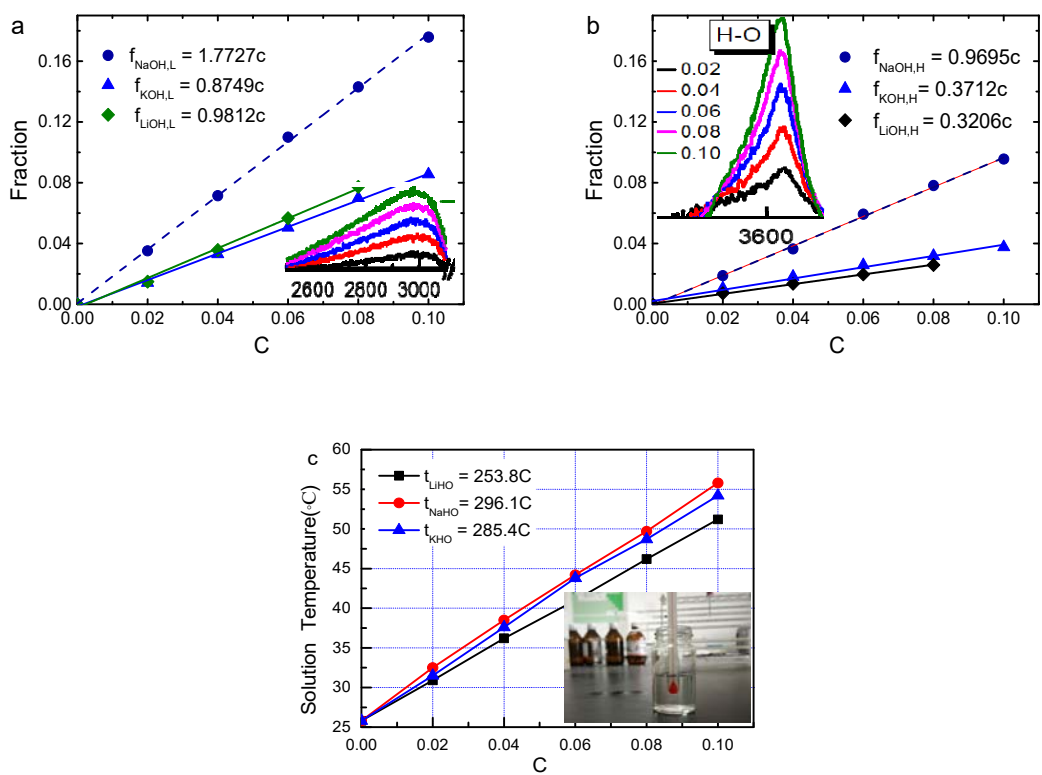

Figure 40. Fraction coefficients for the (a) solvent H–O bonds elongation by O:⇔:O compression, (b) solute H–O bonds contraction by HO⁻ undercoordination, and (c) exothermal solvation of YHO solutions. (reprinted with permission from [7])

Therefore, the $f_x(C)$ and its slope give profound information not only on the solute–solute and solute–solvent interaction but also on the relative number of bonds transiting from the referential mode of water to the hydration, by ionic polarization or O:⇔:O compression. For instance, as shown in Figure 40c, the solution temperature depends linearly on the fractions of the $f_{3100}(C)$ and $f_{3600}(C)$ ion the exothermic reaction. It is has been clear that energy emission of H−O bond elongation by O:⇔:O compression and energy absorption of the solute H−O contraction by undercoordination heat up the solution at solvation [342].

5.4.     Solution Properties versus Hydrogen Bond Transition

Figure 41 a compares the concentration dependence of the contact angle between the Lewis-Hofmeister solutions and glass substrate measured at 298 K. The surface stress is proportional to the contact angle. One can ignore the reaction between the glass surface and the solution, as we want to know the concentration trends of the stress change at the air–solution interface of a specific solution. Ionic polarization and O:⇔:O compression and polarization enhance the stress, but the H↔H point fragilization destructs the stress. Both polarization and undercoordination form the supersolid phase; the former occurs in the hydration shell throughout the bulk, but the latter only takes place in skins. The H↔H fragmentation has the same effect of thermal fluctuation on depressing the surface stress, see Figure 41 b [367]. Thermal excitation weakens the individual O:H bond throughout the bulk water.

In aqueous solutions, solute molecules are taken as Brownian particles drifting randomly under thermal fluctuation by collision of the solvent molecules. The viscosity of salt solutions is one of the important macroscopic parameters often used to classify water–soluble salts into structure making or structure breaking. The drift motion diffusivity $D(\eta, R, T)$ and the solute–concentration–resolved solution viscosity $\eta(C)$ follow the Stokes–Einstein relation [368] and the Jones–Dole expression[369], respectively,

$$\begin{cases} \dfrac{D(\eta,R,T)}{D_0} = \dfrac{k_B T}{6\pi\eta R} & (Drift) \\ \dfrac{\Delta\eta(C)}{\eta(0)} = A\sqrt{C} + BC & (Viscosity) \end{cases}$$

where $\eta$, $R$, and $k_B$ are the viscosity, solute size, and Boltzmann constant, respectively. $D_0$ is the coefficient in pure water. The coefficient A and its nonlinear term is related to the solute mobility and solute–solute interaction. The coefficient B and the linear term reflects the solute–solvent molecular interactions. The $\eta(0)$ is the viscosity of neat water.

SFG measurements [370, 371] revealed that the $SCN^-$ and $CO_2$ solution viscosity increases with solute concentration or solution cooling. The H–O phonon relaxation time increases with the viscosity, and results in molecular motion dynamics. Therefore, ionic polarization stiffens the H–O phonon and slows down the molecular motion in the semirigid or supersolid structures.

One may note that the relative viscosity and the measured surface stress due to salt solvation are in the same manner of the $f_{YX}(C)$. One can adjust the Jones–Dole viscosity coefficients A and B and fit the surface stress to match the measured $f_{YX}(C)$ curve in Figure 41 c and d. The trend consistency clarifies that the linear term corresponds to Y$^+$ hydration shell size and the nonlinear part to the resultant of X$^-$–water and X$^-$–X$^-$ interactions. It is clear now that both the solution viscosity and the surface stress are proportional to the extent of polarization or to the sum of O:H–O bonds in the hydration shells for the monovalent salt solutions, at least. Therefore, polarization raises the surface stress, solution viscosity and rigidity, H–O phonon frequency, and H–O phonon lifetime but decreases the molecular drift mobility, consistently by shortening the H–O bond and lengthening the O:H nonbond.

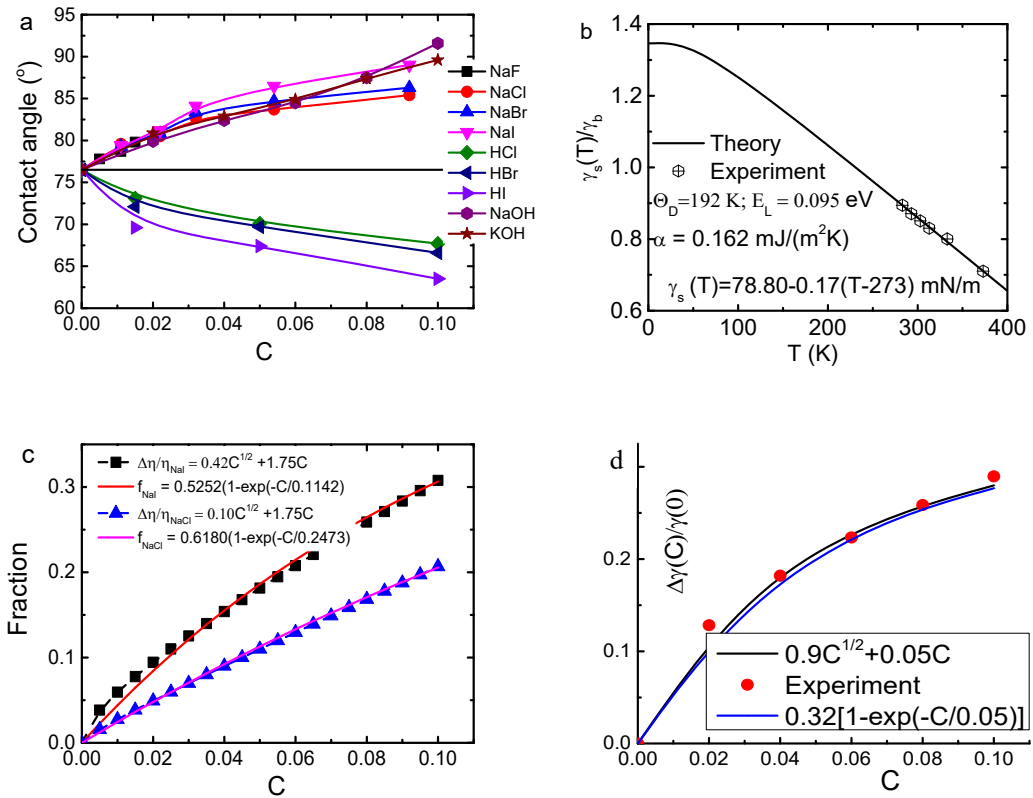

Figure 41. (a) Concentration resolved contact angles between YX, YHO and HX solutions on glass substrate [119] and (b) the thermal decay of the surface stress of liquid water [367]. Reproduction of the (c) $f_{NaCl}(C)$ and $f_{NaI}(C)$ for Na(Cl, I) solutions and (d) contact angle $\theta_{LiBr}(C)$ for LiBr solution by the Jones–Dole's expression of viscosity [369]. Reproduction of the $\gamma_S(T) \propto 1 - U(T/\Theta_{DL})/E_{coh} = 1 - U(T/192)/0.38$ turns out 192 K Debye temperature and the $E_L = 0/38/4 = 0.095$ eV for the O:H nonbond. $U(T/\Theta_{DL})$ is the integral of the Debye specific heat.

## 5.5. Summary

An extension of the conventional phonon spectroscopy to the DPS has enabled resolution of the fraction and stiffness of O:H–O bonds transiting from the mode of ordinary water to their hydration, which amplifies solvation study from the molecular dynamics in temporal and spatial domains to hydration bonding energetic dynamics. Table 13 summarizes observations of the O:H–O segmental cooperative relaxation in length, vibration frequency, and surface stress upon external excitation with respect to standard situation at 277 K.

Table 13. O:H–O segmental cooperative relaxation in length, vibration frequency, and surface stress with respect to $d_{L0}$ = 1.6946 Å, $d_{H0}$ = 1.0004 Å, $\omega_{H0}$ = 3200 cm$^{-1}$, $\omega_{L0}$ = 200 cm$^{-1}$, $\gamma_s$ = 72.5 J·m$^{-2}$ at 277 K upon excitation by heating, compression, molecular undercoordination (skin, cluster, droplet, nanobubble) and acid, base, salt, and organic molecular solvation [124, 157].

|  |  | $\Delta d_H$ | $\Delta \omega_H$ | $\Delta d_L$ | $\Delta \omega_L$ | $\Delta \gamma_s$ | Remark | Ref |
|---|---|---|---|---|---|---|---|---|
| liquid water | Liquid heating | <0 | >0 | >0 | <0 | <0 | $d_L$ elongation; $d_H$ contraction; thermal fluctuation | [158] |
|  | under-coordination |  |  |  |  | >0 | $d_H$ contraction; $d_L$ elongation; polarization; supersolidity | [129] |
|  | compression | >0 | <0 | <0 | >0 | - | $d_L$ compression; $d_H$ elongation | [351] |
| aqueous solution | YX salt | <0 | >0 | >0 | <0 | >0 | Y$^+$ and X$^-$ polarization | [343] |
|  | HX acid |  |  |  |  | <0 | H↔H fragilization; X$^-$ polarization | [341] |
|  | YOH base | >0 | >0 | >0 | >0 | >0 | O:⇔:O compression; Y$^+$ polarization; solute H–O bond contraction | [190] |

The original O:H–O bond premise and exercise have enabled development of new knowledge, resolutions to anomalies, and strategies for engineering surface water ice in resolving the multifield effect on the HB network and properties of water and Lewis-Hofmeister solutions. The DPS has allowed quantitative information on the number fraction and phonon stiffness transition from the mode of ordinary HBs to hydration. The multifield meditation of the O:H–O bonds in the network can thus be discriminated as follows:

1) Surface molecular undercoordination shortens the H–O bond from 1.00 to 0.95 Å and stiffens its

phonon from 3200 to 3450 cm$^{-1}$, meanwhile, lengthens the O:H nonbond from 1.70 to ~1.95 Å and softens the O:H phonon from 200 to 75 cm$^{-1}$. The dangling H–O bond is 0.90 Å long and its phonon frequency is 3610 cm$^{-1}$. The O:H–O bond segmental energies transit from (0.2, 4.0) to (0.1, 4.6) eV when moving from the bulk to the skin in comparison to the gaseous H–O energy of 5.10 eV. The skin mass density drops from the bulk standard of 1.0 to 0.75 g·cm$^{-3}$, which raised the skin thermal diffusivity by a factor of 4/3. Theoretical reproduction of the Mpemba paradox profiles – warm water cools more quickly, evidences the essentiality of the skin high thermal diffusivity.

2) The nonbonding electron polarization furnishes the surface with excessive charge and dipoles. The O:H–O bond relaxation and electronic polarization endowed the aqueous surface with supersolidity that is highly elastic, semi-rigid, electrostatic repulsive, hydrophobic, slippery, thermally more diffusive and stable with higher melting point $T_m$, surface stress, mechanical strength, but lower freezing $T_N$ and evaporation point $T_N$ and mass density. The degree of supersolidity increases with the surface curvature or with further lowing of the molecular coordination numbers. The supersolidity of nanodroplets, nanobubbles, and molecular clusters is higher than that of a flat surface, resulting in the super-heating/cooling/cooling of melting/freezing/evaporating. For instance, a 1.4 nm sized droplet freezes at 205 K [347] instead of the bulk at 258 K [158]. Smaller water supersolid droplet shows superfluidity, superlubricity, and superhydrophobicity, which enables its rapid flow when passing through carbon nanotubes or microchannels.

3) Strikingly, Raman spectroscopy revealed that the skins of 25 °C water and –(15~20) °C ice share H–O bonds of identical length, energy, and vibration frequency of 3450 cm$^{-1}$. X-ray K edge absorption and Raman scattering unveiled that the deformed skin H–O bond is thermally much more stable than it is in the bulk. The supersolidity of the nanodroplet skin hinders vibration energy dissipation to lengthen the H–O phonon lifetime, according to the ultrafast IR spectroscopy.

4) The O:H/H–O segmental phonon frequency softening/stiffening offset their Debye temperatures and disperses outwardly the QS phase boundaries. Hence, the supersolid skin is neither a liquid nor a solid at the ambient but rather an extended QS phase having a higher $T_m$ (~330 K), lower $T_N$ and $T_V$.

5) The skin supersolidity is engineerable by O:H–O relaxation and polarization through programed stimulation. Electrostatic polarization by a parallel electric field enhances the supersolidity, which enables the water floating bridge to form and be sustained between two filled cups. An ion forms a semi-rigid supersolid hydration volume, which separates graphene oxide sheets up to 1.5 nm for the selective ion rejection. However, a superposition of the applied field and the ionic field breaks the floating bridge of acid or salt solutions; a mixture of soil grains and aqueous ions also weakens their fields than either of them alone to accelerate soil wetting by the salt solutions. Molecular undercoordination and electrostatic polarization enhances each other to lengthen and weaken the O:H nonbond whose energy dictates the $T_N$ and $T_V$, being beneficial to evaporation and bio-product cryogenic reservation. Magnetization applies a Lorentz field to flowing water dipoles, which adds a translational/rotational component to the rotational/translational motion and hence raises the mobility of the dipoles to promote the micro-circulation of human body fluid and blood. Both liquid heating and acid solvation depress the skin supersolidity - heating softens and fluctuates the O:H nonbond but acid solvation disrupts the network through H↔H fragmentation. Base solvation creates the O:⇔:O that has the same effect of mechanical compression and polarization; O:⇔:O compression weakens the H–O bond, which emits energy heating up the solution. The H–O bond weakening by base solvation or liquid cooling eases hydrogen generation.

6) Examination of the multifield effect on surface O:H–O bonds and electrons has cultivated new knowledge and strategies of regulating water evaporation and desalination, hydrogen generation, and skin supersolidity modulation for friction and hydrophobicity, etc. Besides, the theory has enabled quantitative resolutions to anomalies such as ice buoyancy, ice regelation, critical energies for phase transition, energy exchange by H–O bond relaxation, hydration shell size and droplet skin thickness determination, etc. The theory also recommends referential means for hypertension medication and prevention, DNA fragmentation, protein denaturation, energy storage and structural stability of the energetic CNHO crystals and *cyclo*-$N_5^-$ complexes, as an extension of the theory. Systematic attainments substantially complement what could be derived from molecular spatial-temporal motion dynamics and classical statistic thermodynamics.

Regarding the aqueous solutions, we have obtained quantitative information that may complement to the premises of continuum thermodynamics, molecular dynamics:

1) $H_3O^+$ hydronium formation in acid solution creates the H↔H anti–HB that serves as a point breaker to disrupt the HBr solution network and the surface stress. The $Br^-$ polarization dictates the O:H (from 200 partially to 110 and 300 cm$^{-1}$) and the H–O phonon (from 3200 to 3480 cm$^{-1}$) frequency cooperative shift. Acid solvation has the same effect of liquid heating on the O:H–O bond network and phonon relaxation through, respectively, H↔H fragilization and thermal fluctuation.

2) $OH^-$ hydroxide forms the O:⇔:O super–HB point compressor to soften the nearest solvent H–O bond (from above 3100 cm$^{-1}$ to below), and meanwhile, the solute H–O bond shortens to its dangling radicals featured at 3610 cm$^{-1}$. The Y$^+$ polarization effect has been annihilated by the O:⇔:O compression and the solute H–O contraction. Base solvation has partly the effect of mechanical compression that lengthens and softens the H–O bond and shortens the O:H nonbond.

3) Y$^+$ and X$^-$ ions serve each as a point polarizer that aligns, stretches, and polarizes the surrounding O:H–O bonds and makes the hydration shell supersolid. The polarization transits the $\omega_L$ from 200 to 100 cm$^{-1}$ and the $\omega_H$ from 3200 to 3480 cm$^{-1}$. Salt solvation has the same effect of molecular undercoordination to form the supersolid states.

4) The solute capability of bond transition follows: $f_H(C) = 0$, $f_Y(C) \propto f_{OH}(C) \propto C$, and $f_X(C) \propto 1-\exp(-C/C_0)$ toward saturation, which indicate the nonparaxiality of protons, invariant hydration volume size of Y$^+$ and HO$^-$, and the variant X$^-$ - X$^-$ repulsion, respectively, and hence evidence the high structure order of H$_2$O molecular in the solvent matrix. The concentration trends consistent among the salt solution viscosity, surface stress, and the $f_{YX}(C)$ suggest their common origin of polarization associated with O:H–O bond transition from water to hydration shells.

5) Solvation and compression have the opposite effect on O:H–O relaxation and the shortened H–O bond by polarization can hardly elongated than the H–O bond in pure water. At constant concentration, the T$_C$ shows the Hofmeister sequences, but the involvement of the anion-anion repulsion alters the pattern of the critical pressure for the phase transition for the concentrated salt solutions.

6) Compression, confinement, and heating of the QS phase compensate one another other on the critical pressure of ice/QS transition at the ambient as compression always shortens the O:H nonbond while heating and confinement do oppositely in the QS phase of negative thermal expansion - H–O undergoes heating and compression elongation and O:H contraction but confinement does oppositely.

## 6. Concluding Remarks

The presented notion of multifield bonding dynamics and the theory-driven phonon spectrometrics have enabled the ever-deep and consistent insight into the perturbation-relaxation-property correlation of structured crystals, as summarized as follows:

1) The LBA approach and the multifield bond oscillation have reconciled the effect of P, T, CN, and charge injection on the bond relaxation and property evolution of the liquid and solid crystals. A stimulus mediates the properties by bond relaxation from one equilibrium to another.

2) Crystal size reduction creates three types of phonons unseen in the bulk – size-reduction induced $E_{2g}$ blueshift, $A_{1g}$ redshift, and the emerging of the LFR mode in the THz frequencies. Nanostructures prefer the core-shell configuration and the performance of bonds in the skin shell dictates the size dependency of liquid and solid crystals.

3) The phonon frequency shift, bandgap, and elastic modulus follows a Debye thermal decay, which offers information of the atomic cohesive energy $E_{con}$ and the Debye temperature; mechanical compression stiffens the phonon frequency nonlinearly and provides information of biding energy density and elastic modulus.

4) Strikingly, the DPS filters the fraction-stiffness transition of bonds from the reference state to the conditioned, which has derived the thickness of skin-shell thickness of liquid and solid nanocrystals up to two atomic diameters.

Progress described in this presentation recommend the following:

1) The notion of the multifield bonding dynamics may overcome limitations of the Gibbs free energy, the molecular dynamics premise, and the Grüneisen parameter, $\partial\omega/\partial x_i$, for materials properties under perturbation. External stimulus mediates the properties by relaxing the interatomic bonding and the associated energetics, densification, localization, entrapment and polarization of electrons. Gibbs energy, $dG(P, T, N, …) = SdT + VdP + \mu dN + ...$, is valid for macroscopic statistic systems and gaseous phases by taking the degrees of freedom as plain variables with S, V, µ being the entropy, volume, and chemical potentials. Molecular dynamics takes the molecules as the basic structure unit with little attention to the coupling of the intramolecular and intermolecular interactions, such as water and molecular crystals.

2) The phonon spectrometrics and the DPS is powerful yet convenient to probe the bond relaxation and transition dynamics without needing the critical high-vacuum conditions. The phonon spectrometrics is suitable for all sorts substance, liquid and solid, varying from conductors to insulators under any applied perturbations, and provides quantitative information about bonding dynamics and consistent insight in to the physics behind observations.

3) As an independent degree of freedom that laid foundation of the defect and surface science, and the nanoscience and engineering, atomic and molecular undercoordination has yet received deserved attention. Atomic CN reduction shortens and stiffens the undercoordinated atoms, which modify electrons in various energy bands, and dictate the performance of substance.

Understanding and strategies may extend to wherever the regular chemical bonds and the segmented hydrogen bonds are involved. It would be very promising for one to keep arm and mind open, and always move and look forward to developing experimental strategies and innovating theories toward resolution to the wonders.

## Nomenclature

| | |
|---|---|
| $\gamma$ | Grüneisen parameter |
| $\eta_x(T)$ | Segmental specific heat featured by the $\Theta_{DX}$ and the $E_x$ (specific heat integral) |
| BOLS | Bond order-length-strength correlation for undercoordinated atoms and molecules |
| BP | Black phosphorus |
| CN | Coordination number, z |
| DFT | Density functional theory |
| DPS | Differential phonon spectrometrics |
| $E_{coh}$ | Cohesive energy |
| $E_{den}$ | Binding energy density |
| $E_g$ | Bandgap; Raman translational vibration mode |
| FTIR | Fourier transformation of infrared spectroscopy |
| FWHM | Full width at half maximum |
| GNR | Graphene nanoribbon |
| LBA | Local bond average |
| LFR | Low-frequency Raman mode |
| MD | Molecular dynamics |
| $P_c/T_c$ | Critical pressure/temperature for phase transition |
| $P_C/T_C$ | Critical pressure/temperature for phase transition |
| SWCNT | Single-walled carbon nanotube |
| TO/LO | Translational/Longitudinal optical wave |
| O:H-O | Hydrogen bond |
| H↔H | anti-hydrogen bond |
| O:⇔:O | super hydrogen bond |

## Acknowledgement

Financial support from the National Natural Science Foundation (Nos. 11602094; 11872052) is all gratefully acknowledged.

295. F.G. Shi, *Size-dependent thermal vibrations and melting in nanocrystals.* Journal of Materials Research, **9**(5): 1307-1313, (1994).
296. Q. Jiang, Z. Zhang, and J.C. Li, *Superheating of nanocrystals embedded in matrix.* Chemical Physics Letters, **322**(6): 549-552, (2000).
297. M.X. Gu, Y.C. Zhou, L.K. Pan, Z. Sun, S.Z. Wang, and C.Q. Sun, *Temperature dependence of the elastic and vibronic behavior of Si, Ge, and diamond crystals.* Journal of Applied Physics, **102**(8): 083524, (2007).
298. H. Herchen and M.A. Cappelli, *1st-order Raman-spectrum of diamond at high-temperatures.* Physical Review B, **43**(14): 11740-11744, (1991).
299. E.S. Zouboulis and M. Grimsditch, *Raman-scattering in diamond up to 1900-K.* Physical Review B, **43**(15): 12490-12493, (1991).
300. D.A. Czaplewski, J.P. Sullivan, T.A. Friedmann, and J.R. Wendt, *Temperature dependence of the mechanical properties of tetrahedrally coordinated amorphous carbon thin films.* Applied Physics Letters, **87**(16): 2108132, (2005).
301. J. Menendez and M. Cardona, *Temperature-dependence of the 1st-order raman-scattering by phonons in Si, Ge, and a-Sn - anharmonic effects.* Physical Review B, **29**(4): 2051-2059, (1984).
302. M.E. Fine, *Elasticity and thermal expansion of germinium between 195 deg-C 275 deg-C.* Journal of Applied Physics, **24**(3): 338-340, (1953).
303. U. Gysin, S. Rast, P. Ruff, E. Meyer, D.W. Lee, P. Vettiger, and C. Gerber, *Temperature dependence of the force sensitivity of silicon cantilevers.* Physical Review B, **69**(4): 045403, (2004).
304. P. Perlin, A. Polian, and T. Suski, *Raman-scattering studies of aluminum nitride at high-pressure.* Physical Review B, **47**(5): 2874-2877, (1993).
305. M. Kuball, J.M. Hayes, A.D. Prins, N.W.A. van Uden, D.J. Dunstan, Y. Shi, and J.H. Edgar, *Raman scattering studies on single-crystalline bulk AlN under high pressures.* Applied Physics Letters, **78**(6): 724-726, (2001).
306. M. Ueno, A. Onodera, O. Shimomura, and K. Takemura, *X-ray-observation of the structural phase-transition of aluminium nitride under high-pressure.* Physical Review B, **45**(17): 10123-10126, (1992).
307. M. Kuball, J.M. Hayes, Y. Shi, J.H. Edgar, A.D. Prins, N.W.A. van Uden, and D.J. Dunstan, *Raman scattering studies on single-crystalline bulk AlN: temperature and pressure dependence of the AlN phonon modes.* Journal of Crystal Growth, **231**(3): 391-396, (2001).
308. M.P. Halsall, P. Harmer, P.J. Parbrook, and S.J. Henley, *Raman scattering and absorption study of the high-pressure wurtzite to rocksalt phase transition of GaN.* Physical Review B, **69**(23): 235207, (2004).
309. P. Perlin, C. Jauberthiecarillon, J.P. Itie, A. San Miguel, I. Grzegory, and A. Polian, *Raman-scattering and x-ray-absorption spectroscopy in gallium nitride under high-pressure.* Physical Review B, **45**(1): 83-89, (1992).
310. S. Limpijumnong and W.R.L. Lambrecht, *Homogeneous strain deformation path for the wurtzite to rocksalt high-pressure phase transition in GaN.* Physical Review Letters, **86**(1): 91-94, (2001).
311. A. Link, K. Bitzer, W. Limmer, R. Sauer, C. Kirchner, V. Schwegler, M. Kamp, D.G. Ebling, and K.W. Benz, *Temperature dependence of the E-2 and A(1)(LO) phonons in GaN and AlN.* Journal of Applied Physics, **86**(11): 6256-6260, (1999).
312. C. Pinquier, F. Demangeot, J. Frandon, J.C. Chervin, A. Polian, B. Couzinet, P. Munsch, O. Briot, S. Ruffenach, B. Gil, and B. Maleyre, *Raman scattering study of wurtzite and rocksalt InN under high pressure.* Physical Review B, **73**(11): 115211, (2006).
313. X.D. Pu, J. Chen, W.Z. Shen, H. Ogawa, and Q.X. Guo, *Temperature dependence of Raman scattering in hexagonal indium nitride films.* Journal of Applied Physics, **98**(3): 2006208, (2005).
314. Y.L. Du, Y. Deng, and M.S. Zhang, *Variable-temperature Raman scattering study on anatase titanium dioxide nanocrystals.* J Phys Chem Solids, **67**(11): 2405-2408, (2006).
315. V. Swamy, A.Y. Kuznetsov, L.S. Dubrovinsky, A. Kurnosov, and V.B. Prakapenka, *Unusual Compression Behavior of Anatase $TiO_2$ Nanocrystals.* Phys Rev Lett, **103**(7): 75505, (2009).
316. A.Y. Kuznetsov, R. Machado, L.S. Gomes, C.A. Achete, and V. Swamy, *Size dependence of rutile $TiO_2$ lattice parameters determined via simultaneous size, strain, and shape modeling.* Appl Phys Lett, **94**: 193117, (2009).
*126*